\documentclass[preprint]{elsarticle}

\usepackage{hyperref}

\journal{New Astronomy Reviews}

\usepackage{amssymb}

\def\deg{\hbox{${}^\circ$}}

\biboptions{authoryear}

\begin{document}

\begin{frontmatter}

\title{High Energy Polarimetry of Prompt GRB Emission\tnoteref{t1}}
\tnotetext[t1]{@2016. This manuscript version is made available under the CC-BY-NC-ND 4.0 license http://creativecommons.org/licenses/by-nc-nd/4.0/} 

\author[unh,swri]{Mark L. McConnell}
\ead{mark.mcconnell@unh.edu}
\address[unh]{Space Science Center, University of New Hampshire, Durham, NH  03824, USA}
\address[swri]{Department of Earth, Oceans, and Space, Southwest Research Institute, Durham, NH  03824, USA}

\begin{abstract}
Evidence of polarized $\gamma$-ray emission ($>$ 50 keV) from Gamma-Ray Bursts (GRBs) has been accumulated in recent years. Measurements have been reported with levels in the range of 30-80\%, typically with  limited statistical significance. No clear picture has yet emerged with regards to the polarization properties of GRBs. Taken at face value, the data suggest that most GRBs have a relatively large level of polarization (typically, $> 50\%$), which may suggest synchrotron emission associated with an ordered magnetic field structure within the GRB jet.  But these results are far from conclusive. Here, we review the observations that have been made, concentrating especially on the instrumental issues and the lessons that might be learned from these data.
\end{abstract}


\end{frontmatter}


\section{Introduction}

The process by which some stellar-mass black holes are thought to form (either from the final stages of a highly evolved, massive star or the merger of two compact objects) results in a release of energy that exceeds anything observed in the Universe since the Big Bang itself. This energy release results in the formation of two oppositely directed jets, which can observationally manifest itself as a Gamma Ray Burst (GRB) and its afterglow emission. The initial burst of $\gamma$-rays, the so-called prompt emission, lasts from a fraction of a second up to a few hundred seconds and is thought to originate in the innermost region of the jets. The longer-lasting afterglow emission, lasting from days to weeks, is believed to originate in the outer part of the jet.  This emission has been well studied across the entire electromagnetic spectrum, providing a better understanding of the late stages of the jet evolution, as it interacts with the surrounding medium. However, a complete picture of the GRB phenomena also requires an understanding of the inner part of the jet, closest to where the black hole is formed. At this time, we have only a limited understanding of the inner jet, as it depends on the short-lived, high-energy prompt emission, which is far more difficult to study given the random nature of these sources. Theoretical modeling argues that a more complete understanding of the inner structure of GRBs, including the geometry and physical processes close to the central engine, requires the exploitation of X-ray and $\gamma$-ray polarimetry. Efforts to measure the high energy polarization have so far met with limited success. As several efforts are currently underway to collect more definitive measurements, it is worthwhile to review the available results  in the context of past experimental efforts.

\section{GRB Theory}

GRBs are distributed isotropically across the sky with a $4\pi$ occurrence rate of roughly two every day, based on observations by both \emph{CGRO/BATSE} and \emph{Fermi/GBM} \citep{1999ApJS..122..465P,2012ApJS..199...18P}. Burst durations range from $<$ 10 ms up to several hundred seconds \citep{1999ApJS..122..465P}. Long-duration bursts ($>$ 2 s) are believed to be associated with the death of massive stars, whereas short-duration bursts ($<$ 2 s) are believed to be associated with the merger of compact star binaries (neutron star-neutron star, neutron star-black hole, etc.). Regardless of the progenitor, a generic ``fireball'' shock model \citep[e.g.,][]{Piran:2005cs,2006RPPh...69.2259M} suggests that a relativistic jet is launched from the center of the explosion. The ``internal'' dissipation within the fireball (e.g., via internal shocks or internal magnetic dissipation processes) leads to emission in the X-ray and $\gamma$-ray band, which corresponds to the observed  prompt emission. Observationally, the canonical  prompt emission spectrum can often be empirically fit by the so-called {\it Band function} \citep{1993ApJ...413..281B}, consisting of a broken power-law with a smooth break at a characteristic energy, commonly referred to as {\it E-peak} ($E_p$). The $E_p$ value corresponds to the peak of the spectrum when plotted in terms of energy output per decade of energy ($E^2 N_E$). The observed distribution of $E_p$  values range from $\sim10$ keV up to at least 1 MeV, with a broad peak near 200 keV. As there is no physical basis for this spectral model, the precise nature of the emission is not well determined. Although synchrotron emission is believed to play a significant role \citep[e.g.,][]{1994ApJ...430L..93R}, many aspects of the emission can also be explained by inverse Compton emission \citep[e.g.,][]{2003ApJ...596L.147E,1995ApJ...447..863S}. Additionally, thermal emission from an expanding photosphere appears to be an important component in some GRBs \cite[e.g.,][]{2014MNRAS.440.3292L}. Eventually, the outflowing jet is decelerated by the circumburst medium, which leads to a long-lasting forward shock. Emission from the external shock is believed to be responsible for the afterglow following the burst. Much has been learned about these afterglows, but little progress has been made in understanding the physical origin of the prompt emission.

Although polarimetry of the prompt $\gamma$-ray emission is expected to provide useful insights, current measurements furnish only very limited constraints for theoretical modeling of the prompt emission. Consequently, there are a large number of models that seek to explain the available polarization measurements. These models can often be characterized as falling into one of two general classes  of models: intrinsic and extrinsic  \citep{2003Natur.423..388W,2006NJPh....8..131L}.

\emph{Intrinsic models} invoke a globally ordered magnetic field in the emission region, with electron synchrotron emission yielding a net linear polarization \citep[e.g.,][]{2003Natur.423..388W,2003ApJ...597..998L,2003ApJ...596L..17G}. In this case, the polarization properties are derived from the intrinsic characteristics (i.e., the magnetic field geometry) of the jet. The model applies for most observer viewing-angle geometries, with typical levels of polarization ($\Pi$) ranging from  $\sim20$\% up to $\sim$60\%. These models are characterized by a highly magnetized jet composition, reconnection as the most possible dissipation mechanism, and synchrotron radiation as the emission mechanism. 

\emph{Geometric Models} models require an optimistic viewing direction (or geometry) to observe a high degree of polarization. The magnetic field structure is random in the emission region, so that no net polarization is detected if the viewing angle is along the jet beam (regardless of radiation mechanism). However, if the viewing direction is near the edge of the jet, in particular about 1/$\Gamma$ outside the jet cone (where $\Gamma$ is the bulk Lorentz factor of the outflow), a high polarization degree results due to loss of emission symmetry \citep[e.g.,][]{1995ApJ...447..863S,2004MNRAS.347L...1L}. This model is characterized by a matter-dominated outflow and shocks as the most likely dissipation mechanism. Both synchrotron and inverse Compton can be the radiation mechanisms. The typical polarization is $\Pi < 20\%$ for most viewing angles, although synchrotron emission can produce $\Pi$ as high as $\sim$70\%, and inverse Compton models  \citep[also known as Compton drag models; ][]{2004MNRAS.347L...1L} can achieve $\Pi$ $\sim$100\% under optimistic geometries.

A statistical study of GRB polarization properties could differentiate between the two classes of models (intrinsic vs. geometric) and, in some cases, distinguish between models within a class, providing a direct diagnostic of the magnetic  field structure, radiation mechanism, and geometric configuration of GRB jets. The distribution of polarization values (assuming random viewing angles) has been studied for three generalized models that characterize the jet physics \citep{Toma:2009fv}. The three principle models include:  a) an intrinsic model for synchrotron emission with ordered B-fields (SO); b) a geometric model for synchrotron emission in random B-fields (SR); and c) a geometric model for Compton-drag (CD). 

\begin{figure}
\begin{center}
\includegraphics[width=3.5in]{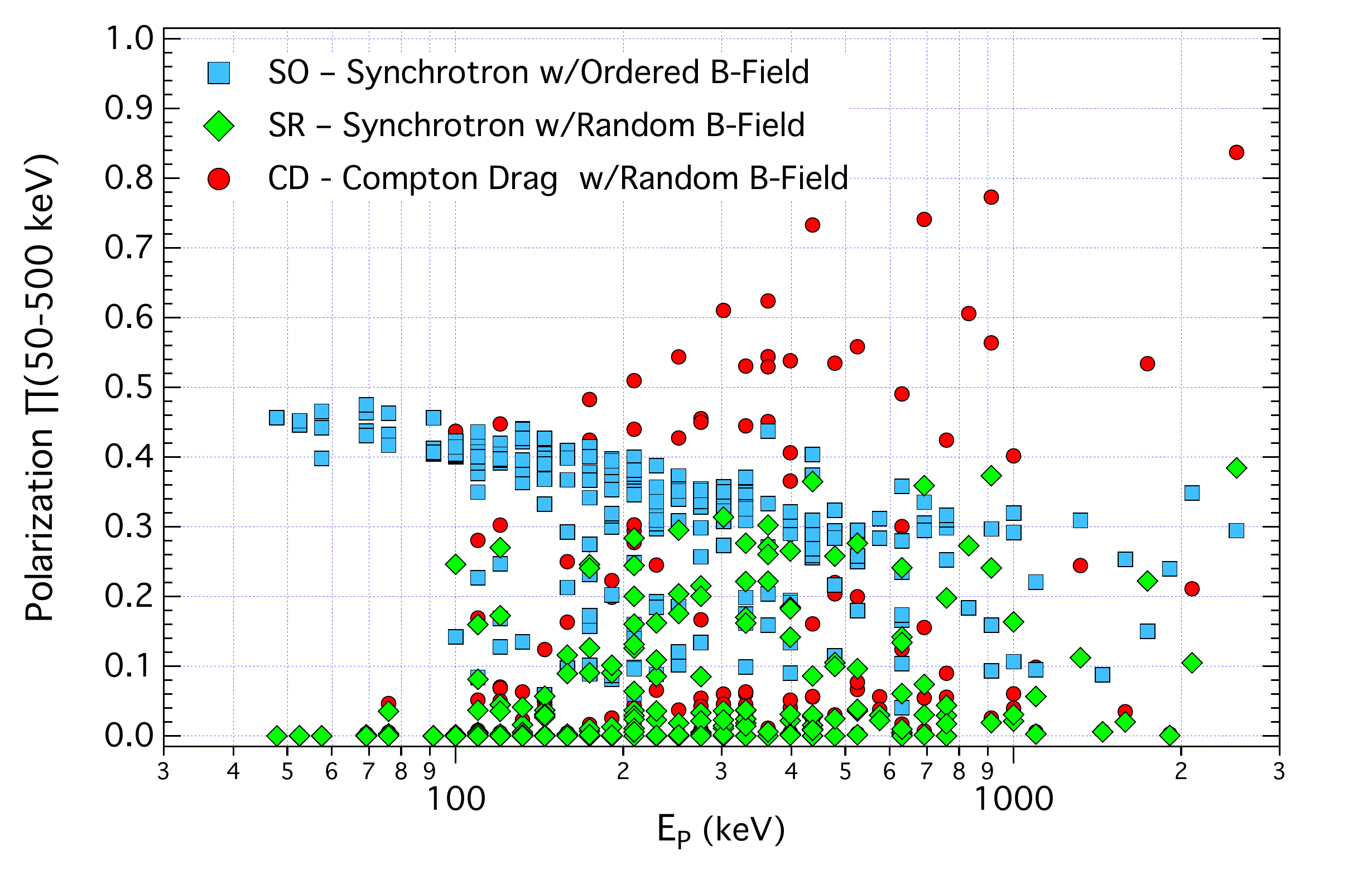}
\caption{Distribution of predicted polarization values (as measured in the 50--500 keV energy range) as a function of $E_p$ for three models for the jet physics. [Figure adapted from \citet{Toma:2009fv}.]}
\label{Toma}
\end{center}
\end{figure}

Each model predicts a different value for the maximum possible polarization ($\Pi_{max}$), so the largest observed values of $\Pi$ already constrains the models. For example, the fraction of bursts exhibiting a high $\Pi$ is significantly smaller in the geometric models than in the intrinsic models. A more powerful diagnostic is the distribution of $\Pi$ as a function of spectral peak energy ($E_p$), as can be seen in Fig. \ref{Toma}. Some models show very distinctive structure in this parameter space, such as the correlation between $E_p$ and $\Pi$ for the SO model. 

The nature of the GRB radiation mechanism(s) can also be derived from the energy-dependence of the polarization. Although it is generally believed that synchrotron emission contributes significantly to the spectrum, other mechanisms, such as thermal blackbody emission from the expanding photosphere \citep{2014MNRAS.440.3292L,2005ApJ...625L..95R} or inverse Compton emission \citep[e.g.,][]{2003ApJ...596L.147E,1995ApJ...447..863S}, may also play an important role. The relative importance of  two mechanisms may be discernible with energy-dependent polarization measurements, since the various components have distinct polarization signatures. 

The temporal evolution of polarization properties also carries essential information with which to diagnose the GRB mechanism. For example, in the Internal-Collision induced MAgnetic Reconnection and Turbulence (ICMART) model of GRBs \citep{2011ApJ...726...90Z}, each broad pulse in the GRB light curve is related to one event that destroys the ordered magnetic  fields to produce radiation. One therefore expects a decrease of the polarization with time across the broad pulse. For bursts with multiple broad pulses, one expects a possible variation of $\Pi$ with time that is broadly correlated with the light curve. For the geometric models, the polarization degree is determined by the viewing angle direction, and is not expected to vary significantly, but may undergo a $180\deg$  flip in the photosphere model \citep{2014MNRAS.440.3292L}, or show a periodic pattern if the jet precesses.

\section{Compton Polarimetry}

In the energy range that corresponds to the peak output of GRBs (10 keV up to several MeV), Compton scattering is the dominant mechanism for photon detection. Compton scattering is also an effective means for polarization measurements, which takes advantage of the fact that photons will tend to scatter at right angles to the incident electric field (polarization) vector. The experimental challenge is to accurately measure the azimuthal distribution of scatter  angles about the incident photon direction. Any net polarization results in a nonuniform distribution of azimuthal scattering angles. The cross-section for Compton scattering is given by the Klein-Nishina formula, 

	\begin{equation}
	\label{eq:1}
d\sigma = {r_{0}^{2} \over 2} d\Omega \left({\nu' \over \nu_o}\right)^2 
\left({\nu_o \over \nu'} + {\nu' \over \nu_o} - 2 \sin^2\theta \cos^2\eta \right)
	\end{equation}

\noindent where,

	\begin{equation}
	\label{eq:2}
{\nu' \over \nu_o} = {1 \over 1 + \left(h\nu_o \over mc^2 \right) \left(1 - \cos\theta \right)}
	\end{equation}

\noindent Here $\nu_o$ is the frequency of the incident photon,   $\nu'$ is the frequency of the scattered photon, $\theta$ is the Compton scatter angle of the scattered photon measured from the direction of the incident photon, and $\eta$ is the azimuthal scatter angle, measured from the plane containing the electric vector of the incident photon.  For a given value of $\theta$, the scattering cross section for polarized radiation reaches a minimum at $\eta = 0^{\circ}$ and a maximum at $\eta = 90^{\circ}$.  In other words, photons tend to be scattered at a right angle with respect to the incident electric field vector. In the case of an unpolarized beam of incident photons, there will be no net positive electric field vector and therefore no preferred azimuthal scattering angle ($\eta$); the distribution of scattered photons with respect to $\eta$ will be uniform.  However, in the polarized case, the incident photons will exhibit a net positive electric field vector and the distribution in $\eta$ will be asymmetric.  The asymmetry will be most pronounced for Compton scatter angles near $\theta = 90^{\circ}$.  
(Strictly speaking, the Klein-Nishina cross section applies to free electrons. In an actual detector, scattering on bound electrons introduces small corrections to the scattering angles. Since these affects are minor and do not impact  the discussion on  polarization,  we ignore them here.)

A Compton scatter polarimeter typically consists of two distinct detectors to determine the energies of both the scattered photon and the scattered electron \citep[e.g.,][]{1997SSRv...82..309L}. One detector, the scattering detector, provides the medium for the Compton interaction to take place.  The primary purpose of the second detector, the calorimeter, is to absorb the full energy of the scattered photon. The relative location of the two detectors provides a measure of the scatter angle.

\begin{figure}
\begin{center}
\includegraphics[width=3in]{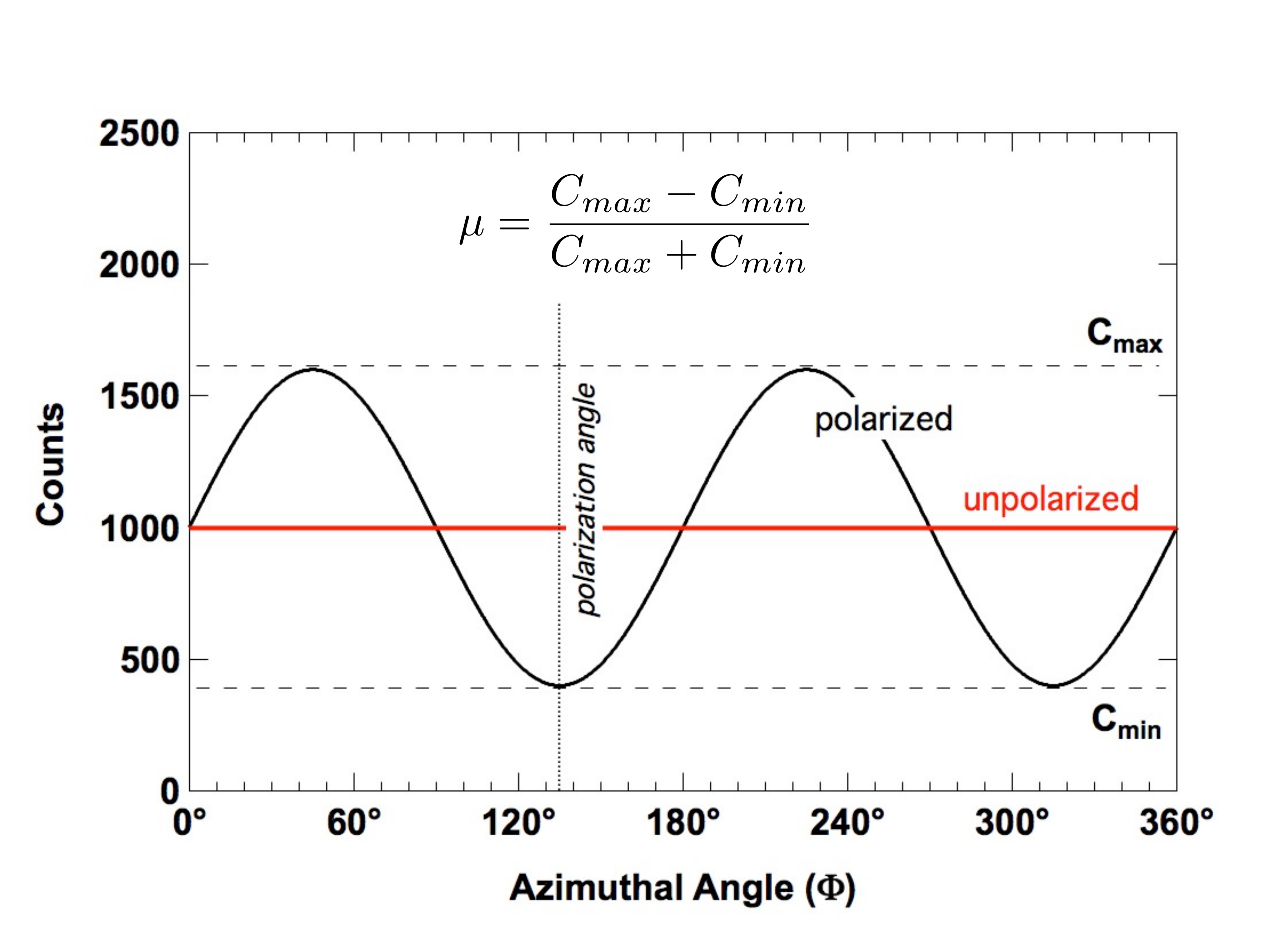}
\caption{Idealized modulation patterns measured by a Compton polarimeter for both a polarized and an unpolarized incident beam.}
\label{ModPattern}
\end{center}
\end{figure}

The ultimate goal of a Compton scatter polarimeter is to measure the azimuthal modulation pattern of the scattered photons.  From Eqn. \ref{eq:1}, we see that the azimuthal modulation follows a $\cos^{2}\eta$ distribution.  More specifically, we can write the integrated azimuthal distribution of the scattered photons as,

	\begin{equation}
	\label{eq:3}
C(\eta) = A + B \cos^{2} (\eta - \phi )
	\end{equation}

\noindent where $\phi$ is the polarization angle of the incident photons; $A$ and $B$ are constants used to fit the modulation pattern (Fig.~\ref{ModPattern}). In practice, a measured distribution must also be corrected for geometrical effects based on the corresponding distribution for an unpolarized beam \citep{1997SSRv...82..309L}.  The quality of the polarization signal is quantified by the polarization modulation factor.  For a given energy and incidence angle for an incoming photon beam, this can be expressed as \citep[e.g.,][]{1997SSRv...82..309L},

	\begin{equation}
	\label{eq:4}
\mu_p = { C_{p,max} - C_{p,min} \over C_{p,max} + C_{p,min}}  = {B_p \over {B_p + 2 A_p}}
	\end{equation}

\noindent where $C_{p,max}$ and $C_{p,min}$ refer to the maximum and minimum number of counts registered in the polarimeter, respectively, with respect to $\eta$; $A_p$ and $B_p$ refer to the corresponding parameters in Eqn. \ref{eq:3}. In this case the $p$ subscript denotes that this refers to the measurement of a source with unknown polarization. In order to determine the polarization of the measured flux, we need first to know how the polarimeter would respond to a similar flux with 100\% polarization.  This can be done using Monte Carlo simulations. We then define a corresponding modulation factor for an incident flux that is 100\% polarized,
	
	\begin{equation}
	\label{eq:5}
\mu_{100} = { C_{100,max} - C_{100,min} \over C_{100,max} + C_{100,min}}  = {B_{100} \over {B_{100} + 2 A_{100}}}
	\end{equation}

This result, in conjunction with the observed modulation factor ($\mu_p$), is used to determine the level of polarization in a measured beam \citep{1975SSRv...18..389N},
 
	\begin{equation}
	\label{eq:6}
P = { \mu_p \over \mu_{100} }
	\end{equation}

\noindent where $P$ is the measured polarization. At the 99\% confidence level, the minimum detectable polarization (MDP) can be expressed as \citep[e.g.,][]{2011APh....34..784K,2010SPIE.7732E..11W},

	\begin{equation}
	\label{eq:7}
MDP(\%) = {4.29 \over \mu_{100} R_{src} } \sqrt{R_{src} + R_{bgd} \over T}
	\end{equation}

\noindent where  $R_{src}$ is the total source counting rate, $R_{bgd}$ is the total background counting rate and $T$ is the total observation time. This defines the polarization sensitivity for a given measurement, based on the  characteristics of the instrument (as defined by $\mu_{100}$), the instrumental background ($R_{bgd}$), the source counting rate ($R_{src}$), and the observation time interval ($T$).

As an aside, this formalism can be related to the more traditional Stokes parameters,

\begin{equation}
I = A + {B \over 2}
\end{equation}  

\begin{equation}
Q = {A  \over 2} cos(2\phi)
\end{equation}  

\begin{equation}
U = {A  \over 2} sin(2\phi)
\end{equation}  

\noindent where $A$ and $B$ correspond to the definition in Eqn. \ref{eq:3}. Note that Compton polarimetry is sensitive to only linear polarization, so that the Stokes parameter $V$ is zero.

\section{Instrumentation for $\gamma$-Ray Polarimetry}
\label{instruments}

In this section, we describe the experiments that have so far made important contributions to the study of GRB polarization. The intent here is to provide the necessary background for a better understanding of the available measurements, which will be reviewed in section \ref{measurements}.

\subsection{Compton Gamma Ray Observatory (CGRO)}

CGRO was a mission designed to study the sky over a broad range of  $\gamma$-ray energies using a suite of four separate instruments. CGRO was launched by the Space Shuttle Atlantis on April 5, 1991 and was de-orbited (due to concerns related to the loss of spacecraft gyroscopes) on June 4, 2000. Two of the four instruments on CGRO were, in principle, capable of making polarization measurements.

\begin{figure}
\begin{center}
\includegraphics[width=4in]{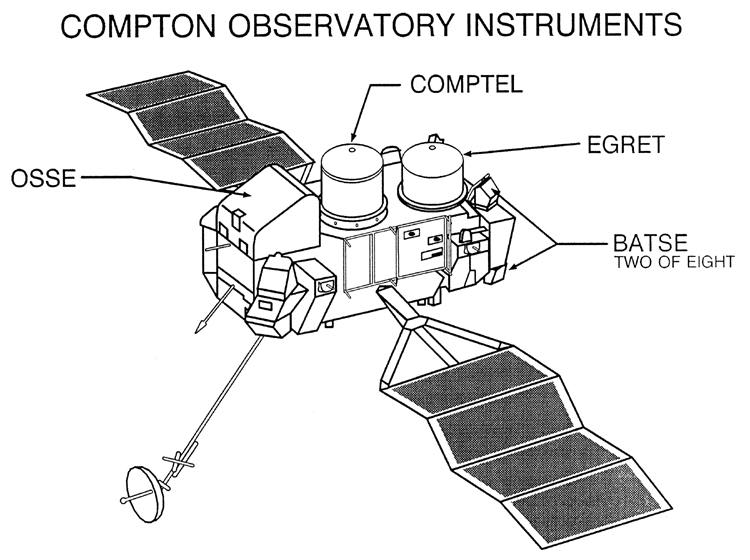}
\caption{The Compton Gamma Ray Observatory (CGRO) carried a suite of four instruments designed to cover a wide range of $\gamma$-ray energies.  Two of these instruments (COMPTEL and BATSE) could be used for polarization measurements. }
\label{CGRO}
\end{center}
\end{figure}

The Compton Imaging Telescope (COMPTEL) was designed to image  $\gamma$-rays in the 1--30 MeV energy range \citep{1993ApJS...86..657S}. It consisted of two detector layers.  The upper (D1) detector layer consisted of seven large volume liquid scintillator (NE 213A) tanks (28.5 cm in diameter by 8.5 cm thick), each read out by an array of eight PMTs.  The lower (D2) detector layer consisted of fourteen large volume NaI(Tl) scintillator crystals (28.2 cm in diameter and 7.5 cm thick) read out by an array of seven PMTs. In each case, signals from the PMT array were used to localize the event interactions to within 1-2 cm. The two detector layers (each surrounded by plastic anticoincidence shielding) were separated by distance of 1.5 m. Both pulse shape measurements in D1 and time-of-flight measurements between D1 and D2 were used to help identify source photons.

\begin{figure}
\begin{center}
\includegraphics[width=3in]{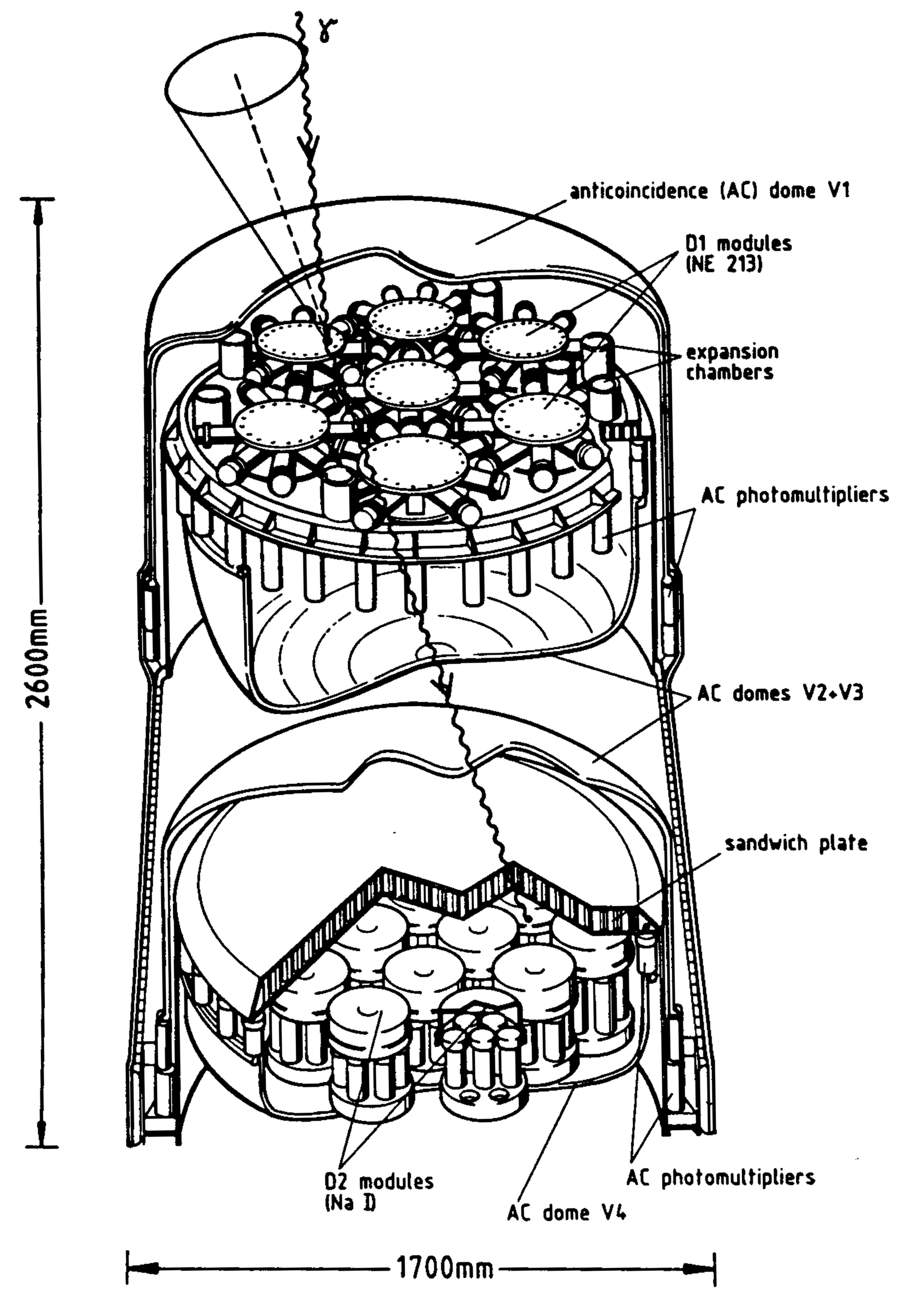}
\caption{The COMPTEL instrument used Compton imaging for 1-30 MeV $\gamma$-rays, but had limited polarization sensitivity.}
\label{COMPTEL}
\end{center}
\end{figure}

Although it employed Compton scattering as a detection mechanism, and was therefore inherently sensitive to polarization, the separation of the D1 and D2 layers was chosen to facilitate the use of time-of-flight for background reduction. The detector geometry was not optimized for polarization measurements and the resulting polarization sensitivity was quite poor. Nonetheless, an initial effort was made to measure GRB polarization , but without success \citep{1996A&AS..120C.695L}. A more thorough  polarization study of all COMPTEL GRB  data is currently in progress \citep{2016HEAD...1511204M}.

The Burst and Transient Source Experiment (BATSE) was specifically designed as an all-sky monitor for GRBs \citep{1989SPIE.1159..156P}. It consisted of eight detector assemblies, one on each corner of the CGRO spacecraft. Each detector assembly included two uncollimated detectors -- a Large Area Detector (LAD) and a  small area Spectroscopy Detector (SD). Each LAD contained a single NaI(Tl) scintillator that was 50.8 cm in diameter by 1.27 cm thick and read out by three 12.7 cm PMTs.  The large diameter-to-thickness ratio was intended to provide an angular response that, when coupled with the response of adjacent LADs, would enable a localization of the burst direction. Each SD detector contained a much smaller (but thicker) NaI(Tl) scintillator that was 12.7 cm in diameter and 7.62 cm thick. These detectors were intended to provide spectroscopy extending to much higher energies than the LADs. 

\begin{figure}
\begin{center}
\includegraphics[width=4in]{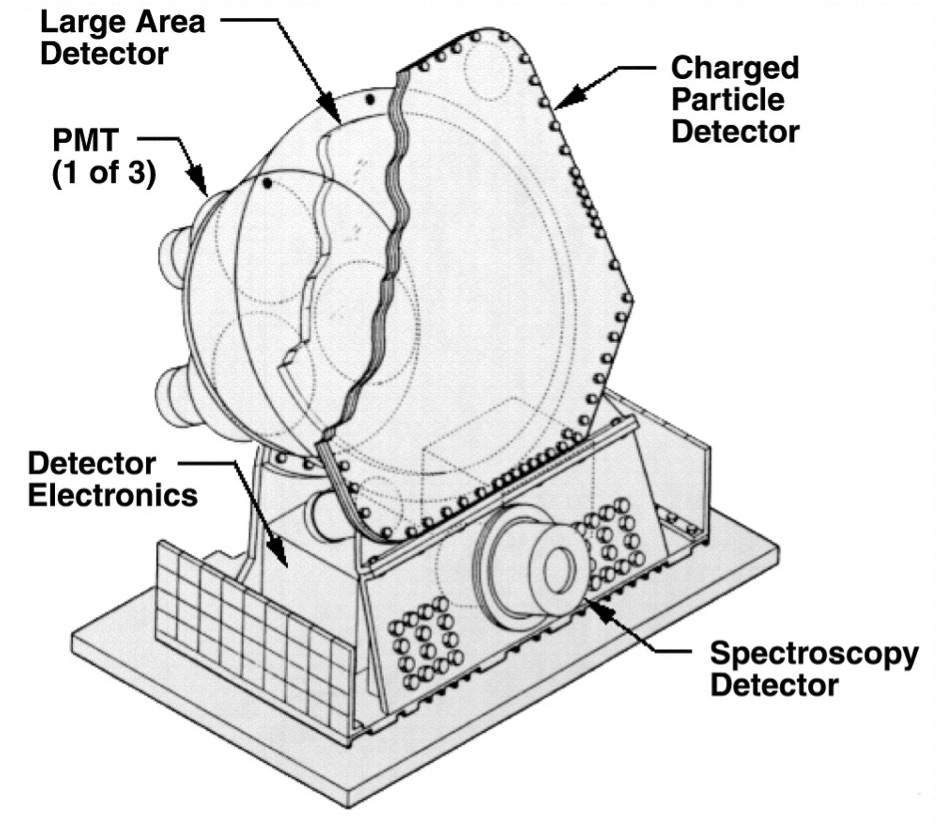}
\caption{A BATSE detector assembly was mounted at each corner of the the CGRO spacecraft, providing a full coverage of the un-occulted sky at all times.}
\label{BATSE}
\end{center}
\end{figure}

The BATSE detectors were not directly sensitive to the polarization of the incident flux. Instead, a rather unique approach to polarimetry was considered. Instead of using the BATSE detectors looking directly at the GRB, the downward looking BATSE detectors were used to  measure the albedo flux of photons scattered off  the Earth's atmosphere \citep{1996AIPC..384..851M}. Since the intrinsic polarization of the GRB flux will influence the photon scattering direction, the observed distribution of albedo flux is sensitive to the polarization parameters of the incident flux. Therefore, a measurement of the albedo distribution can provide a measure of the source polarization. The observational challenge is exacerbated by the fact that the time delay between the direct GRB flux and the scattered albedo flux is insufficient to distinguish experimentally. Ideally, the distribution of albedo flux across the disk of the Earth would be measured with an imaging instrument. The uncollimated BATSE detectors could provide only a very crude measure of this distribution, but one that was used to place constraints on GRB polarization parameters.

\subsection{RHESSI}

\begin{figure}
\centering
\includegraphics[height=1.7in]{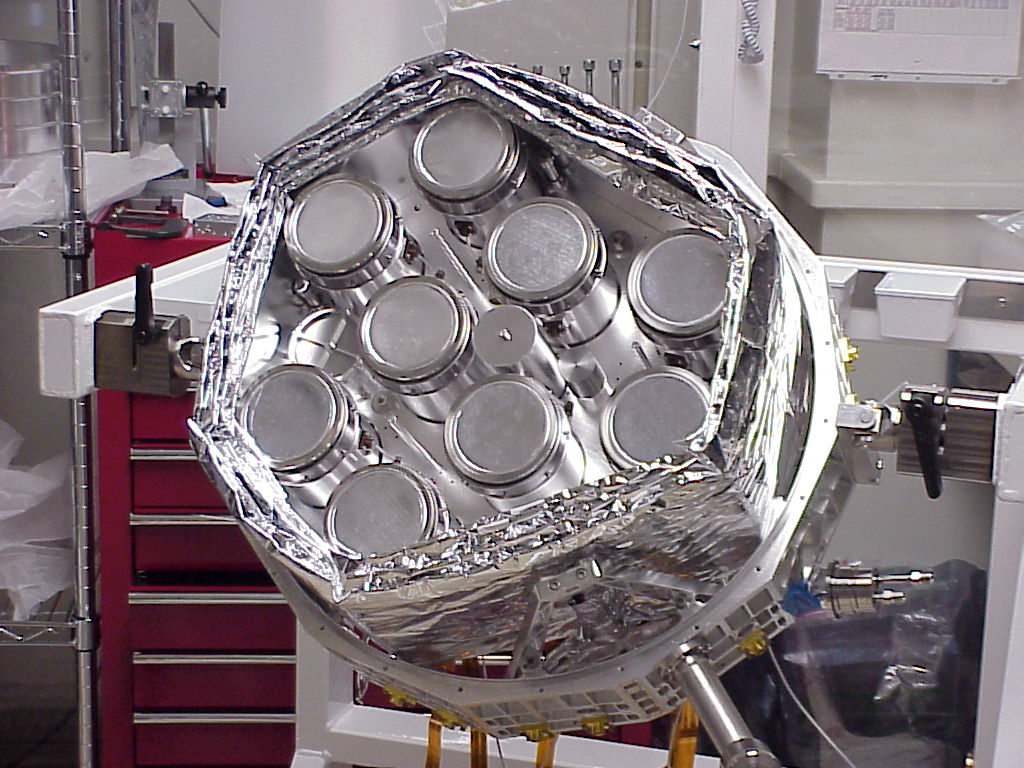}
\includegraphics[height=1.7in]{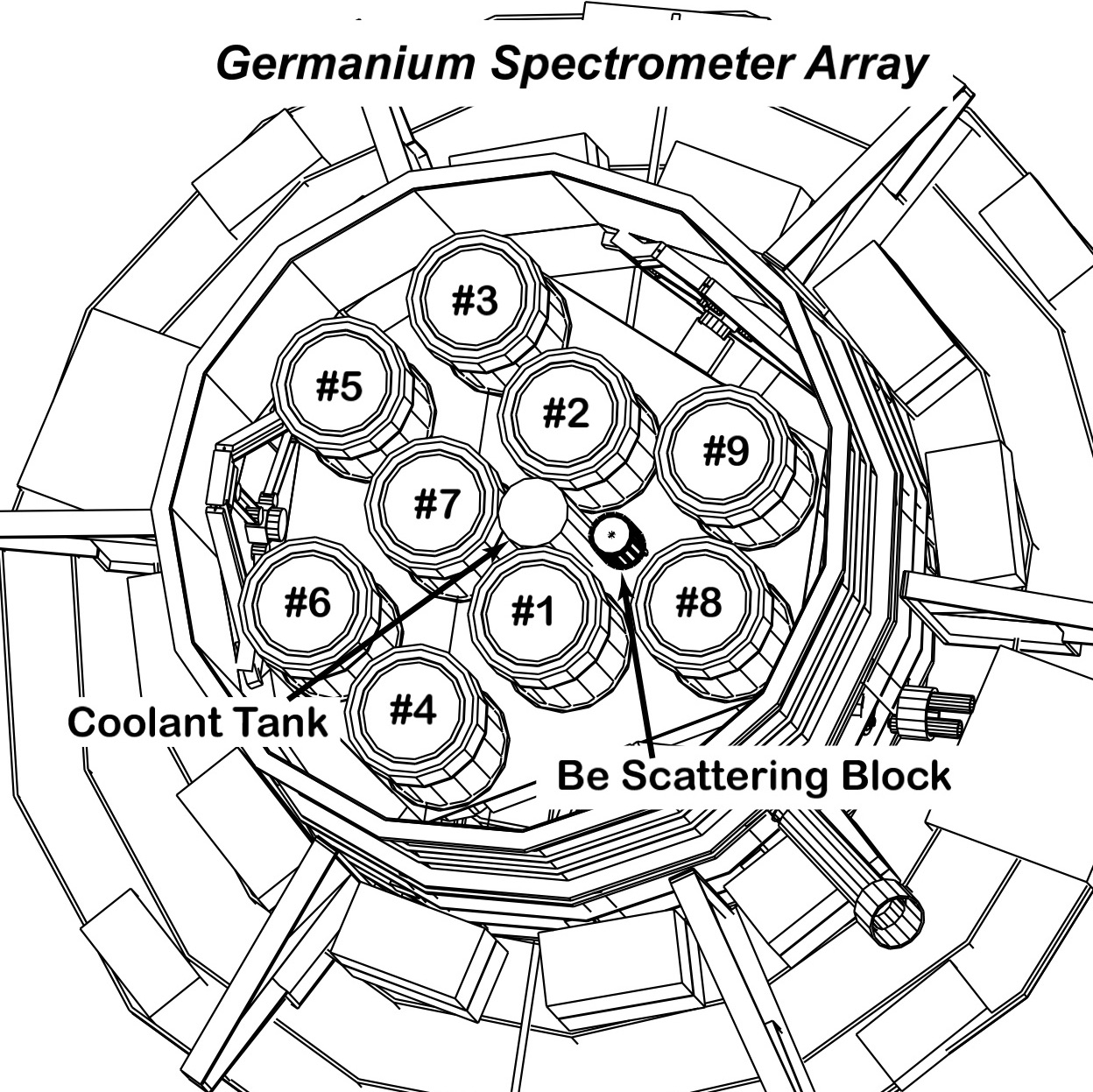}
\caption{GRB polarimetry measurements have been made using events that involve photon scattering between Ge detector elements within the detector array (seen here).  [Figures from \cite{2002SoPh..210..125M}.]}
\label{RHESSI} 
\end{figure}

The Ramaty High-Energy Solar Spectroscopic Imager (RHESSI) was designed to image solar photons covering the energy range of 3 keV to 17 MeV \citep{2002SoPh..210....3L}. Launched in 2002, it remains operational. Arcsec imaging of the solar disk is achieved with an imaging system made up of nine Rotating Modulation Collimators (RMCs), each consisting of a pair of widely separated grids mounted on the rotating spacecraft. The spectrometer (Figure~\ref{RHESSI}) consists of nine segmented Germanium detectors, one behind each RMC \citep{2002SoPh..210...33S}. Each Ge detector is 7.1 cm in diameter and 8.5 cm long). The detectors are cooled to $\sim75^{\circ}$ K by a space-qualified long-life mechanical cryocooler.  Each detector is made from a single germanium crystal, which is electrically divided into independent front and rear segments to provide an optimum response for low and high energy photons, respectively. This provides the equivalent of a $\sim1$ cm thick planar Ge detector (the front segment) in front of a thick $\sim7$ cm coaxial Ge detector (the rear segment). 

The front segment thickness is chosen to stop photons incident from the front (solar-facing side) of the instrument up to $\sim100$ keV, where photoelectric absorption dominates, while minimizing the active volume for background. Front-incident photons that Compton-scatter, and background photons or particles entering from the rear, are rejected by anticoincidence with the rear segment.  A passive, graded-Z ring around the front segment (a laminate of Ta/Sn/Fe) absorbs hard X-rays incident from the side, to provide the low background of a phoswich-type scintillation detector.

RHESSI is a spinning spacecraft, with a spin rate of $\sim15$ rpm. The energy and arrival time of every photon, together with spacecraft orientation data, are used to generate X-ray/$\gamma$-ray images with an angular resolution of $\sim2$ arcseconds and a FoV of $\sim1^{\circ}$.  For sources within the imaging FoV (such as solar flares), a small Be scattering block (seen in Fig. \ref{RHESSI}) scatters photons into the rear segments of nearby Ge detectors. The distribution of these scattered photons could be used to measure polarization \citep{2002SoPh..210..125M}, but this has proven to be problematic. For sources outside the imaging FoV (such as GRBs), the array of (largely) un-shielded Ge detectors is used  for polarimetry by measuring photons that scatter between Ge detectors \citep[e.g.,][]{2003Natur.423..415C}.

\subsection{INTEGRAL}

The International Gamma-Ray Astrophysics Laboratory (INTEGRAL) was launched in 2002 and remains operational. It carries a suite of four instruments, two of which provide spectroscopy and imaging of $\gamma$-rays.  The {\it Imager on Board the INTEGRAL Satellite} (IBIS) is a coded mask imaging instrument using two different detection planes that collectively cover the energy range from 15 keV to 10 MeV with high angular resolution and moderate spectral resolution \citep{2003A&A...411L.131U}. The {\it Spectrometer on INTEGRAL} (SPI) is also a coded mask imaging instrument, designed to operate from 20 keV to 8 MeV, but it is designed with high spectral resolution and moderate angular resolution \citep{2003A&A...411L..63V}. Although not optimized for polarization measurements, both IBIS and SPI have had polarimetry as one of their secondary science objectives. Neither was calibrated with polarized photons before launch. Both IBIS and SPI have reported measurements of polarization from the Crab and from Cyg X-1 \cite{2012int..workE...5L}. During the first 10 years of the INTEGRAL mission, 89 GRBs were detected within the IBIS field-of-view.

\subsubsection {IBIS}

The IBIS instrument consists of two separate detection planes that share the same coded mask. The top detection plane is referred to as the INTEGRAL Soft Gamma-Ray Imager (ISGRI) \citep{2003A&A...411L.141L}. ISGRI is optimized for the energy range from 15 keV to 1 MeV using a detection plane that consists of a large (128 $\times$ 128) array of CdTe detectors, each of which is 2 mm thick and 4 $\times$ 4 mm$^2$ in area. The 16384 CdTe detectors provide an active area of $\sim$2600 cm$^2$. The lower detection plane is referred to as the Pixellated Imaging CsI Telescope (PICsIT) \citep{2003A&A...411L.149L}. PICsIT is optimized for the energy range of 175 keV to 10 MeV using an array of CsI(Tl) scintillation crystals.  Each  crystal is 8.4 $\times$ 8.4 mm in area and 3 cm deep, and is read out by a PIN photodiode. The 4096 CsI detectors provide an active area of $\sim$2890 cm$^2$. Both ISGRI and PICsIT are designed to operate independently in conjunction with the shared coded mask (which 16 mm thick and composed of a W-Cu alloy). However, events that trigger both detection planes within some preset time window (3.8 $\mu$s) can be treated as Compton events that scatter from one layer to the next. In this way, IBIS can combine the advantages of both a Compton telescope and a coded mask imager \citep{2007ApJ...668.1259F}.  This Compton mode is most effective from a few hundred keV up to a few MeV.

\begin{figure}
\centering
\includegraphics[height=1.7in]{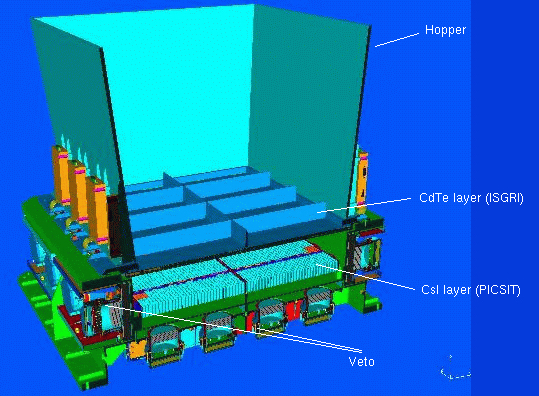}
\includegraphics[height=1.7in]{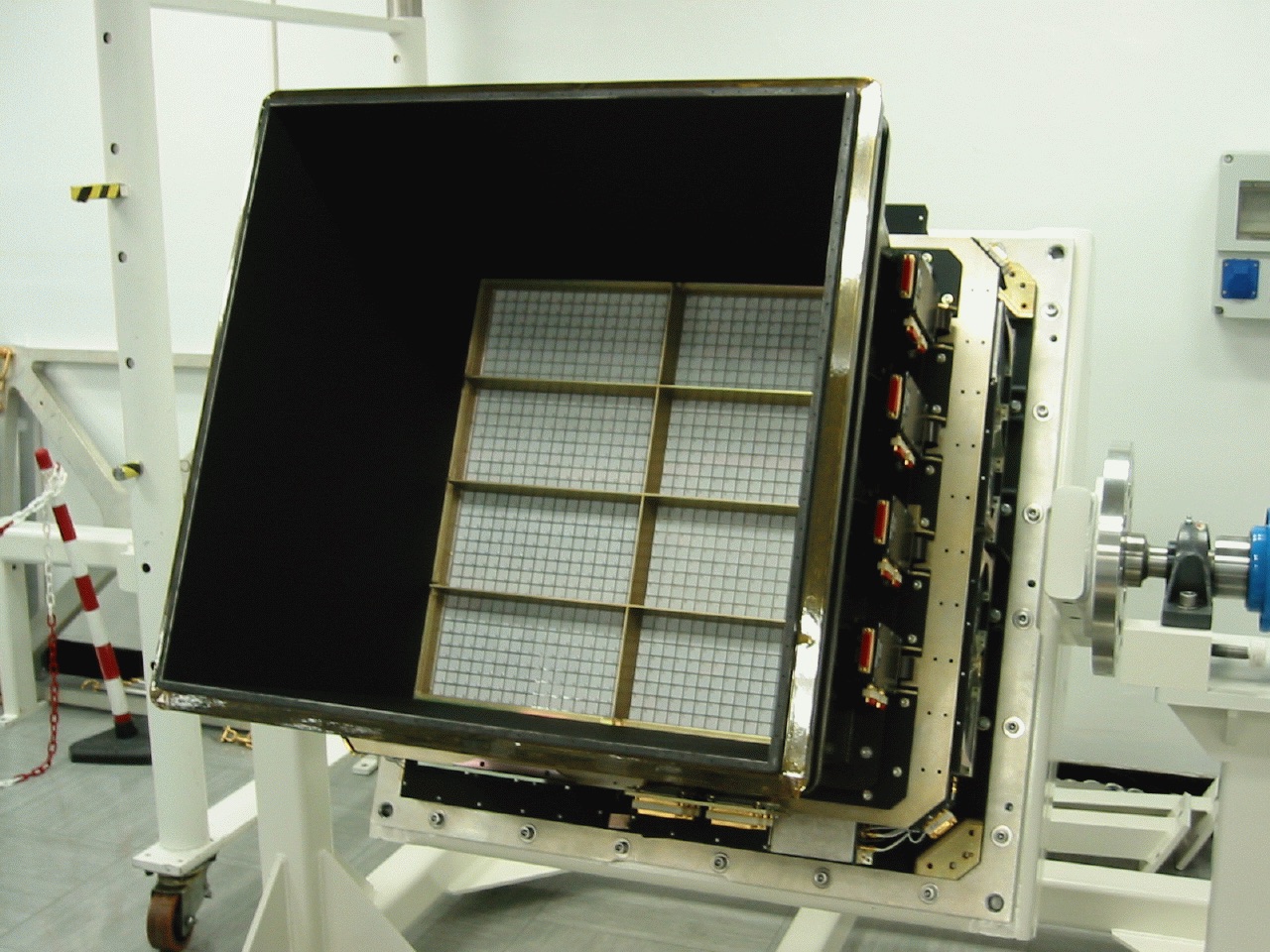}
\caption{Schematic (left) of the IBIS detector assembly, showing the ISGRI and PICsIT detection planes and some of the the passive shielding used to minimize background reaching the detectors. The photo (right) shows the flight unit as it was being prepared for flight.}
\label{fig:IBIS} 
\end{figure}

Although Compton scattering events {\it within} a detector layer could be used for polarimetry, the on-board electronics rejects multiple triggers within the ISGRI detection plane and the instrument mode required to collect proper PICsIT data is never employed.  Polarimetry with IBIS can only be achieved  through the use of the Compton events  \citep{2007ApJ...668.1259F} that scatter between ISGRI and PICsIT.  Most of the events tagged as Compton events are due to background events that are removed by the coded aperture deconvolution, thus making use of the shadowing effects of the mask. Compton kinematics is also used to help insure that the selected events are consistent with the source direction.

\subsubsection {SPI}

\begin{figure}
\centering
\includegraphics[height=2.25in]{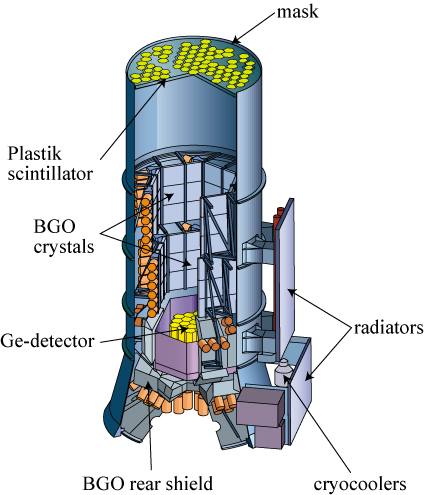}
\includegraphics[height=1.9in]{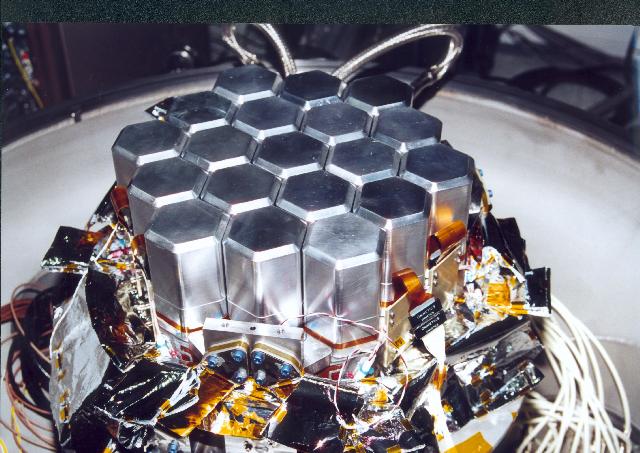}
\caption{Schematic (left) of SPI instrument, showing the location of the Ge detector array with respect to the coded mask and surrounding BGO anticoincidence shielding.  The photo (right) shows the Ge detector during assembly.}
\label{SPI} 
\end{figure}

SPI is a coded aperture imaging instrument that uses an array of 19 large volume Ge detectors as the detection plane, with a 3 cm thick W mask. Each Ge detector is hexagonal in shape measuring 5.6 cm flat-to-flat and 7 cm deep. The detectors are arranged into a close-packed tessellated array, as seen in Fig. \ref{SPI}. There is no information on the photon interaction site within a detector. Spatial information is provided only be the location of the triggered Ge detector. Polarimetry with SPI relies on photons scattering between Ge detectors \citep{2013ApJ...769..137C,2004ESASP.552..859K}.  Coincidence events between detectors are defined by a hardware time window of 350 ns. Since INTEGRAL's launch, four detectors have failed: detector 2 (2003), detector 17 (2004), detector 5 (2009), and detector 1 (2010). 

The analysis of SPI data is hindered by the fact that there are a limited number of scatter angles that are measurable, given the limited number of pixels in the detector array and the fact that the INTEGRAL spacecraft does not spin. If the photon interaction sequence of the photon can be determined, then six independent scatter angles can be defined. For photons with incident energy below $\sim500$ keV, the first interaction site typically corresponds to the detector with the lowest energy deposit. At higher energies, the ambiguity about the interaction sequence can generally not be resolved and only three independent scatter angles can be defined.

\subsection{IKAROS / GAP}

\begin{figure}
\centering
\includegraphics[height=3.0in]{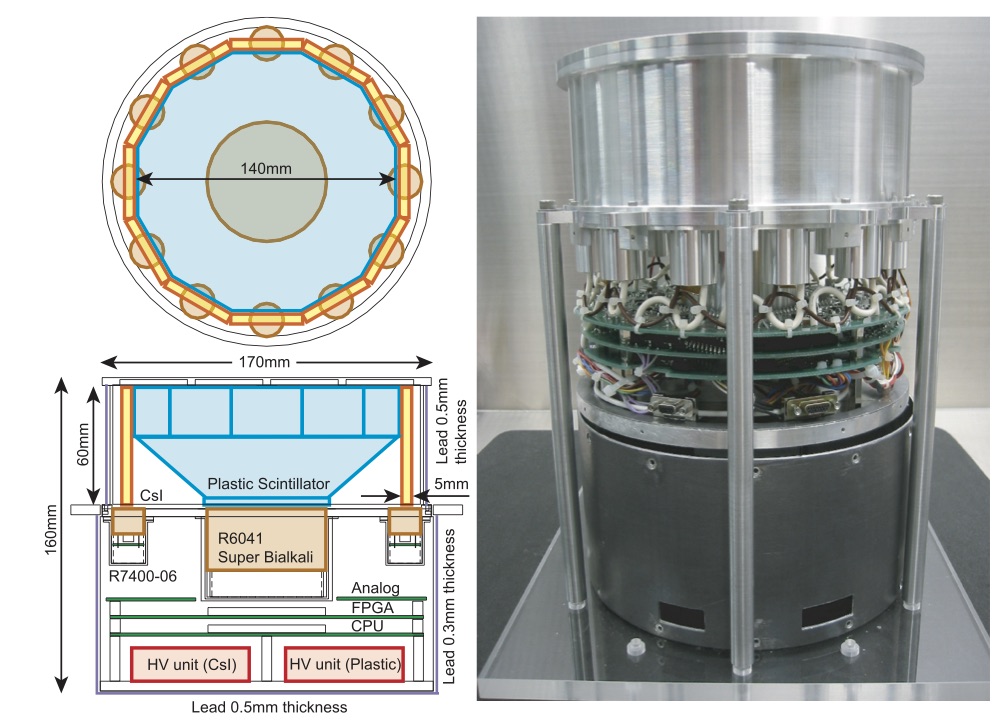}
\caption{Schematic view of the GAP detector (left) showing the various instrument components, and a photo (right) of the assembled instrument with the lead shielding removed. [Figure from \citet{2011PASJ...63..625Y}.]}
\label{fig:GAP_Schematic} 
\end{figure}

The GAmma-ray Polarimeter (GAP) represents the first flight instrument designed specifically for GRB polarization measurements \citep{2011PASJ...63..625Y,2010AIPC.1279..227M}. As shown in Figure~\ref{fig:GAP_Schematic}, it consists of a single, large plastic scintillator surrounded by an array of 12 CsI(Tl) scintillators. The plastic is shaped as a dodecagon polygon measuring 14 cm face-to-face. With a full height of 6 cm, the lower side of the plastic scintillator is tapered to match the face of a 2-inch (5.08 cm) PMT. The 12 CsI(Tl) scintillators, each measuring 6 cm high and 5 mm thick and read out by a single PMT, is intended to absorb photons scattered from the central plastic scintillator. The location of the hit CsI(Tl) detector gives a measure of the photon scatter direction. The instrument is sensitive to polarization in the 70--300 keV energy range. Inflight gain corrections are made possible by the use of onboard $^{241}$Am calibration sources.

\begin{figure}
\centering
\includegraphics[height=3.0in]{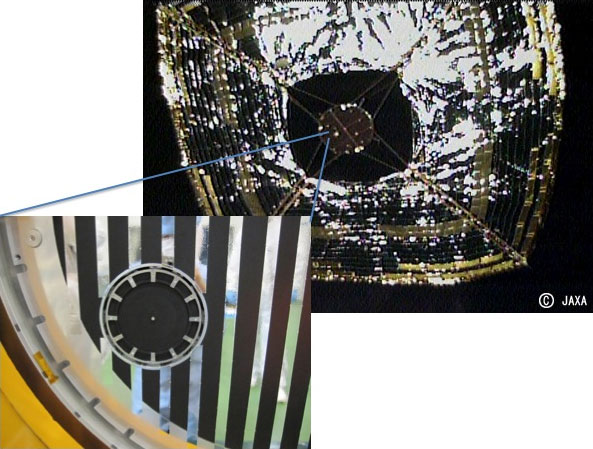}
\caption{A view of the IKAROS spacecraft shortly after deployment of the solar sail (taken by a spin-loaded camera released by the spacecraft).  The inset shows the location of the GAP instrument.}
\label{fig:GAP-IKAROS} 
\end{figure}

The GAP experiment was flown as part of a Japanese solar power sail demonstration mission known as IKAROS (Interplanetary Kite-craft Accelerated by the Radiation Of the Sun; \cite{2013AcAau..82..183T}). The IKAROS spacecraft measures 1.58 m in diameter and 0.95 m in height, with a mass of 307 kg.  A 20 m diameter solar sail surrounds the spacecraft and provides propulsion through interplanetary space. The spacecraft rotates at a rate of $\sim$1--2 rotations per minute, utilizing centrifugal force to help keep the sail properly deployed.  The IKAROS spacecraft was launched as a piggyback payload with the Venus Climate Orbiter on May 21, 2010. The solar sail was successfully deployed on June 9, 2010 (Figure~\ref{fig:GAP-IKAROS}). The primary mission lasted until December of 2010.  Since that time, IKAROS has experienced several  hibernation periods due to lack of solar power. As of May 2014, IKAROS was in a ten-month orbit around the Sun, spending roughly seven months of each orbit in hibernation mode.  It is still considered to be operational, so data from the GAP continues on an intermittent basis.

\subsection{Astrosat}

The Indian Astrosat mission was launched on September 28, 2015 \citep{2014SPIE.9144E..1SS}. One of the instruments on that mission is the CZT Imager (CZTI), a coded mask telescope that uses a pixelated array of CdZnTe (CZT) detectors. The CZT detection plane consists of 2.5 mm $\times$ 2.5 mm pixels, each of which has a thickness of 5 mm. CZTI images photons from 10-100 keV over a $6\deg \times 6\deg$ FoV.  The CZT detectors maintain sensitivity up to about 250 keV, but the coded mask becomes transparent at energies above $\sim 100$ keV. 

\begin{figure}
\centering
\includegraphics[height=2.5in]{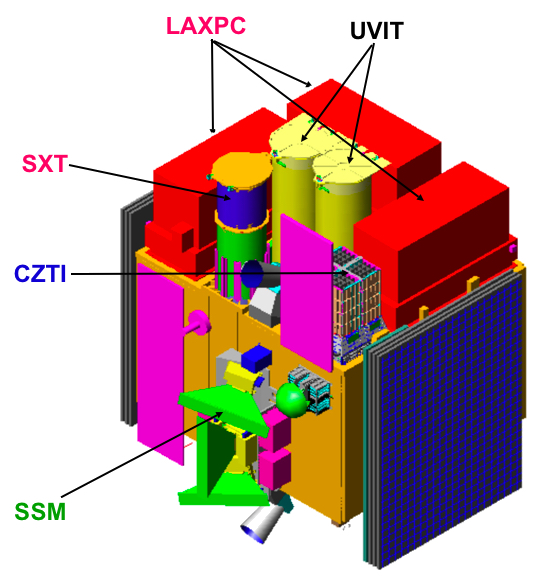}
\includegraphics[height=2.5in]{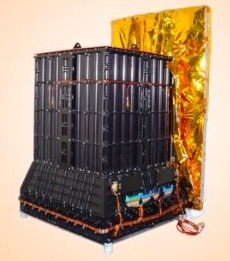}
\caption{The instrument suite of the ASTROSAT satellite (left), and a photo of the CZTI instrument (right). [Figures from \citet{2014SPIE.9144E..1SS}.]}
\label{ASTROSAT} 
\end{figure}

The pixellated nature of CZTI lends itself to polarization studies \citep{Vadawale:2015bk,2014ExA....37..555C}. Each event is recorded with a time resolution of 20 $\mu$sec. Compton scattered events can be identified by looking for coincidence events between nearby pixels, based on the time signatures of individual events (two events must be within 40 $\mu$sec of each other).  True Compton events also require that the the sum and ratio of deposited energies must be consistent with Compton scattering.  As Compton scattering in CZT becomes dominant at energies above $\sim$100 keV, polarimetry is effective at energies where the coded mask is transparent.  Imaging polarimetry is therefore not likely.  But polarization of transient sources (like GRBs) does not require  imaging.

\section{GRB Polarization Measurements}
\label{measurements}

In this section, we summarize the observations to date in  order of initial publication date.  This order was chosen to help illustrate how the field has evolved. 

\begin{table}[htp]
\tiny
\caption{GRB Polarization Measurements}
\begin{center}
\begin{tabular}{llllll}
\hline
Pub  &            &                      & Energy &             &            \\
Date &  GRB  &  Instrument   &  (keV)    &  $\Pi$  &  Refs  \\
\hline
\hline
2004 & GRB 021206  & RHESSI              & 150 -- 2000 & 80\% $\pm$ 20\%   & \cite{2003Natur.423..415C}    \\
2004 & GRB 021206  & RHESSI              & 150 -- 2000 &      $<4.1$\%            &  \cite{2004MNRAS.350.1288R}  \\
2004 &  GRB 021206  & RHESSI              & 150 -- 2000  &  $41^{+57}_{-44}$\%                              &   \cite{2004ApJ...613.1088W}   \\
 2005 & GRB 930131  & CGRO/BATSE    & 20 -- 1000   & (35--100\%)\textsuperscript{a}           &  {\cite{2005A&A...439..245W}}   \\
2005 & GRB 960924  & CGRO/BATSE    & 20 -- 1000   & (50--100\%)\textsuperscript{a}          &  {\cite{2005A&A...439..245W}}   \\
2007 & GRB 041219a & INTEGRAL/SPI  & 100 -- 350   &  98\% $\pm$ 33\%  &  \cite{2007ApJS..169...75K}   \\
2007 & GRB 041219a & INTEGRAL/SPI  & 100 -- 350   &  96\% $\pm$ 40\%  &  {\cite{2007A&A...466..895M}}   \\
2009 & GRB 041219a & INTEGRAL/IBIS & 200 -- 800   &  43\% $\pm$ 25\%\textsuperscript{b}  &  \cite{2009ApJ...695L.208G}   \\
2009 & GRB 061122   & INTEGRAL/SPI  & 100 -- 1000 & $<60$\%                 & {\cite{2009A&A...499..465M}}   \\
2011 & GRB 100826a & IKAROS/GAP     &  70 -- 300    &  27\% $\pm$ 11\%\textsuperscript{c}  &  \cite{2011ApJ...743L..30Y}   \\
2012 & GRB 110301a & IKAROS/GAP     &  70 -- 300    &  70\% $\pm$ 22\%  &   \cite{2012ApJ...758L...1Y}   \\
2012 & GRB 110721a & IKAROS/GAP     &  70 -- 300    &  80\% $\pm$ 22\%  &   \cite{2012ApJ...758L...1Y}    \\
2013 & GRB 061122   & INTEGRAL/IBIS & 250 -- 800   &  $>60\%$                &  \cite{2013MNRAS.431.3550G}    \\
2014 & GRB 140206a & INTEGRAL/IBIS & 200 -- 800   & $>48\%$                 &   \cite{2014MNRAS.444.2776G}    \\
2016 & GRB 151006a & Astrosat/CZTI     & 100 -- 300   &  --                            & \cite{Rao:2016tt} \\
\hline
\hline
\multicolumn{6}{l}{ } \\
\multicolumn{6}{l}{\textsuperscript{a}\footnotesize{ albedo polarimetry}} \\
\multicolumn{6}{l}{\textsuperscript{b}\footnotesize{ variable $\Pi$ }} \\
\multicolumn{6}{l}{\textsuperscript{c}\footnotesize{ variable position angle }} \\

\end{tabular}
\end{center}
\label{GRBMeasurements}
\end{table}

\subsection{GRB 021206}

This burst was detected by RHESSI at 22:49 UT on December 6, 2002.  The emission was localized by the IPN (Ulysses, Konus, and Mars Odyssey) to lie about $18\deg$  from the Sun (and therefore $\sim18\deg$ from the RHESSI pointing direction), with a total 25--100 keV fluence of $1.6 \times 10^{-4}$ erg cm$^{-2}$.

\begin{figure}
\begin{center}
\includegraphics[width=4in]{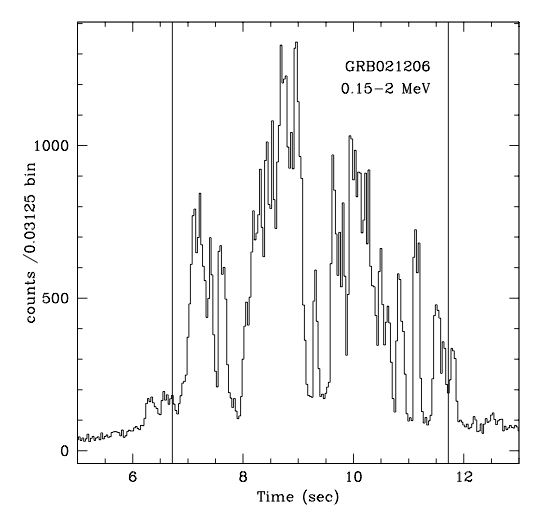}
\caption{The light curve of GRB 021206 as recorded by RHESSI shows emission in the energy range of 150 keV to 2 MeV. The vertical lines delineate the 5 s time period used in the polarization analysis \citep{2004MNRAS.350.1288R,2003Natur.423..415C}. [Figure from \citet{2004MNRAS.350.1288R}.]}
\label{GRB021206_LightCurve}
\end{center}
\end{figure}

The first analysis of these data led to the first published report of polarization from a GRB \citep{2003Natur.423..415C}. The analysis concentrated on a 5 s time interval shown in Figure \ref{GRB021206_LightCurve}. This time interval corresponds to 1-1/4 spacecraft rotations (not an integral number of spacecraft rotations). The biggest challenge in the analysis of these data was that the RHESSI hardware was not designed to trigger on coincidence events between two Ge detectors.  Coincidence events were identified in the data stream simply by comparing time stamps of each event. The measured scatter angle distribution for coincidence events between two detectors was compared with the distribution expected for unpolarized radiation. The unpolarized distribution was derived from simulations. For a burst location that lies $18\deg$ from the RHESSI optical axis (as was the case here), there is considerable attenuation of the flux by the mass distribution on the spacecraft equipment deck. The simulations took into account this mass distribution and the rotation of the spacecraft. The simulations also took into account the flux variability during the burst.  The difference between the measured scatter angle distribution and the simulated (unpolarized) distribution showed a signal that was characteristic of radiation with a linear polarization value of $\Pi_m = 80 \pm 20\%$.

The unexpectedly large value for the burst polarization led others to conduct their own independent analysis of the data \citep{2004ApJ...613.1088W,2004MNRAS.350.1288R}.  Unfortunately, these studies were unable to confirm the initial result, thus leading to widespread skepticism of the initial polarization measurement. 

The second analysis \citep{2004MNRAS.350.1288R} focused on the problem of identifying true coincidence events.  The analysis was unusual in that it did not include any simulations to take into account the response of the instrument and the effects of mass surrounding the detector array.  Rather it relied largely on a statistical analysis of the event data. The event selection process resulted in fewer events than in the initial analysis \citep{2003Natur.423..415C}.  They also noted an asymmetry in the distribution of events within the detector array, and suggested that this may be due to photons scattering off the atmosphere, implying that this might also induce a spurious result in the analysis.  This analysis was unable to provide meaningful constraints on the polarization.  

The third analysis \citep{2004ApJ...613.1088W} concentrated on a 4 s period of the burst, corresponding to exactly one rotation of the spacecraft. It included more stringent constraints on the event selection process, including a more extensive analysis of coincidence events and the addition of a kinematic constraint to insure that selected events were consistent with a single Compton scatter. Far fewer events passed these more stringent criteria than in either of the two previous studies. Simulations that incorporated polarization effects were also incorporated into the analysis for the first time. A formal linear polarization value of $\Pi = 41^{+57}_{-44}\%$ was obtained, a value that was insufficient to claim a detection of polarization. It was concluded that the initial analysis \citep{2003Natur.423..415C} included a significant number of accidental coincidence events that skewed the result. The same analysis technique was applied to the data for GRB 030519B, but the data were insufficient to provide any constraint on the polarization for that event \citep{2004ApJ...613.1088W}. 

\begin{figure}
\begin{center}
\includegraphics[width=4in]{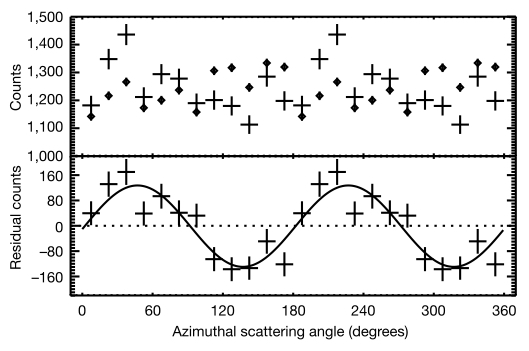}
\caption{RHESSI Observation of GRB 021206.  The top panel shows both the measured scattered angle distribution (crosses) and the simulated scatter angle distribution (diamonds) for unpolarized radiation. The bottom panel shows the difference between the two distributions, which is consistent with polarized radiation from the GRB. [Figure from \citet{2003Natur.423..415C}.]}
\label{GRB021206_Polarization}
\end{center}
\end{figure}

\subsection{GRB 930131}

This event was detected at 18:57:11 UT on 1993 January 31 by BATSE. It took place only 8.4$^{\circ}$ from the local zenith.  The time history showed two intense peaks within the first 1.2 s, followed by a long tail lasting about 50 s.  The measured $T_{90}$ was 19.2 s. Most of the flux was emitted during the very brief time interval of the first peak, which lasted  $\lesssim$ 0.1 s.  Although the total fluence was not exceptional (with a 50--300 keV fluence of $6.54 \times 10^{-5}$ ergs cm$^{-2}$ s$^{-1}$), this was one of the most intense GRBs (in terms of peak flux) observed by CGRO \citep{1994ApJ...422L..59K}. It was detected by BATSE \citep{1994ApJ...422L..59K}, COMPTEL \citep{1994ApJ...422L..67R} and EGRET \citep{1994ApJ...422L..63S}. Photons with energies up to 1 GeV were detected. Since the GRB location was close to the pointing axis of CGRO, it was imaged by both COMPTEL and EGRET.  

This was one of two GRBs selected for study by analyzing the atmospheric albedo flux with CGRO/BATSE \citep{2005A&A...439..245W}.  It was chosen for analysis based on several factors, including its short duration (to minimize instrumental background changes during the burst), its intensity (to maximize the signal), and its large distance from the geocenter (i.e., it was close to the local zenith, providing an ideal geometry for albedo polarization measurements).  The BATSE data suffered from severe deadtime effects during the intense peak of the event, especially those detectors that were facing in the direction of the source. Those detectors facing towards the Earth were shielded from the direct flux and thereby recorded a much lower counting rate, a counting rate that was dominated by scattered albedo flux.

\begin{figure}
\begin{center}
\includegraphics[width=4in]{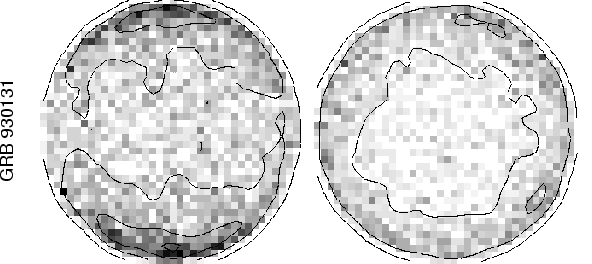}
\caption{Simulated angular distributions (using GEANT4) for flux scattered off the Earth's atmosphere for GRB 930131. The fully polarized GRB case is on the left; the unpolarized GRB case is on the right. The unpolarized case shows a more symmetric distribution for this source located near the local zenith. At the altitude of CGRO, the Earth's disk subtends a half-angle of about 70$^{\circ}$.  [Figure from \citet{2005A&A...439..245W}.]}
\label{GRB930131_Distribution}
\end{center}
\end{figure}

The analysis was based on the idea that the observed distribution of flux scattered off the atmosphere will depend on both the level of polarization and the polarization angle \citep{1996AIPC..384..851M}. Fig. \ref{GRB930131_Distribution} shows the simulated distribution of flux from GRB 930131 for both the polarized and unpolarized case. The simulations of atmospheric scattering made use of GEANT4 software, which, at the time, was known to have some deficiencies in the polarization physics \citep{2005NIMPA.540..158M}. The output of the atmospheric scattering simulations (e.g., Fig. \ref{GRB930131_Distribution}) was used as input to a GEANT3-based simulation of the BATSE response to the scattered flux. Constraints on the angular distribution of the scattered flux were determined using the total counts of the four most Earthward-facing BATSE detectors and comparing the measured data with simulations.

In the case of GRB 930131, the data were found to be consistent with polarization values ranging between 35\% and 100\%. Although several possible sources of systematic error were considered in the analysis, it could not be ruled out that some additional systematic effect (or an error in the simulations) might be impacting the results.

\subsection{GRB 960924}

The albedo polarization analysis employed for GRB 930131 was also applied to the analysis of GRB 960924 \citep{2005A&A...439..245W}. This event took place on 1996 September 24 at 11:41:51 UT. It was located $17.6^{\circ}$ from the local zenith. The measured $T_{90}$ was 5.3 s and the  total 50--300 keV fluence was $2.47 \times 10^{-4}$ ergs cm$^{-2}$ s$^{-1}$.  The simulated distribution of scattered flux from both the fully polarized and unpolarized cases is shown in Fig. \ref{GRB021206_Polarization}. The polarization analysis found that the data once again suggested very high levels of polarization. More specifically, the data were found to be consistent with polarization values ranging from 50\% to 100\%.

\begin{figure}
\begin{center}
\includegraphics[width=4in]{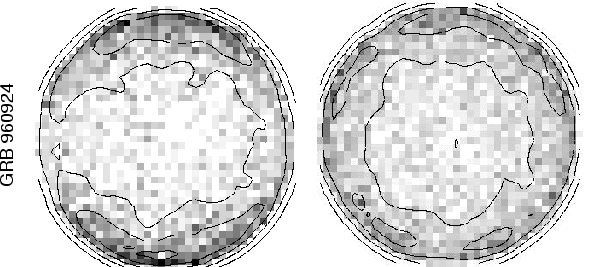}
\caption{Simulated angular distributions (using GEANT4) for flux scattered off the Earth's atmosphere for GRB 9960924. The fully polarized GRB case is on the left; the unpolarized GRB case is on the right. The unpolarized case shows a more symmetric distribution for this source located near the local zenith. At the altitude of CGRO, the Earth's disk subtends a half-angle of about 70$^{\circ}$. [Figure from \citet{2005A&A...439..245W}.]}
\label{GRB021206_Polarization}
\end{center}
\end{figure}

\subsection{GRB 041219A}

This burst was detected by INTEGRAL at 01:42:18 UTC on December 19, 2004 \citep{2006A&A...455..433M}.  Located at an off-axis angle of only $3.2\deg$, it was well within the imaging FoV of both IBIS and SPI. The prompt emission was also detected at high energies by both SWIFT/BAT and  RXTE/ASM. Rapid followup by ground-based observers resulted in the detection of the prompt emission in both the optical and near infrared, but no redshift is available for this burst. 

\begin{figure}
\begin{center}
\includegraphics[width=4in]{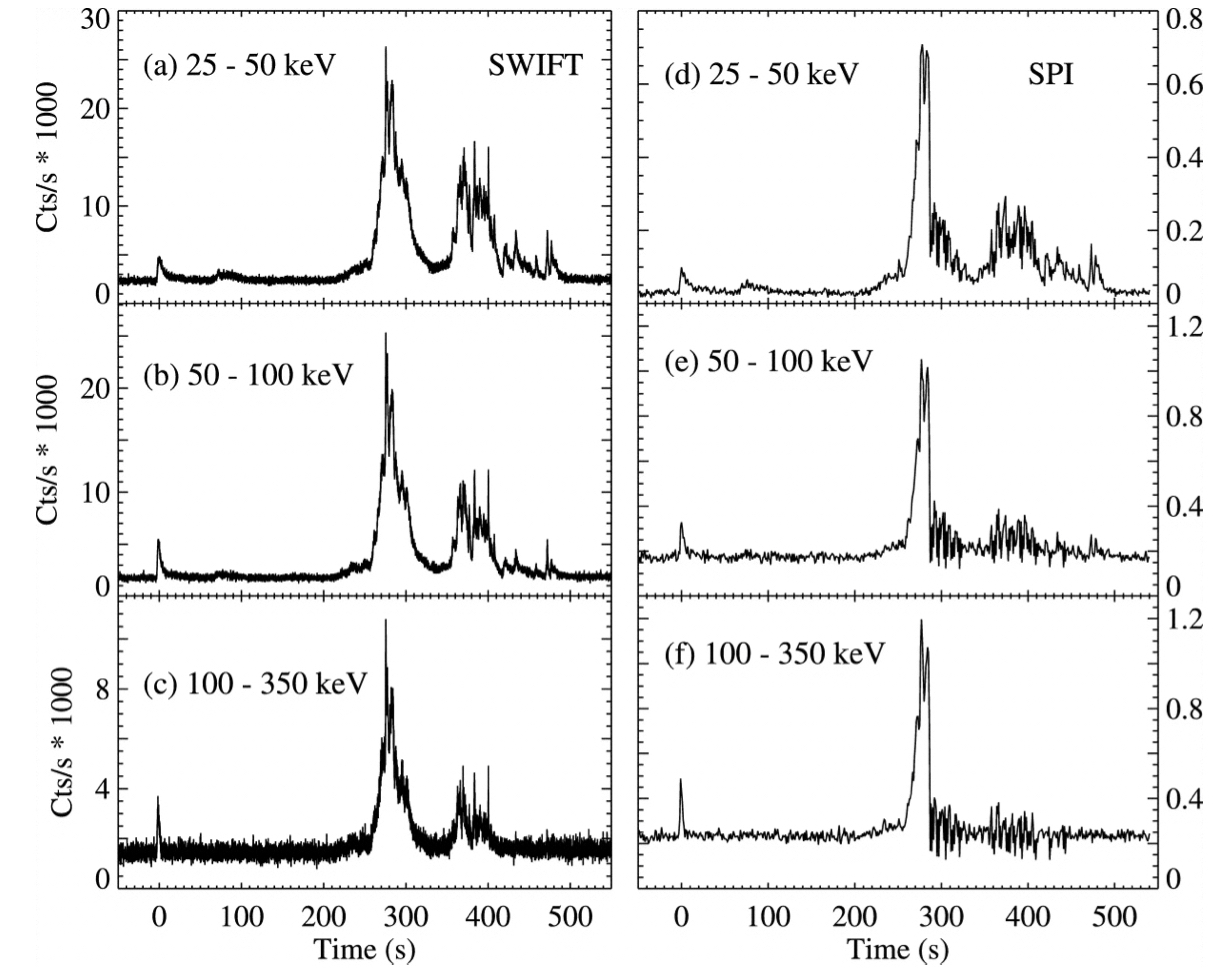}
\caption{The light curve of GRB 041219A as recorded by SWIFT/BAT (panels a-c) and by INTEGRAL/SPI (singles events, panels d-f) \citep{2007ApJS..169...75K}. The small precursor at time $t = 0$ allowed sufficient time for ground-based observations during the more intense emission starting near $t = 250 s$. SPI (and also IBIS) encountered significant  data loss during the later stages of the event due to the telemetry limitations of INTEGRAL.  [Figure from \citet{2007ApJS..169...75K}.]}
\label{GRB041219A_LightCurve}
\end{center}
\end{figure}

Polarization studies of this event were reported by three different groups.  Two of the reports involved analysis of SPI data \citep{2007A&A...466..895M,2007ApJS..169...75K}.  A third report utilized data from IBIS \citep{2009ApJ...695L.208G}. The brightest part of the burst saturated the available telemetry of INTEGRAL, resulting in a significant loss of data during the later stages of the burst (as can be seen in Figure \ref{GRB041219A_LightCurve}).

The first report \citep{2007ApJS..169...75K},  based on the use of SPI data, considered only nearest-neighbor detectors. At the time, two of the Ge detectors in the SPI detector array were no longer operational and therefore excluded from the analysis. The effect of these detector losses was included in the simulations, which were generated using MGEANT  \citep{2003A&A...411L..81S} with the GLEPS package for handling polarization\citep{2004HEAD....8.4101M}. The energy range used in the analysis covered 100 -- 350 keV, but excluded a Ge background line near 198 keV.  Three separate time intervals were considered, the first of which took place before the onset of telemetry loses. Livetime data was used to correct the remaining time intervals for telemetry gaps. Azimuthal scattering angle distributions were derived both for the first time interval and for the combination of all three time intervals. The two different polarization values derived from these data gave consistent results  ($\Pi = 100\% \pm36\%$ for the first time interval and $\Pi = 99\% \pm33\%$ for the combined time intervals). Slightly different energy limits gave less significant results. The analysis could not rule out the possibility that instrumental systematics dominated the measured modulation.  No statistically-significant claim of polarization was made.

\begin{figure}
\begin{center}
\includegraphics[height=2in]{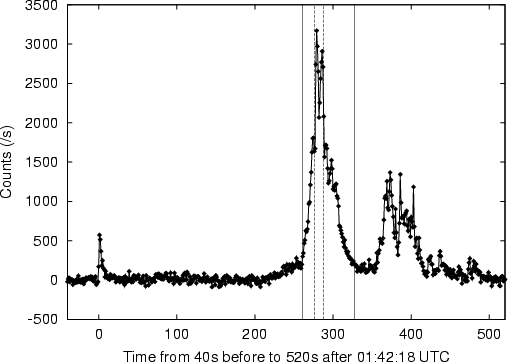}
\caption{Background-subtracted single event lightcurve of GRB 041219a, summed over all SPI detectors in the energy range 20 keV--8 MeV.  The analysis intervals used in Ref.  \citep{2007A&A...466..895M} are shown. [Figure from \citet{2007A&A...466..895M}.]}
\label{GRB041219a_McGlynn}
\end{center}
\end{figure}

The second report \citep{2007A&A...466..895M} was also based on the use of SPI data. The analysis was similar to that in  \cite{2007ApJS..169...75K}, but included more extensive simulations using GEANT4.  In this case, the GEANT4 polarization code was modified to correct for errors in the polarization physics \citep{2005NIMPA.540..158M}. The polarization analysis considered only the first peak of the event (the most intense part of the burst emission), a time interval of 66 s starting at about 250 s after the burst trigger.  In addition, a 12 s analysis interval within that time period was chosen to capture the brightest part of the emission. These time intervals are shown on the background-subtracted single event lightcurve in Fig. \ref{GRB041219a_McGlynn}. The analysis generated results for several energy intervals within each of these two time windows. For the 100--350 keV energy band, the measured polarization in the 66 s time window was $\Pi = 63^{+31}_{-30}\%$ at a polarization angle of $70^{+14}_{-11}\%$ degrees.  For the 100--350 keV energy band, the measured polarization in the 12 s time window was $\Pi = 96^{+39}_{-40}\%$ at a polarization angle of $60^{+12}_{-14}\%$ degrees. The results derived from various energy bands in the two time intervals are all consistent with a polarization value of $60\pm35\%$. It was concluded that a systematic effect capable of mimicking the weak ($2\sigma$) result could not be excluded.

 The third analysis of this event was based on IBIS Compton mode data \citep{2009ApJ...695L.208G}. The polarization analysis considered data in the 200--800 keV energy interval covering several different time windows. No evidence for polarization was seen in data integrated over the first peak ($\Pi < 4\%$), but there is evidence for polarization in the data integrated over the second peak ($\Pi = 43\% \pm 25\%$). Analysis of finer time bins provided evidence within each peak time period for variations in both the level of polarization and polarization angle. In fact, the variations seen in  the finer time resolution analysis explains the null average polarization in the first peak as a result of rapid variations observed in the polarization angle and degree.

\subsection{GRB 061122}

GRB 061122 was detected by instruments on INTEGRAL at 07:56:45 on 22 November 2006. Its location 8.2$\deg$ off-axis allowed imaging by both IBIS and SPI. The 20--200 keV fluence was determined to be $3 \times 10^{-6}$ erg cm$^{-2}$. It had a $T_{90}$ duration of 12 s and a peak energy of $E_{p} = 165$ keV.  At the time, it was the second most intense burst observed by INTEGRAL, second only to GRB 041219A. The prompt emission was also detected by Konus-WIND.  A fading X-ray afterglow was detected about 7 hours after the prompt phase by the XRT on Swift. Precise localization of the afterglow permitted a determination of the redshift, which was found to be $z = 1.33$.

The SPI light curve of this event (20 keV -- 8 MeV), shown in the left-hand panel of Fig. \ref{GRB061122_SPI}, exhibits a single, symmetric pulse. The polarization analysis of the SPI data followed that of Ref. {\citep{2007A&A...466..895M}.  The results, shown in the right-hand panel of Fig. \ref{GRB061122_SPI}. At the 68\% confidence level, the polarization level is constrained to be $< 60\%$.

\begin{figure}
\centering
\includegraphics[height=1.65in]{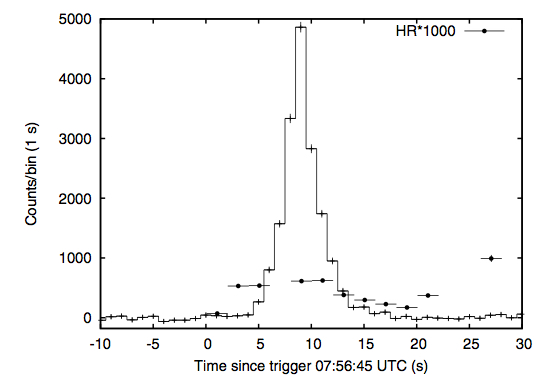}
\includegraphics[height=1.6in]{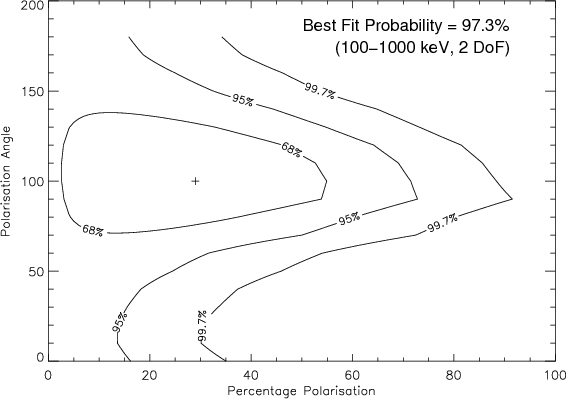}
\caption{The SPI lightcurve (left) of GRB 061122 shows a single symmetric pulse in the 20 keV -- 8 MeV energy range. (Also shown here are hardness ratios derived from IBIS data.) The polarization results (right) for GRB 061122 from an analysis of SPI data. [Figures from \citet{2009A&A...499..465M}.]}
\label{GRB061122_SPI} 
\end{figure}

The IBIS lightcurve of this event (200 -- 800 keV) is shown in the left-hand panel of Fig. \ref{GRB061122_IBIS}. This is similar to that of SPI, but the data suffer from telemetry limitations during the later part of the event. The polarization analysis was performed for different energy intervals and different time selections. The best signal-to-noise was obtained during the 8 s time interval starting at 07:56:50 UT. The polarization results for this time interval are shown in the right-hand panel of Fig. \ref{GRB061122_IBIS}. At the 68\% confidence level, these results are consistent with a polarization level of $> 60\%$ in all energy bands. This is in stark contrast to the SPI result, which indicated a polarization level of $< 60\%$.

\begin{figure}
\centering
\includegraphics[height=1.65in]{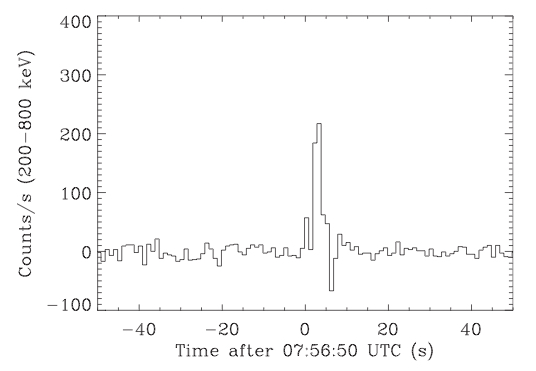}
\includegraphics[height=1.6in]{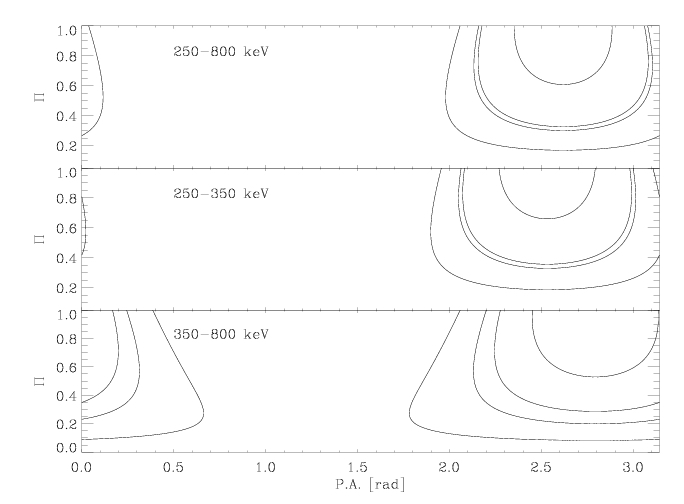}
\caption{The IBIS light curve (left) of GRB 061122 in the 200 -- 800 keV energy range, derived from Compton events, suffers from data loss during the later part of the event due to telemetry limitations. The polarization results (right) for GRB 061122 from an analysis of IBIS data. The contours correspond to confidence levels of 68, 90, 95, and 99 percent. [Figures from \citet{2013MNRAS.431.3550G}.]}
\label{GRB061122_IBIS} 
\end{figure}

\subsection{GRB 100826A}

The GAP detected GRB 100826A on 2010 August 26 at 22:57:20.8 (UT) on the way to Venus at about 0.21 AU away from the Earth \citep{2011ApJ...743L..30Y}. The burst location was determined by the IPN  to be $20.0\deg$ from the GAP pointing axis.  KONUS data for this event gave a total 20 keV -- 10 MeV fluence of ($3.0 \pm 0.3) \times 10^{-4}$ erg cm$^{-2}$. No afterglow observation was reported, so its redshift is unknown. 

The GAP lightcurve for this event is shown in Fig. \ref{GRB100826A_LightCurve}. The background for the analysis comes from integrating over a 36.7 hr period before and after the GRB. Simulations with GEANT4 incorporated off-axis angles, instrument spin angle, and spectral parameters (based on KONUS data).  A matrix of simulations for various degrees of polarization and polarization angles were compared with the measured data to determine constraints on the polarization parameters.

Initially, the full burst time interval (100 s) was used in the analysis, with no discernible polarization. An upper limit of $\Pi < 30\%$ was determined.   Subsequently, in order to investigate the possibility of time variable polarization, two  independent time intervals were chosen for independent analysis (as shown in Fig. \ref{GRB100826A_LightCurve}). The azimuthal scatter angle distributions for these two separate time intervals are shown in Fig. \ref{GRB100826A_Polarization}, along with the corresponding polarization parameter constraints derived from a combined fit of these data. Final results for time interval 1 were $\Pi_1 = 25\% \pm 15\%$ and $\phi_{p1} = 159\deg \pm 18\deg$.  Final results for time interval 2 were $\Pi_2 = 31\% \pm 21\%$ and $\phi_{p2} = 75\deg \pm 20\deg$.  Although there is no evidence for a change in the level of polarization (with an average of $\Pi = 27\% \pm 11\%$), there is evidence in these data for a change in the polarization angle (at the $99.9\%$ or $3\sigma$ level). The significance of the polarization measurement is at the $99.4\%$ ($2.9\sigma$) level.

\begin{figure}
\centering
\includegraphics[height=2in]{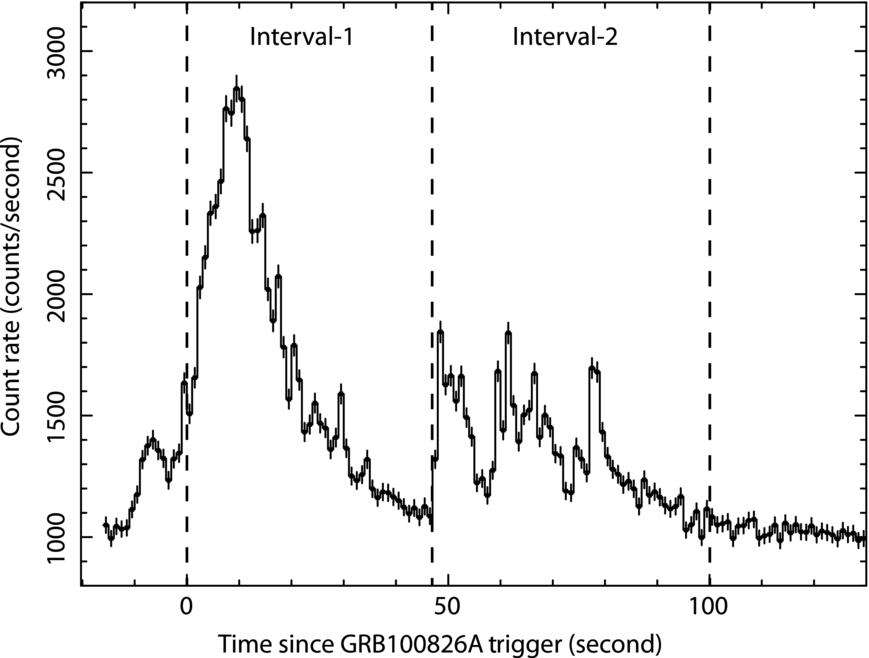}
\caption{The GAP lightcurve of GRB 100826A  for 70 -- 300 keV. Two analysis time intervals are shown. [Figure from \cite{2011ApJ...743L..30Y}.]}
\label{GRB100826A_LightCurve} 
\end{figure}

\begin{figure}
\centering
\includegraphics[height=2in]{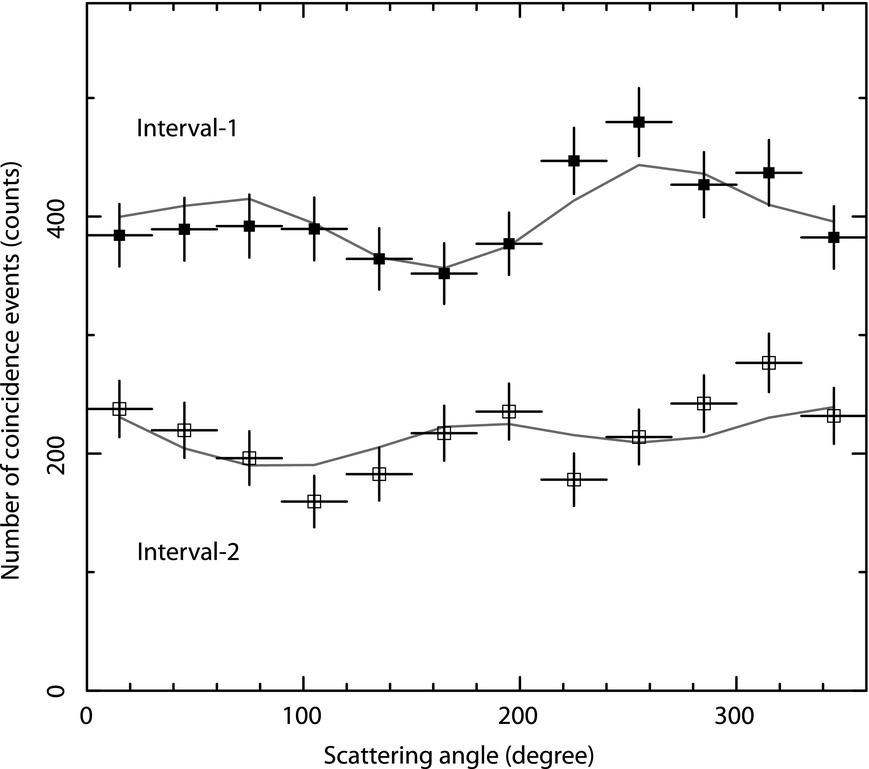}
\includegraphics[height=2in]{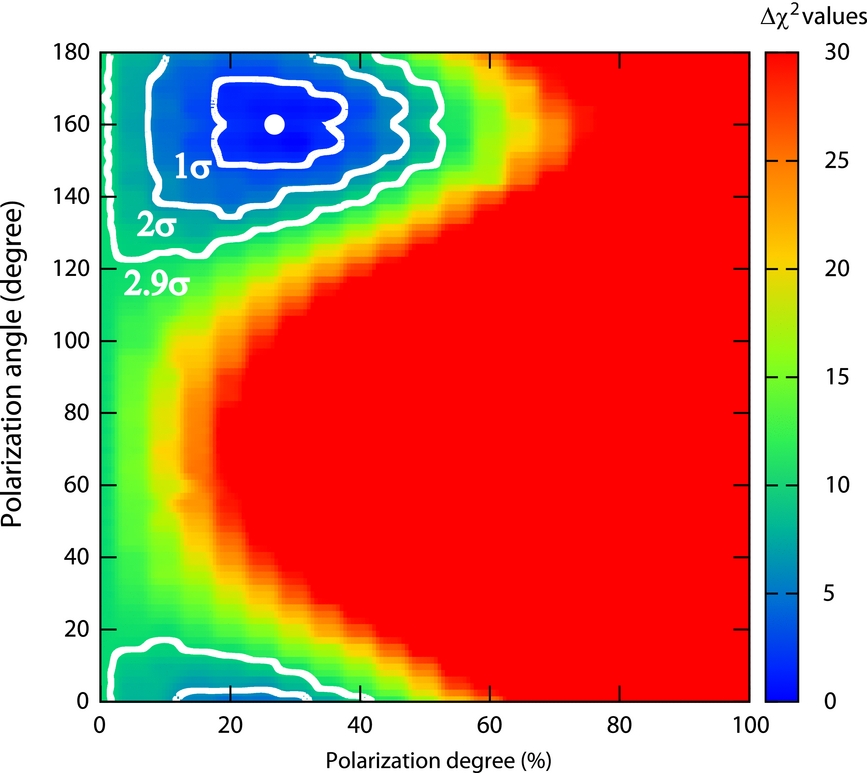}
\caption{The azimuthal scatter angle histogram of GAP data for GRB 100826A for two separate time intervals (left) and the polarization constraints (1, 2, and 3 $\sigma$ contours) for the combined fit of the data for intervals 1 and 2 (right). [Figures from \cite{2011ApJ...743L..30Y}.]}
\label{GRB100826A_Polarization} 
\end{figure}

\subsection{GRB 110301A}

The GAP instrument detected GRB 110301A on 2011 March 1 at 05:05:34.9 UT.  At the time, GAP  was 0.946 AU away from the Earth \citep{2012ApJ...758L...1Y}.  The burst location was determined by Fermi/GBM to be $48\deg \pm3\deg$ from the GAP pointing axis. 

The GAP lightcurve is shown  in the top panel of  Fig. \ref{GRB110301A_LightCurve}. The $T_{90}$ for this event, as measured by BATSE-GBM, was 5.69 s.  The dashed lines in the lightcurve indicate the time interval for GAP polarization analysis. The burst background level was estimated using data just before and after the event. Estimated systematic uncertainties considered the spacecraft rotation rate (1.61 rpm) and the off-axis location of the burst. Results from the polarization analysis are shown in Fig. \ref{GRB110301A_Polarization}. The polarization parameters for the full time interval were $\Pi = 70\% \pm 22\%$ and $\phi_p = 73\deg \pm 11\deg$.  A time-resolved analysis showed no evidence of time variability in the polarization parameters.

\begin{figure}
\centering
\includegraphics[height=2in]{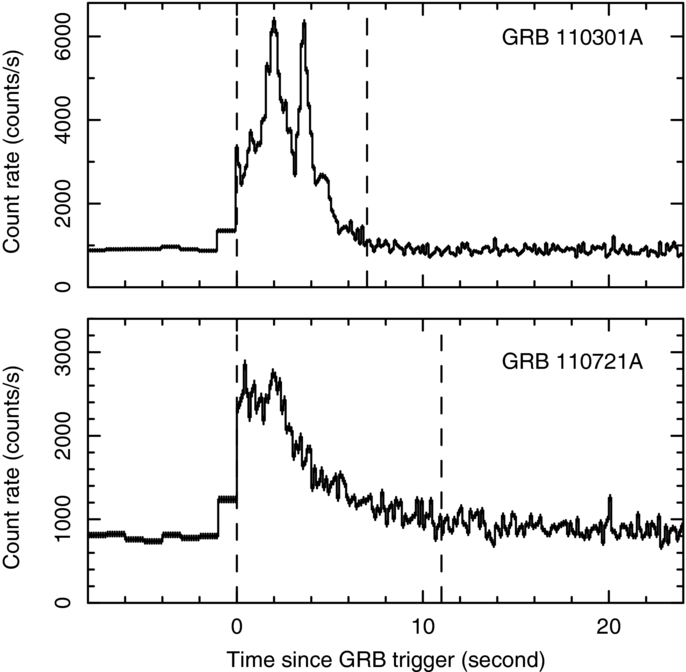}
\caption{The GAP lightcurves of GRB 110301A (top) and GRB 110721A (bottom)  for 70 -- 300 keV.  [Figures from \citet{2012ApJ...758L...1Y}.]}
\label{GRB110301A_LightCurve} 
\end{figure}

\begin{figure}
\centering
\includegraphics[height=2in]{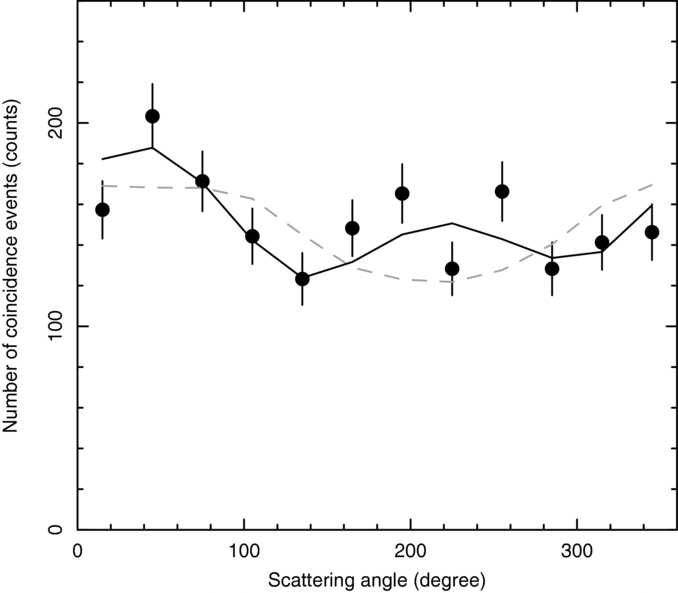}
\includegraphics[height=2in]{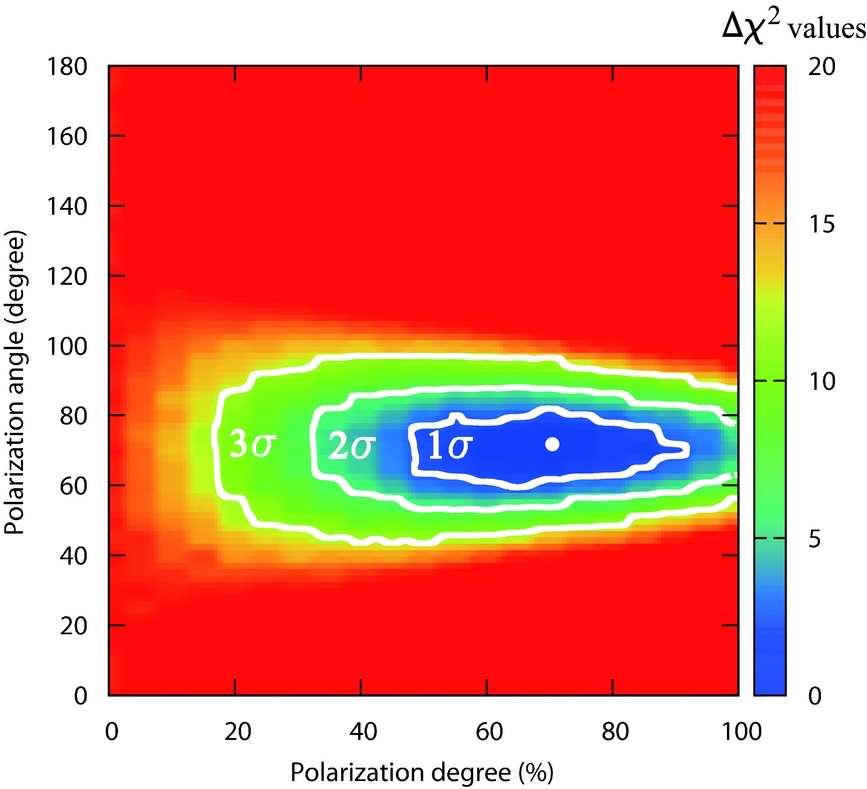}
\caption{The azimuthal scatter angle histograms of GAP data for GRB 110301A (left) and the corresponding constraints (1, 2, and 3 $\sigma$ contours) on the polarization parameters (right). [Figures from \citet{2012ApJ...758L...1Y}.]}
\label{GRB110301A_Polarization} 
\end{figure}

\subsection{GRB 110721A}

The GAP instrument detected GRB 110721A on 2011 July 21 at 04:47:38.9 UT.  At the time, GAP  was 0.699 AU away from the Earth \cite{2012ApJ...758L...1Y}.  The burst was observed by both Fermi/GBM and Fermi/LAT. Followup observations of the X-ray afterglow by Swift-XRT led to subsequent optical observations and a redhshift determination of $z = 0.38$.  The spectral analysis showed an $E_p$ value near 1130 keV at the time of maximum intensity, but the time integrated value of $E_p$ was much lower (about 390 keV), suggesting significant time variability. Its location was determined  to be  $30\deg$ from the GAP pointing axis. 

The GAP lightcurve is shown in the bottom panel of Fig. \ref{GRB110301A_LightCurve}. The $T_{90}$ for this event, as measured by Fermi-GBM, was 21.82 s.   The dashed lines in the lightcurve indicates the time interval for GAP polarization analysis. The burst background level was estimated using data just before and after the event. Estimated systematic uncertainties considered the spacecraft rotation rate (0.22 rpm) and the off-axis location of the burst. Results from the polarization analysis are shown in Fig. \ref{GRB110721A_Polarization}. The polarization parameters for the full time interval were $\Pi = 84^{+16}_{-28}\%$ and $\phi_p = 160\deg \pm 11\deg$.  A time-resolved analysis showed no evidence of time variability in the polarization parameters.

\begin{figure}
\centering
\includegraphics[height=2in]{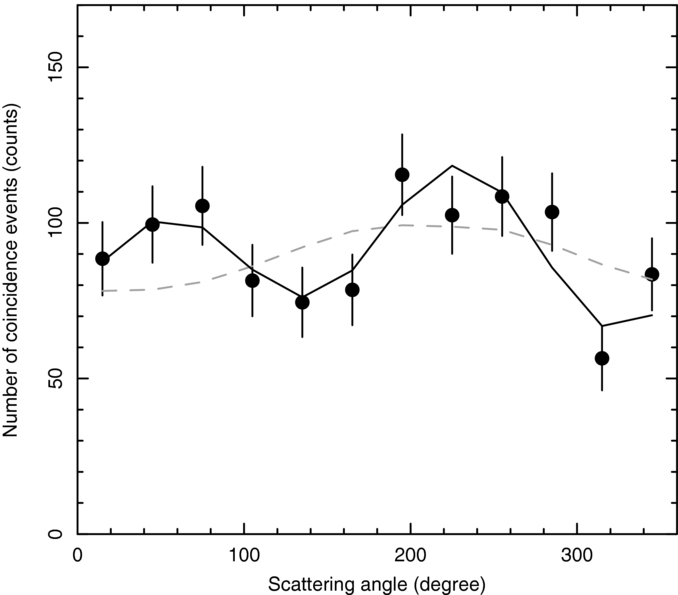}
\includegraphics[height=2in]{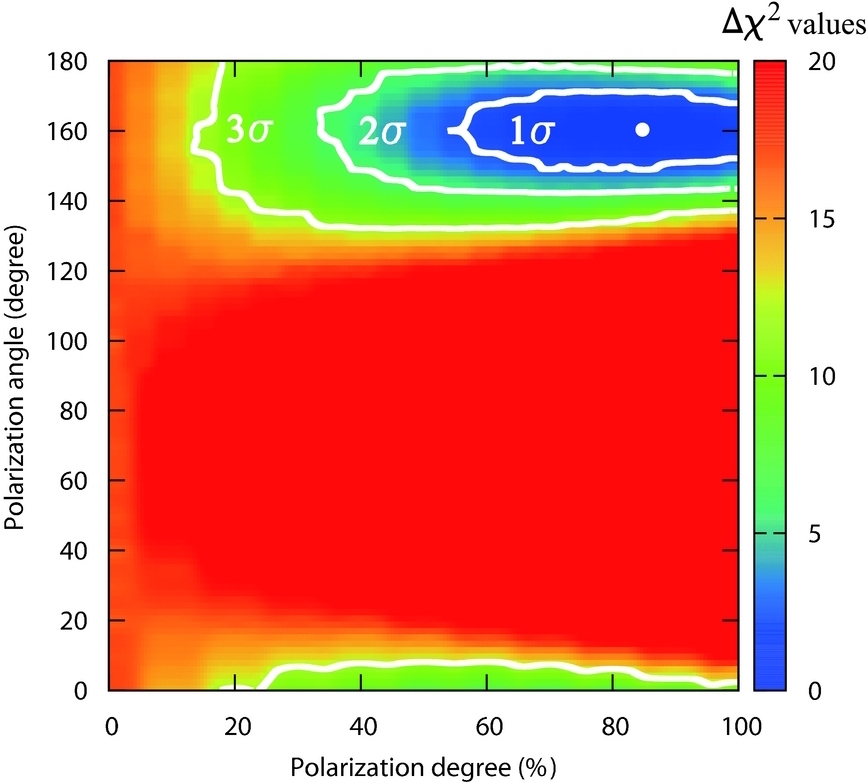}
\caption{The azimuthal scatter angle histograms of GAP data for GRB 110721A (left) and the corresponding constraints (1, 2, and 3 $\sigma$ contours) on the polarization parameters (right). [Figures from \citet{2012ApJ...758L...1Y}.]}
\label{GRB110721A_Polarization} 
\end{figure}

\subsection{GRB 140206A}

This burst was seen within the INTEGRAL field-of-view starting at  7:17:26 UT on February 2, 2014 \citep{2014MNRAS.444.2776G}. As shown in Fig. \ref{GRB140206A_LightCurve}, the burst consisted of two pulses (separated by about 60 s), the second of which was the most intense.  The burst was also seen by both Swift and Fermi/GBM.  A relatively bright optical afterglow (15th magnitude) was reported by many observers, which ultimately led to a redshift measurement ($z \sim 2.7$).

At the time of the burst, the INTEGRAL spacecraft had just passed through the radiation belts and was experiencing a relatively high level of background. This resulted in the saturation of the available telemetry, which limited the ability to transmit all of the IBIS data.   Fortunately, the Compton mode IBIS data, which is the primary mode used for polarization studies, is placed into the telemetry stream with the highest priority. As can be seen in Fig. \ref{GRB140206A_LightCurve} (second panel), there was a significant loss of singles ISGRI data, but there was no loss of Compton mode data. The burst was also seen by Fermi/GBM, but the first peak was obscured by Earth occultation (which explains why only the second peak is visible in the Fermi data shown in Fig \ref{GRB140206A_LightCurve}). 

The polarization of the second peak was studied using IBIS Compton mode data from the 200--400 keV energy interval.  The results, shown in Fig. \ref{GRB140206A_Polarization}, give a lower limit to the polarization (at the 68\% confidence level) of 48\% and a polarization angle of $\phi_p = 80\deg \pm 15\deg$.

\begin{figure}
\centering
\includegraphics[height=3.5in]{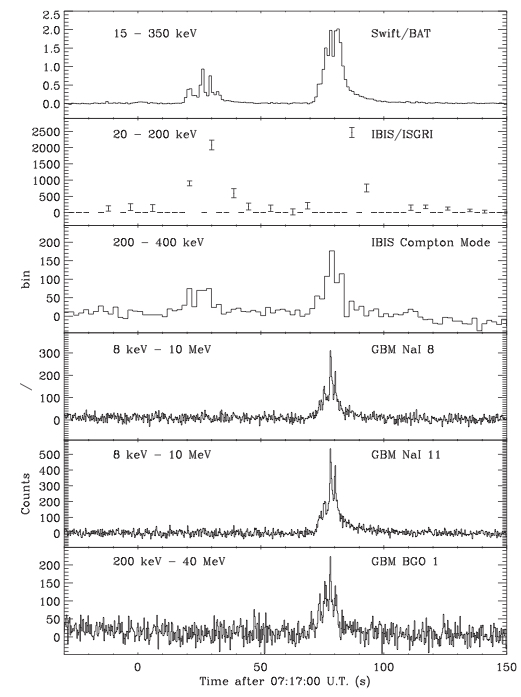}
\caption{The  lightcurves of GRB 140206A for various energy intervals from Swift/BAT, INTEGRAL/IBIS, and Fermi/GBM. Telemetry saturation resulted in a loss of ISGRI data and the first peak was not seen by Fermi due to Earth occultation. [Figure from \citet{2014MNRAS.444.2776G}.]}
\label{GRB140206A_LightCurve} 
\end{figure}

\begin{figure}
\centering
\includegraphics[height=1.65in]{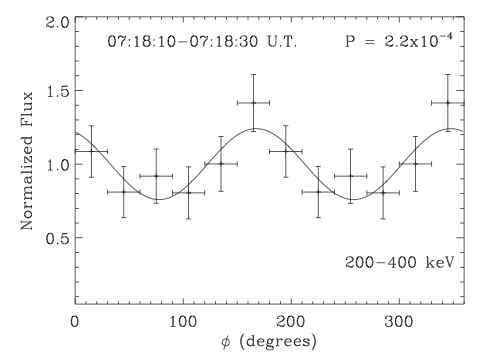}
\includegraphics[height=1.65in]{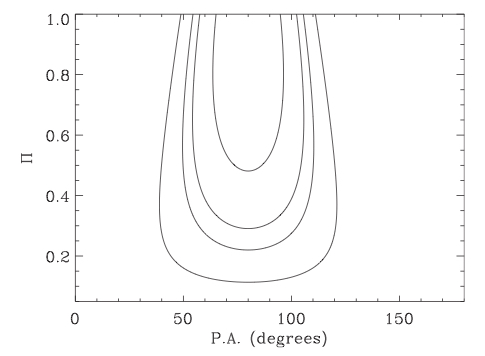}
\caption{The azimuthal scatter angle histograms of INTEGRAL/IBIS data for GRB 140206A in the 200--400 keV energy range (left) and the corresponding constraints on the polarization parameters (right).  [Figures from \citet{2014MNRAS.444.2776G}.]}
\label{GRB140206A_Polarization} 
\end{figure}

\subsection{GRB 151006A}

There have already been reports of possible detections of GRB polarization with the CZTI instrument on Astrosat, with GRB 151006A \citep{2015GCN..18422...1B} and GRB 160131A \citep{2016GCN..19011...1V}.  In both cases, preliminary reports suggested high levels of polarization.  The first detailed results for GRB 151006A have just been published \citep{Rao:2016tt}.  This event took place on the very first day of CZTI operations and was also seen by Swift, Fermi, and the CALET Gamma-ray Burst Monitor (CGBM) on the ISS.  Fermi/GBM reported a $T_{90}$ of $\sim$ 84 s.  The burst was located $\sim60^{\circ}$ from the Astrosat pointing direction. The burst was clearly seen in coincidence events that were identified as being consistent with Compton scattering. Analysis of these data indicated evidence for polarization at a significance level of $\sim1.5\sigma$.  More complete analysis of these data is awaiting detailed simulations of the CZTI collimator and surrounding spacecraft mass.  To date, only a zeroth order mass model has been employed. An understanding of the effects of the surrounding mass is crucial to the analysis, in that any asymmetric scattering and/or absorption effects can distort the polarization signal or mimic a polarization signal when none is present. The true significance of this observation must therefore await further analysis.

\section{Future Prospects}

RHESSI, INTEGRAL, and Astrosat all remain operational and offer the possibility of additional GRB polarization measurements. IKAROS/GAP is also still operational, but only intermittently. Several other instruments  are actively under development and may soon be able to provide additional data.  These active programs are described in this section.

\subsection{TSUBAME}

Tsubame is a microsatellite developed at the Tokyo Institute of Technology that was placed into orbit in November, 2014 \citep{2014SPIE.9144E..0LY,2011SPIE.8145E..08Y,2008AIPC.1000..607A}.  The satellite uses a nearly cubic platform measuring 50 cm $\times$ 50  cm $\times$ 47 cm. It carries two instruments.  One is a Wide field-of-view Burst Monitor (WBM), consisting of five CsI scintillators mounted on five faces of the satellite structure. When a GRB is detected, the relative counting rate in these five detectors is used to determine the GRB location.  For bright GRBs, location accuracy of 5$\deg$ is possible. Once the location is determined, the spacecraft is designed to rapidly slew to point its Hard X-ray Compton Polarimeter (HXCP) to the burst direction. Pointed observations can begin 15 s after the burst trigger. 

\begin{figure}
\centering
\includegraphics[height=3.0in]{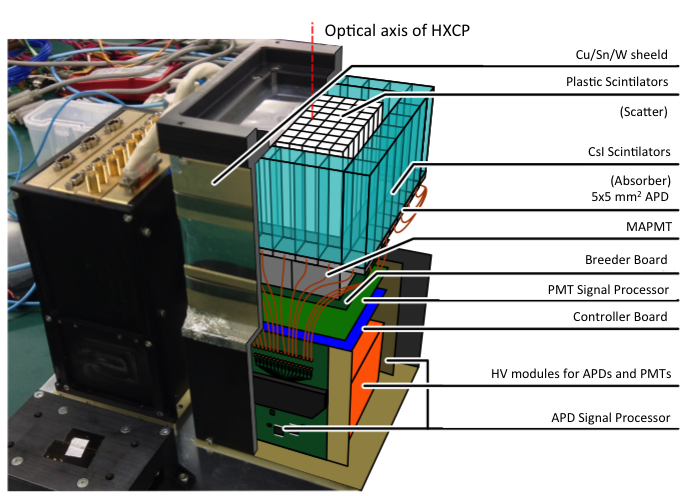}
\caption{The components of the TSUBAME Hard X-ray Compton Polarimeter. [Figure from \citet{2015arXiv150301975K}.]}
\label{TSUBAME} 
\end{figure}

As the name implies, the HXCP is a Compton scatter polarimeter. The central part of the instrument is an array of  8 $\times$ 8 plastic scintillator scattering elements, each  of which is 6.5 mm $\times$ 6.5 mm $\times$ 49 mm in size. The plastic array is read out by four 16-channel MAPMTs. Surrounding the scattering elements are 28 CsI scintillators, each 6.5 mm $\times$ 10 mm $\times$ 49 mm in size. The detector covers the energy range of 30-200 keV with an effective area of $\sim7$ cm$^2$ and a  15$\deg \times$ 15$\deg$ field-of-view. Since its launch in 2014, there have been no reports of GRB observations. Technical problems with the spacecraft telemetry have so far prevented successful operations.

\subsection{POLAR}

POLAR is an experiment that was launched on the Chinese space station Tiagong 2 in September, 2016 \citep{2016APh....83....6X,2011NIMPA.648..139O,2005NIMPA.550..616P}. Designed to measure polarization in the 50-500 keV energy range, it is based on the use of small, plastic scintillating bars (6 mm $\times$ 6 mm $\times$ 200 mm) that are read out out by 64-channel multi-anode photomultiplier tubes (MAPMTs). Each MAPMT provides a single anode readout for 64 scintillator elements. A 5 $\times$ 5 array of MAPMTs results in a 40 $\times$ 40 array of scintillator elements. The low-Z plastic scintillator is optimized for Compton scattering, but the energy response is limited due to the low probability of total photon absorption, resulting in a very non-diagonal response function. The instrument response has been characterized by extensive calibration efforts \citep[e.g.,][]{2016APh....83....6X}. Passive shielding will be used to limit the on-orbit background. The detector will provide $\sim50$ cm$^2$ of effective area at 100 keV and $\sim90$ cm$^2$ of effective area at 300 keV (Produit, private communication) with a modulation factor ranging from $\sim30\%$ to  $\sim50\%$ \cite{2011NIMPA.648..139O}. The field-of-view is about 2 steradian (1/3 of the visible sky). The expected lifetime is about three years.

\begin{figure}
\centering
\includegraphics[height=2.in]{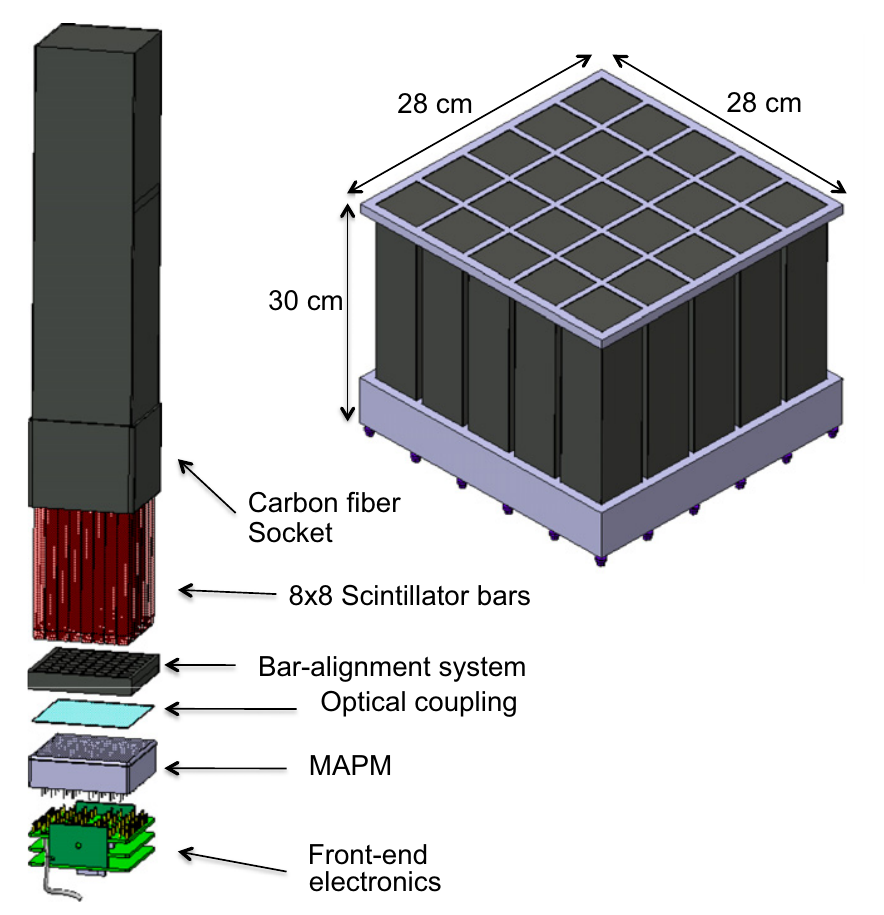}
\includegraphics[height=2.in]{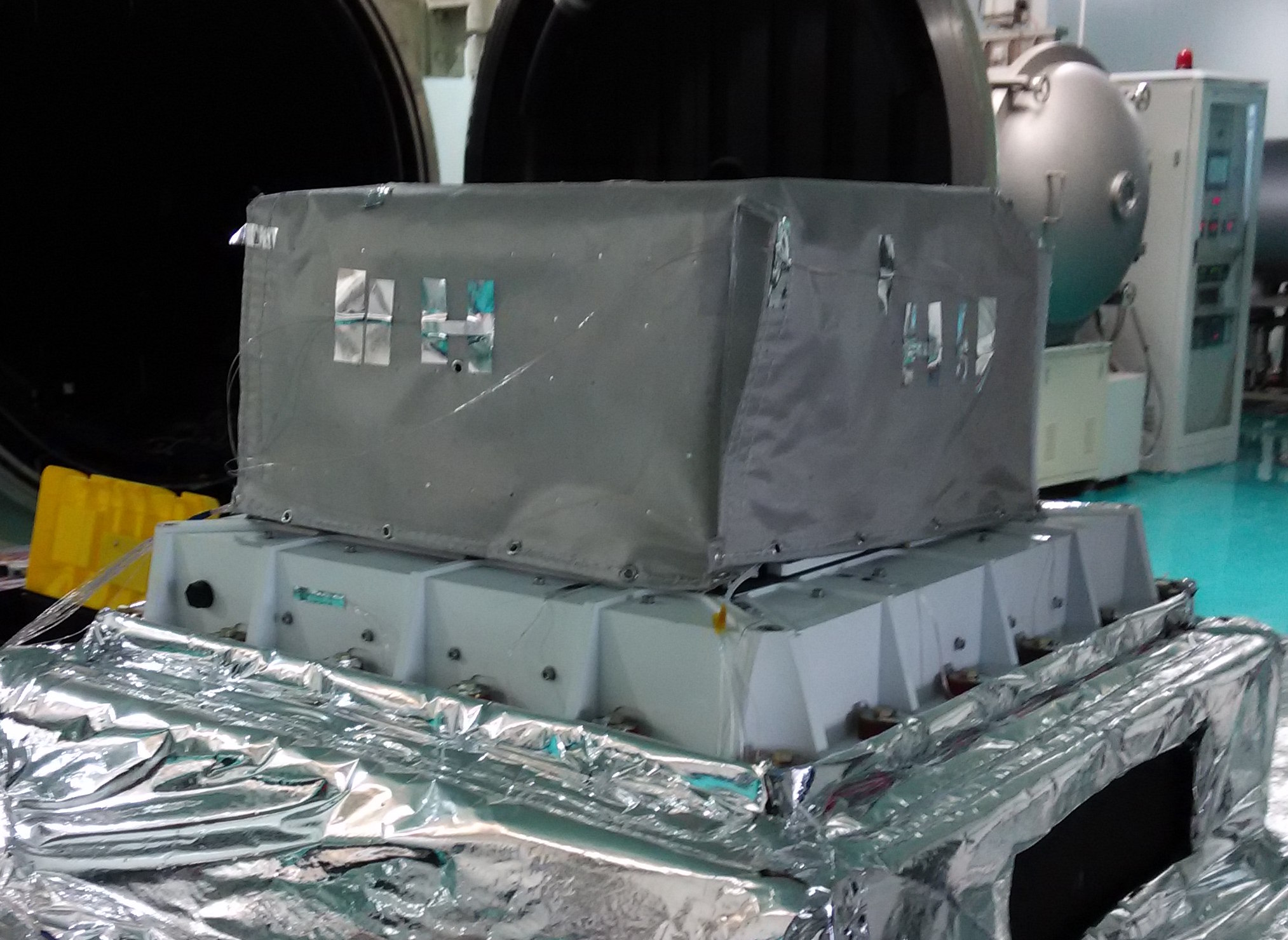}
\caption{A schematic view of the POLAR instrument (left) showing one of the 25 detector modules and the $5 \times 5$ array. [Figure from \cite{2010NIMPA.624..624S}.] The photo (right) shows POLAR during recent thermal-vacuum testing. [Photo courtesy Nicolas Produit.]}
\label{POLAR} 
\end{figure}

\subsection{NCT / COSI}

The Compton Spectrometer and Imager (COSI) \citep{2015NIMPA.784..359C,2014SPIE.9144E..3MK}, previously known as the Nuclear Compton Telescope (NCT), is a balloon-borne, Ge-based Compton telescope that operates in the 200 keV -- 5 MeV energy band, with a field-of-view that covers about $25\%$ of the sky.  COSI uses an array of 12 cross-strip germanium detectors to track the path of incident photons, where position and energy deposits from Compton interactions allow for a reconstruction of the source position in the sky, an inherent measure of the linear polarization, and significant background reduction. The size of each of the 12 HPGe crystals is 8 cm $\times$ 8 cm $\times$ 1.5 cm.  Orthogonal strip electrodes (with a pitch of 2 mm) are deposited on both the anode and cathode surface of each detector. The detectors have very good position resolution (1.6 mm$^3$) and provide excellent spectral resolution  ($<1\%$).  Although the effective area is rather small, ranging from $\sim18$ cm$^2$ at 300 keV to $\sim6$ cm$^2$ at 2 MeV, the imaging capability, coupled with the low background nature of the design, provides significant source sensitivity. 

GRB polarimetry is one of its many science goals of this experiment, which is designed to fly as an ultra-long duration balloon (ULDB) payload. Given an anticipated ULDB flight time of 100 days, it was expected to measure the polarization of $\sim8$ GRBs and set an upper limit on the degree of polarization for $\sim8$ more. COSI was  launched from Wanaka, New Zealand on May 16, 2016.  The ensuing flight lasted 46 days, although problems with the stability of the balloon resulted in the payload spending much of its time at altitudes too low for effective measurements. During that time it observed at least one intense GRB (GRB 160530A), for which a polarization measurement should be feasible. No results have been reported as yet.

\begin{figure}
\centering
\includegraphics[height=2.75in]{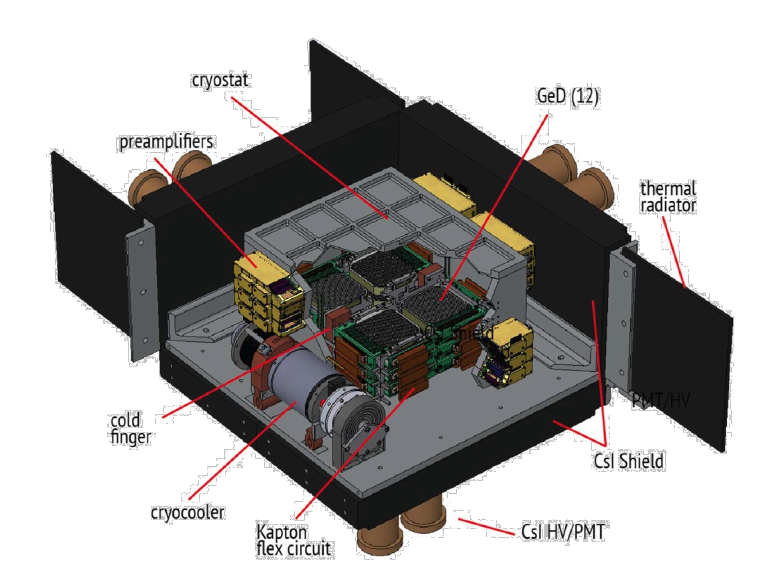}
\caption{A schematic view of of the COSI cryostat (with the Ge detectors) and the surrounding CsI shields. The mechanical cryocooler is seen in the foreground.  [Figure from \citet{2014SPIE.9144E..3MK}.]}
\label{COSI} 
\end{figure}

\subsection{GRAPE}

The Gamma RAy Polarimeter Experiment (GRAPE) is a balloon program, whose goal is to conduct GRB observations on long duration balloon flights. GRAPE is a large field-of-view Compton polarimeter operating in the 50--500 keV energy range. The instrument design is a modular one,  based on the use of a 64-channel MAPMT to read out a hybrid array of small scintillator elements, each of which has a volume of 5 mm $\times$ 5 mm $\times$ 50 mm. GRAPE uses two scintillator types (plastic and CsI(Tl)), which provides moderate energy resolution and low susceptibility to neutron background. The scintillators are arranged with a central array of 6 $\times$ 6 plastic scintillators surrounded by CsI(Tl) scintillators. Ideal events are those that scatter from a plastic element to a CsI(Tl) element. Each module has an effective area of $\sim2$ cm$^2$ at 150 keV. 

To date, the GRAPE program has conducted two balloon flights intended to validate the instrument design by performing a measurement of the Crab polarization. In 2011, the payload was flown with an array of 16 polarimeter modules. Higher-than-expected background levels prevented a successful Crab measurement \citep{2014SPIE.9144E..3PM}. A subsequent flight in 2014 was made with improved shielding and a larger array of 24 modules, but the flight did not last long enough for a full observation of the Crab. The program is currently working on a revised design in preparation for a series of long duration balloon flights \citep{McConnell:2016gd}.

\begin{figure}
\centering
\includegraphics[height=2.75in]{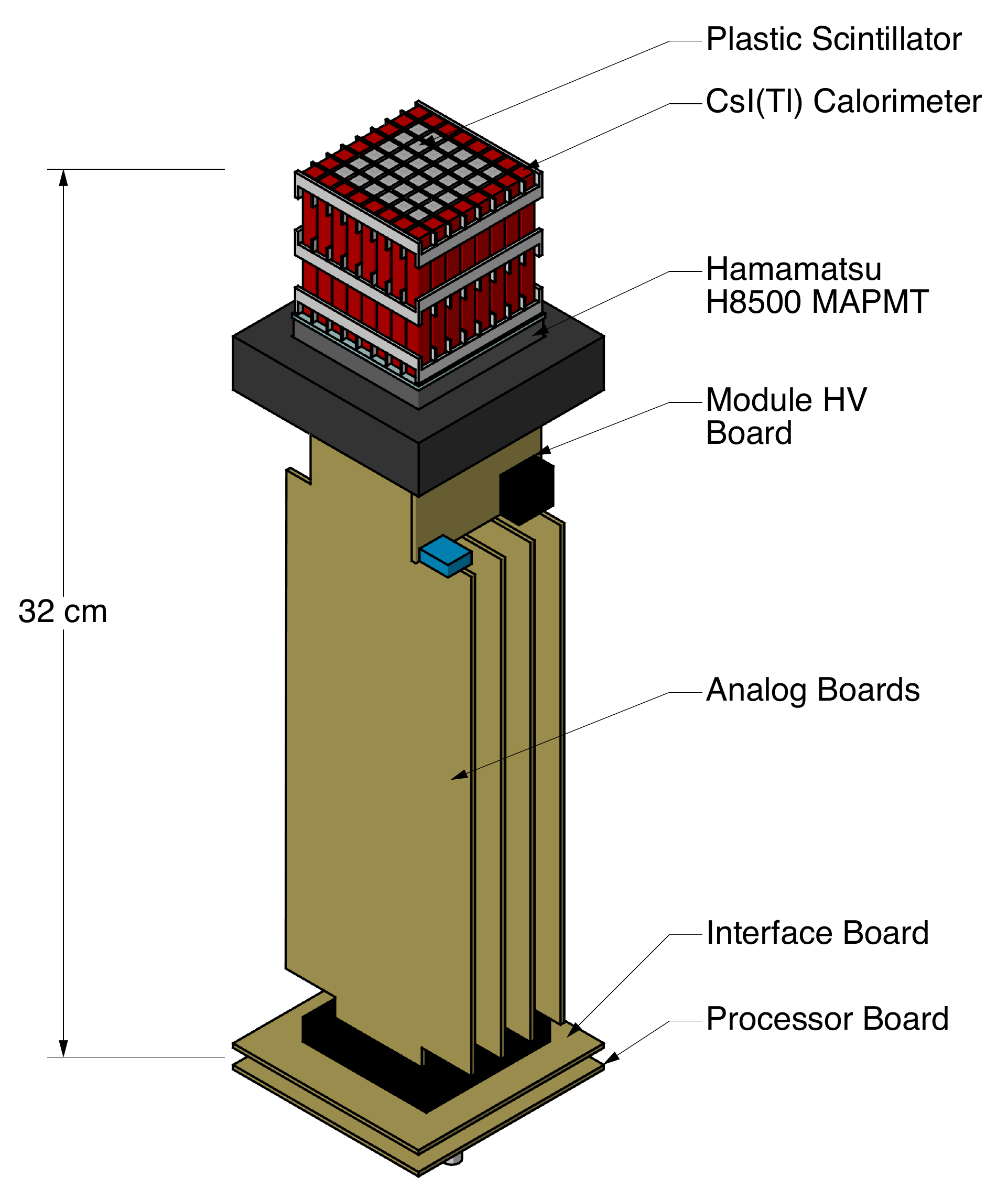}
\includegraphics[height=2.75in]{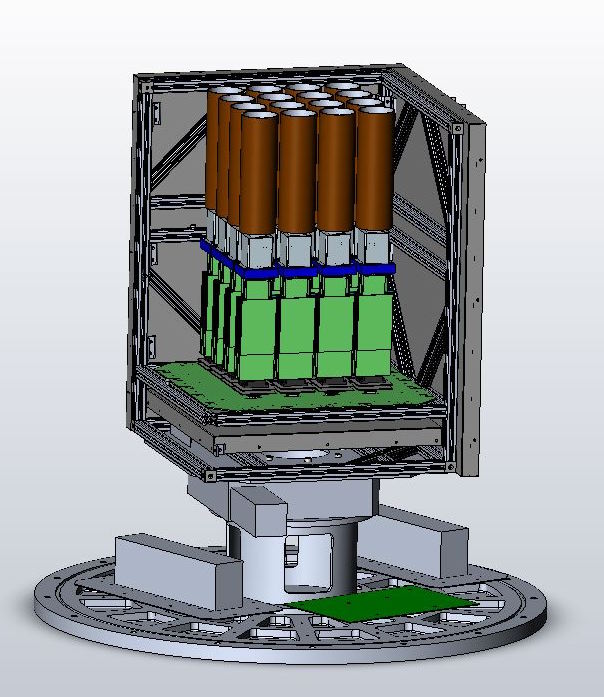}
\caption{A schematic view of single GRAPE module (left) showing the layout of the scintillator and elements and the discrete electronics boards. The configuration for the 2011 validation balloon flight (right) included an array of 16 collimated modules optimized for observations of the Crab. [Figures from  \citet{2014SPIE.9144E..3PM}.]}
\label{GRAPE} 
\end{figure}

This project has provided the basis for several space-borne GRB polarimeter concepts. POET (POlarimeters for Energetic Transients) was proposed in both 2008 \citep{2007SPIE.6686E..29H,2009AIPC.1133...64M} and 2014 \citep{2014SPIE.9144E..1IB,2014SPIE.9144E..0OM} as a NASA Small Explorer (SMEX) mission. In both cases, the payload consisted of a high energy (50--500 keV) Compton scatter polarimeter and a low energy (2--15 keV) photoelectric polarimeter.  PETS (Polarimetry of Energetic Transients in Space) was proposed in 2012, to fly a high energy polarimeter as an external payload on the International Space Station (ISS).  A similar proposal concept, known as LEAP (LargE Area burst Polarimeter), is currently being prepared for submission in late 2016.  In order to avoid optical crosstalk issues, more recent designs for the high energy polarimeter have moved away from the use of MAPMTs to the use of individual PMT-scintillator detector elements.

\section{Discussion}

To date, there have been reports of polarization measured in ten GRBs.  The following general remarks can be made with regards to these published results.

\begin{itemize}
  \item Polarizaton measurements of  two GRBs (GRB 930131 and GRB 960924) were based on the use of albedo polarimetry.
  \item Polarizaton measurements of  eight GRBs (GRB 021206, GRB 041219a, GRB 061122, GRB 100826a, GRB 110301a, GRB 110721a, GRB 120206a, and GRB 151006a) were based on the use of Compton scattering polarimetry.
  \item Three GRBs (GRB 100826a, GRB 110301a, and GRB 110721a) have been studied using instrument an instrument designed specifically for GRB polarization measurements.
  \item In one case (GRB 021206), the same data has been used in three independent studies.  The results were inconsistent and led to significant controversy.
  \item In two cases (GRB 041219a and GRB 061122), observations were made with two different instruments (INTEGRAL/SPI and INTEGRAL/IBIS). Some of the inconsistencies seen in the analysis of these data were ascribed to time variability.
\item All results are of limited statistical significance ($\sim1.5-4 \sigma$).  
\end{itemize}

Collectively, these data do not  provide a consistent picture of $\gamma$-ray polarization in GRBs.  The limited statistical significance of these results makes it difficult to draw any hard conclusions about the underlying physics.  The data {\em suggest} that GRB polarization levels are generally quite high, thus arguing for ordered magnetic field structures within the GRB jet.  But even this conclusion requires additional high-quality data  before the community can reach consensus. 

What is really needed here is a collection of high significance results to determine the true nature of GRB polarization.   Experimentally, we can identify several  important issues that should be considered in the design of future experiments.

\begin{description}

 \item[Instrument Rotation]  It has long been argued that polarimeter experiments can best handle systematic effects in the data (such as spatial nonuniformities in the background counting rate) by continuously rotating the instrument about the direction axis to the source, something which is impossible for a transient source with an unpredictable location.  To be truly effective,  however, the observation must be made over an integral number of rotations.   For long pointed observations of a persistent source, where the observation time is much longer than the rotation period, this is certainly the case. For transient sources, such as GRBs, this will never be the case.  Rotation is therefore not a viable solution for effectively handling systematic effects in GRB polarimeters. Of the observations that have been made to date, four have involved rotating spacecraft (RHESSI and IKAROS/GAP). The experience with these data does not provide compelling justification for instrument rotation. Indeed, one can even argue that rotation complicates the analysis of the data without providing any clear benefit.

  \item[Earth Albedo] A significant fraction of the flux from a GRB can be scattered off the Earth's atmosphere, especially at lower energies (below $\sim100$ keV).  This is the basis for albedo polarization measurements. The same phenomena can also hinder observations with Compton polarimeters. For instruments in Low Earth Orbit (LEO), this can potentially induce an asymmetric counting rate in the instrument and may lead to systematic effects in the polarization analysis.  Shielding (passive and/or active) can help to reduce the impact of scattered flux, but not all instruments considered here have been shielded. In fact, Earth albedo was mentioned as one potential systematic effect in the RHESSI data \citep{2004MNRAS.350.1288R}.

  \item[Saturated Counting Rates] The flux from a GRB can be quite intense.  This can result in significant instrumental deadtime, especially for large area detectors and/or detectors with relatively slow response (such as scintillators with long decay time or with delayed phosphorescence effects). The use of faster detectors (such as some of the faster scintillators that are now available) is one way to ameliorate this problem, by providing greater throughput. It is also better to have a larger number of smaller detectors than a smaller number of larger detectors, since smaller detectors result in lower counting rates and are thus less prone to deadtime effects. 
  
  \item[Telemetry Throughput] Another related issue (that has been seen many times with INTEGRAL) is that a high event rate can sometime result in loss of data due to telemetry limitations. This can be handled by the availability of sufficient on-board data storage, so that data can be sent down during periods of lower telemetry needs.

  \item[Coincidence Timing] Compton polarimeters typically require coincident trigger between two detector elements.  A small coincidence window is preferred to minimize the chance of accidental coincidences, especially during intense flux levels. Reliance on software coincidence can introduce systematic uncertainties to the analysis. Hardware coincidence techniques are preferred.
  
  \item[Pre-flight Calibrations] Many of the instruments considered here (including RHESSI and those on INTEGRAL) were never calibrated with  beams of known polarization. Knowledge of their instrument response has relied entirely on simulations that have never been validated with experimental data. Likewise, the albedo polarimetry approach is difficult (if not impossible) to calibrate directly; it must also rely entirely on simulations for the instrument response. Proper calibration data always serves as a means to identify potentially important instrumental issues.
  
  \item[Proper Modeling] Detailed Monte Carlo simulations play an important role in the analysis of polarization data.  Instrument asymmetries (mass distribution, detection efficiencies, etc.) can serve to either distort a true polarization signal or, more seriously, to generate a spurious signal when none exists. External effects, such as scattering from the Earth's atmosphere, can also impact the analysis. In some cases, instrument rotation can be used to mitigate these effects, but for transient events (such as GRBs), this approach is not viable. In this case, systematic effects can best be understood through extensive modeling with detailed instrument mass models.

\end{description}

It is widely believed that high energy polarization measurements of  prompt GRB emission will provide the next big step in our understanding of  GRBs.  Experimentally, this is a challenging endeavor.  Fortunately, several groups are actively pursuing additional measurements, both with currently operating instruments, and with new experimental designs that will be deployed in the not-too-distant future. With so much interest in the field, it is likely that  important breakthroughs will be made within the next few years, shedding new light on the nature of GRBs.

\section{Acknowledgements}

The author would like to thank his numerous collaborators over the past several years, including Matthew Baring, Peter Bloser, Mark Finger, Jessica Gaskin, Shuichi Gunji, Dieter Hartmann, Joanne Hill, R. Marc Kippen, Jason Legere, Rob Preece, James Ryan, David Smith, Kenji Toma, W. Thomas Vestrand, Colleen Wilson-Hodge, and Bing Zhang.  This work was supported by NASA Grants NNX13AB96G, NNX16AC05G, and NNX15AI70G. The author would also like to thank the anonymous referee for several useful comments.

\vspace*{0.5cm}

\section*{References}

\bibliography{GRB_Review}

\begin{thebibliography}{81}
\expandafter\ifx\csname natexlab\endcsname\relax\def\natexlab#1{#1}\fi
\providecommand{\url}[1]{\texttt{#1}}
\providecommand{\href}[2]{#2}
\providecommand{\path}[1]{#1}
\providecommand{\DOIprefix}{doi:}
\providecommand{\ArXivprefix}{arXiv:}
\providecommand{\URLprefix}{URL: }
\providecommand{\Pubmedprefix}{pmid:}
\providecommand{\doi}[1]{\href{http://dx.doi.org/#1}{\path{#1}}}
\providecommand{\Pubmed}[1]{\href{pmid:#1}{\path{#1}}}
\providecommand{\bibinfo}[2]{#2}
\ifx\xfnm\relax \def\xfnm[#1]{\unskip,\space#1}\fi
\bibitem[{Arimoto et~al.(2008)Arimoto, Tsubuku, Toizumi, Kobayashi, Yatsu,
  Shimokawabe, Kataoka, Kawai, Omagari, Fujiwara, Konda, Tanaka, Maeno,
  Yamanaka, Ashida, Nishida, Fujihashi, Ikeda, Inagawa, Miura and
  Matunaga}]{2008AIPC.1000..607A}
\bibinfo{author}{Arimoto, M.}, \bibinfo{author}{Tsubuku, Y.},
  \bibinfo{author}{Toizumi, T.}, \bibinfo{author}{Kobayashi, M.},
  \bibinfo{author}{Yatsu, Y.}, \bibinfo{author}{Shimokawabe, T.},
  \bibinfo{author}{Kataoka, J.}, \bibinfo{author}{Kawai, N.},
  \bibinfo{author}{Omagari, K.}, \bibinfo{author}{Fujiwara, K.},
  \bibinfo{author}{Konda, Y.}, \bibinfo{author}{Tanaka, Y.},
  \bibinfo{author}{Maeno, M.}, \bibinfo{author}{Yamanaka, T.},
  \bibinfo{author}{Ashida, H.}, \bibinfo{author}{Nishida, J.},
  \bibinfo{author}{Fujihashi, K.}, \bibinfo{author}{Ikeda, T.},
  \bibinfo{author}{Inagawa, S.}, \bibinfo{author}{Miura, Y.},
  \bibinfo{author}{Matunaga, S.}, \bibinfo{year}{2008}.
\newblock \bibinfo{title}{{X-ray polarimetry small satellite TSUBAME}}.
\newblock \bibinfo{journal}{AIP Conference Proceedings} \bibinfo{volume}{1000},
  \bibinfo{pages}{607--610}.
\bibitem[{Band et~al.(1993)Band, Matteson, Ford, Schaefer, Palmer, Teegarden,
  Cline, Briggs, Paciesas, Pendleton, Fishman, Kouveliotou, Meegan, Wilson and
  Lestrade}]{1993ApJ...413..281B}
\bibinfo{author}{Band, D.L.}, \bibinfo{author}{Matteson, J.L.},
  \bibinfo{author}{Ford, L.}, \bibinfo{author}{Schaefer, B.},
  \bibinfo{author}{Palmer, D.M.}, \bibinfo{author}{Teegarden, B.},
  \bibinfo{author}{Cline, T.L.}, \bibinfo{author}{Briggs, M.},
  \bibinfo{author}{Paciesas, W.S.}, \bibinfo{author}{Pendleton, G.N.},
  \bibinfo{author}{Fishman, G.J.}, \bibinfo{author}{Kouveliotou, C.},
  \bibinfo{author}{Meegan, C.A.}, \bibinfo{author}{Wilson, R.},
  \bibinfo{author}{Lestrade, P.}, \bibinfo{year}{1993}.
\newblock \bibinfo{title}{{BATSE observations of gamma-ray burst spectra. I -
  Spectral diversity}}.
\newblock \bibinfo{journal}{Astrophysical Journal} \bibinfo{volume}{413},
  \bibinfo{pages}{281--292}.
\bibitem[{Bhalerao et~al.(2015)Bhalerao, Bhattacharya, Rao and
  Vadawale}]{2015GCN..18422...1B}
\bibinfo{author}{Bhalerao, V.}, \bibinfo{author}{Bhattacharya, D.},
  \bibinfo{author}{Rao, A.R.}, \bibinfo{author}{Vadawale, S.},
  \bibinfo{year}{2015}.
\newblock \bibinfo{title}{{GRB 151006A: Astrosat CZTI detection.}}
\newblock \bibinfo{journal}{GCN Circular} \bibinfo{volume}{18422}.
\bibitem[{Bloser et~al.(2014)Bloser, McConnell, Legere, Ertley, Hill, Kippen
  and Ryan}]{2014SPIE.9144E..1IB}
\bibinfo{author}{Bloser, P.F.}, \bibinfo{author}{McConnell, M.L.},
  \bibinfo{author}{Legere, J.S.}, \bibinfo{author}{Ertley, C.D.},
  \bibinfo{author}{Hill, J.E.}, \bibinfo{author}{Kippen, R.M.},
  \bibinfo{author}{Ryan, J.M.}, \bibinfo{year}{2014}.
\newblock \bibinfo{title}{{A high-energy Compton polarimeter for the POET SMEX
  mission}}.
\newblock \bibinfo{journal}{Proceedings of SPIE} \bibinfo{volume}{9144},
  \bibinfo{pages}{91441I--91441I--6}.
\bibitem[{Chattopadhyay et~al.(2014)Chattopadhyay, Vadawale, Rao, Sreekumar and
  Bhattacharya}]{2014ExA....37..555C}
\bibinfo{author}{Chattopadhyay, T.}, \bibinfo{author}{Vadawale, S.V.},
  \bibinfo{author}{Rao, A.R.}, \bibinfo{author}{Sreekumar, S.},
  \bibinfo{author}{Bhattacharya, D.}, \bibinfo{year}{2014}.
\newblock \bibinfo{title}{{Prospects of hard X-ray polarimetry with
  Astrosat-CZTI}}.
\newblock \bibinfo{journal}{Experimental Astronomy} \bibinfo{volume}{37},
  \bibinfo{pages}{555--577}.
\bibitem[{Chauvin et~al.(2013)Chauvin, Roques, Clark and
  Jourdain}]{2013ApJ...769..137C}
\bibinfo{author}{Chauvin, M.}, \bibinfo{author}{Roques, J.P.},
  \bibinfo{author}{Clark, D.J.}, \bibinfo{author}{Jourdain, E.},
  \bibinfo{year}{2013}.
\newblock \bibinfo{title}{{Polarimetry in the Hard X-Ray Domain with INTEGRAL
  SPI}}.
\newblock \bibinfo{journal}{Astrophysical Journal} \bibinfo{volume}{769},
  \bibinfo{pages}{137}.
\bibitem[{Chiu et~al.(2015)Chiu, Boggs, Chang, Tomsick, Zoglauer, Amman, Chang,
  Chou, Jean, Kierans, Lin, Lowell, Shang, Tseng, von Ballmoos and
  Yang}]{2015NIMPA.784..359C}
\bibinfo{author}{Chiu, J.L.}, \bibinfo{author}{Boggs, S.E.},
  \bibinfo{author}{Chang, H.K.}, \bibinfo{author}{Tomsick, J.A.},
  \bibinfo{author}{Zoglauer, A.}, \bibinfo{author}{Amman, M.},
  \bibinfo{author}{Chang, Y.H.}, \bibinfo{author}{Chou, Y.},
  \bibinfo{author}{Jean, P.}, \bibinfo{author}{Kierans, C.},
  \bibinfo{author}{Lin, C.H.}, \bibinfo{author}{Lowell, A.},
  \bibinfo{author}{Shang, J.R.}, \bibinfo{author}{Tseng, C.H.},
  \bibinfo{author}{von Ballmoos, P.}, \bibinfo{author}{Yang, C.Y.},
  \bibinfo{year}{2015}.
\newblock \bibinfo{title}{{The upcoming balloon campaign of the Compton
  Spectrometer and Imager (COSI)}}.
\newblock \bibinfo{journal}{Nuclear Inst. and Methods in Physics Research}
  \bibinfo{volume}{784}, \bibinfo{pages}{359--363}.
\bibitem[{Coburn and Boggs(2003)}]{2003Natur.423..415C}
\bibinfo{author}{Coburn, W.}, \bibinfo{author}{Boggs, S.E.},
  \bibinfo{year}{2003}.
\newblock \bibinfo{title}{{Polarization of the prompt $\gamma$-ray emission
  from the $\gamma$-ray burst of 6 December 2002}}.
\newblock \bibinfo{journal}{Nature} \bibinfo{volume}{423},
  \bibinfo{pages}{415--417}.
\bibitem[{Eichler and Levinson(2003)}]{2003ApJ...596L.147E}
\bibinfo{author}{Eichler, D.}, \bibinfo{author}{Levinson, A.},
  \bibinfo{year}{2003}.
\newblock \bibinfo{title}{{Polarization of Gamma-Ray Bursts via Scattering off
  a Relativistic Sheath}}.
\newblock \bibinfo{journal}{Astrophysical Journal} \bibinfo{volume}{596},
  \bibinfo{pages}{L147--L150}.
\bibitem[{Forot et~al.(2007)Forot, Laurent, Lebrun and
  Limousin}]{2007ApJ...668.1259F}
\bibinfo{author}{Forot, M.}, \bibinfo{author}{Laurent, P.},
  \bibinfo{author}{Lebrun, F.}, \bibinfo{author}{Limousin, O.},
  \bibinfo{year}{2007}.
\newblock \bibinfo{title}{{Compton Telescope with a Coded Aperture Mask:
  Imaging with the INTEGRAL/IBIS Compton Mode}}.
\newblock \bibinfo{journal}{Astrophysical Journal} \bibinfo{volume}{668},
  \bibinfo{pages}{1259--1265}.
\bibitem[{G{\"o}tz et~al.(2013)G{\"o}tz, Covino, Fernandez-Soto, Laurent and
  Bosnjak}]{2013MNRAS.431.3550G}
\bibinfo{author}{G{\"o}tz, D.}, \bibinfo{author}{Covino, S.},
  \bibinfo{author}{Fernandez-Soto, A.}, \bibinfo{author}{Laurent, P.},
  \bibinfo{author}{Bosnjak, Z.}, \bibinfo{year}{2013}.
\newblock \bibinfo{title}{{The polarized gamma-ray burst GRB 061122}}.
\newblock \bibinfo{journal}{Monthly Notices of the Royal Astronomical Society}
  \bibinfo{volume}{431}, \bibinfo{pages}{3550--3556}.
\bibitem[{G{\"o}tz et~al.(2014)G{\"o}tz, Laurent, Antier, Covino, D'Avanzo,
  D'Elia and Melandri}]{2014MNRAS.444.2776G}
\bibinfo{author}{G{\"o}tz, D.}, \bibinfo{author}{Laurent, P.},
  \bibinfo{author}{Antier, S.}, \bibinfo{author}{Covino, S.},
  \bibinfo{author}{D'Avanzo, P.}, \bibinfo{author}{D'Elia, V.},
  \bibinfo{author}{Melandri, A.}, \bibinfo{year}{2014}.
\newblock \bibinfo{title}{{GRB 140206A: the most distant polarized gamma-ray
  burst}}.
\newblock \bibinfo{journal}{Monthly Notices of the Royal Astronomical Society}
  \bibinfo{volume}{444}, \bibinfo{pages}{2776--2782}.
\bibitem[{G{\"o}tz et~al.(2009)G{\"o}tz, Laurent, Lebrun, Daigne and Bo{\v
  s}njak}]{2009ApJ...695L.208G}
\bibinfo{author}{G{\"o}tz, D.}, \bibinfo{author}{Laurent, P.},
  \bibinfo{author}{Lebrun, F.}, \bibinfo{author}{Daigne, F.},
  \bibinfo{author}{Bo{\v s}njak, {\v Z}.}, \bibinfo{year}{2009}.
\newblock \bibinfo{title}{{Variable Polarization Measured in the Prompt
  Emission of GRB 041219A Using IBIS on Board INTEGRAL}}.
\newblock \bibinfo{journal}{The Astrophysical Journal Letters}
  \bibinfo{volume}{695}, \bibinfo{pages}{L208--L212}.
\bibitem[{Granot(2003)}]{2003ApJ...596L..17G}
\bibinfo{author}{Granot, J.}, \bibinfo{year}{2003}.
\newblock \bibinfo{title}{{The Most Probable Cause for the High Gamma-Ray
  Polarization in GRB 021206}}.
\newblock \bibinfo{journal}{Astrophysical Journal} \bibinfo{volume}{596},
  \bibinfo{pages}{L17--L21}.
\bibitem[{Hill et~al.(2007)Hill, Barthelmy, Black, Deines-Jones, Jahoda,
  Sakamoto, Kaaret, McConnell, Bloser, Macri, Legere, Ryan, Smith and
  Zhang}]{2007SPIE.6686E..29H}
\bibinfo{author}{Hill, J.E.}, \bibinfo{author}{Barthelmy, S.D.},
  \bibinfo{author}{Black, J.K.}, \bibinfo{author}{Deines-Jones, P.},
  \bibinfo{author}{Jahoda, K.}, \bibinfo{author}{Sakamoto, T.},
  \bibinfo{author}{Kaaret, P.E.}, \bibinfo{author}{McConnell, M.L.},
  \bibinfo{author}{Bloser, P.F.}, \bibinfo{author}{Macri, J.R.},
  \bibinfo{author}{Legere, J.S.}, \bibinfo{author}{Ryan, J.M.},
  \bibinfo{author}{Smith, B.R.J.}, \bibinfo{author}{Zhang, B.},
  \bibinfo{year}{2007}.
\newblock \bibinfo{title}{{A burst chasing x-ray polarimeter}}.
\newblock \bibinfo{journal}{Proceedings of SPIE} \bibinfo{volume}{6686},
  \bibinfo{pages}{29}.
\bibitem[{Kalemci et~al.(2007)Kalemci, Boggs, Kouveliotou, Finger and
  Baring}]{2007ApJS..169...75K}
\bibinfo{author}{Kalemci, E.}, \bibinfo{author}{Boggs, S.E.},
  \bibinfo{author}{Kouveliotou, C.}, \bibinfo{author}{Finger, M.H.},
  \bibinfo{author}{Baring, M.G.}, \bibinfo{year}{2007}.
\newblock \bibinfo{title}{{Search for Polarization from the Prompt Gamma-Ray
  Emission of GRB 041219a with SPI on INTEGRAL}}.
\newblock \bibinfo{journal}{The Astrophysical Journal Supplement Series}
  \bibinfo{volume}{169}, \bibinfo{pages}{75--82}.
\bibitem[{Kalemci et~al.(2004)Kalemci, Boggs, Wunderer and
  Jean}]{2004ESASP.552..859K}
\bibinfo{author}{Kalemci, E.}, \bibinfo{author}{Boggs, S.E.},
  \bibinfo{author}{Wunderer, C.}, \bibinfo{author}{Jean, P.},
  \bibinfo{year}{2004}.
\newblock \bibinfo{title}{{Measuring Polarization with SPI on INTEGRAL}}, in:
  \bibinfo{booktitle}{Proceedings of the 5th INTEGRAL Workshop on the INTEGRAL
  Universe (ESA SP-552). 16-20 February 2004}, p. \bibinfo{pages}{859}.
\bibitem[{Kierans et~al.(2014)Kierans, Boggs, Lowell, Tomsick, Zoglauer, Amman,
  Chiu, Chang, Lin, Jean, von Ballmoos, Yang, Shang, Tseng, Chou and
  Chang}]{2014SPIE.9144E..3MK}
\bibinfo{author}{Kierans, C.A.}, \bibinfo{author}{Boggs, S.E.},
  \bibinfo{author}{Lowell, A.}, \bibinfo{author}{Tomsick, J.},
  \bibinfo{author}{Zoglauer, A.}, \bibinfo{author}{Amman, M.},
  \bibinfo{author}{Chiu, J.L.}, \bibinfo{author}{Chang, H.K.},
  \bibinfo{author}{Lin, C.H.}, \bibinfo{author}{Jean, P.}, \bibinfo{author}{von
  Ballmoos, P.}, \bibinfo{author}{Yang, C.Y.}, \bibinfo{author}{Shang, J.R.},
  \bibinfo{author}{Tseng, C.H.}, \bibinfo{author}{Chou, Y.},
  \bibinfo{author}{Chang, Y.H.}, \bibinfo{year}{2014}.
\newblock \bibinfo{title}{{Calibration of the Compton Spectrometer and Imager
  in preparation for the 2014 balloon campaign}}.
\newblock \bibinfo{journal}{Proceedings of SPIE} \bibinfo{volume}{9144},
  \bibinfo{pages}{91443M--91443M--16}.
\bibitem[{Kouveliotou et~al.(1994)Kouveliotou, Preece, Bhat, Fishman, Meegan,
  Horack, Briggs, Paciesas, Pendleton, Band, Matteson, Palmer, Teegarden and
  Norris}]{1994ApJ...422L..59K}
\bibinfo{author}{Kouveliotou, C.}, \bibinfo{author}{Preece, R.D.},
  \bibinfo{author}{Bhat, N.}, \bibinfo{author}{Fishman, G.J.},
  \bibinfo{author}{Meegan, C.A.}, \bibinfo{author}{Horack, J.M.},
  \bibinfo{author}{Briggs, M.S.}, \bibinfo{author}{Paciesas, W.S.},
  \bibinfo{author}{Pendleton, G.N.}, \bibinfo{author}{Band, D.},
  \bibinfo{author}{Matteson, J.}, \bibinfo{author}{Palmer, D.M.},
  \bibinfo{author}{Teegarden, B.}, \bibinfo{author}{Norris, J.P.},
  \bibinfo{year}{1994}.
\newblock \bibinfo{title}{{BATSE observations of the very intense gamma-ray
  burst GRB 930131}}.
\newblock \bibinfo{journal}{Astrophysical Journal} \bibinfo{volume}{422},
  \bibinfo{pages}{L59--L62}.
\bibitem[{Krawczynski(2011)}]{2011APh....34..784K}
\bibinfo{author}{Krawczynski, H.S.}, \bibinfo{year}{2011}.
\newblock \bibinfo{title}{{Analysis of the data from Compton X-ray polarimeters
  which measure the azimuthal and polar scattering angles}}.
\newblock \bibinfo{journal}{Astroparticle Physics} \bibinfo{volume}{34},
  \bibinfo{pages}{784--788}.
\bibitem[{Kurita et~al.(2015)Kurita, Ohuchi, Arimoto, Yatsu, Kawai, Ohta, Koga,
  Kim, Tawara, Suzuki, Miyasato, Nagasu, Kawajiri, Matsushita, Matunaga,
  Moriyama, Kimura and team}]{2015arXiv150301975K}
\bibinfo{author}{Kurita, S.}, \bibinfo{author}{Ohuchi, H.},
  \bibinfo{author}{Arimoto, M.}, \bibinfo{author}{Yatsu, Y.},
  \bibinfo{author}{Kawai, N.}, \bibinfo{author}{Ohta, K.},
  \bibinfo{author}{Koga, M.}, \bibinfo{author}{Kim, E.},
  \bibinfo{author}{Tawara, K.}, \bibinfo{author}{Suzuki, S.},
  \bibinfo{author}{Miyasato, K.}, \bibinfo{author}{Nagasu, T.},
  \bibinfo{author}{Kawajiri, S.}, \bibinfo{author}{Matsushita, M.},
  \bibinfo{author}{Matunaga, S.}, \bibinfo{author}{Moriyama, N.},
  \bibinfo{author}{Kimura, S.}, \bibinfo{author}{team, T.},
  \bibinfo{year}{2015}.
\newblock \bibinfo{title}{{Development of a micro-satellite TSUBAME for X-ray
  polarimetry of GRBs}}.
\newblock \bibinfo{journal}{arXiv.org} ,
  \bibinfo{pages}{arXiv:1503.01975}\href{http://arxiv.org/abs/1503.01975}{\tt
  arXiv:1503.01975}.
\bibitem[{Labanti et~al.(2003)Labanti, Di~Cocco, Ferro, Gianotti, Mauri, Rossi,
  Stephen, Traci and Trifoglio}]{2003A&A...411L.149L}
\bibinfo{author}{Labanti, C.}, \bibinfo{author}{Di~Cocco, G.},
  \bibinfo{author}{Ferro, G.}, \bibinfo{author}{Gianotti, F.},
  \bibinfo{author}{Mauri, A.}, \bibinfo{author}{Rossi, E.},
  \bibinfo{author}{Stephen, J.B.}, \bibinfo{author}{Traci, A.},
  \bibinfo{author}{Trifoglio, M.}, \bibinfo{year}{2003}.
\newblock \bibinfo{title}{{The Ibis-Picsit detector onboard Integral}}.
\newblock \bibinfo{journal}{Astronomy and Astrophysics} \bibinfo{volume}{411},
  \bibinfo{pages}{L149--L152}.
\bibitem[{Laurent et~al.(2012)Laurent, G{\"o}tz, Gouiff{\`e}s, Grinberg, Moran,
  Rodriguez and Wilms}]{2012int..workE...5L}
\bibinfo{author}{Laurent, P.}, \bibinfo{author}{G{\"o}tz, D.},
  \bibinfo{author}{Gouiff{\`e}s, C.}, \bibinfo{author}{Grinberg, V.},
  \bibinfo{author}{Moran, P.}, \bibinfo{author}{Rodriguez, J.},
  \bibinfo{author}{Wilms, J.}, \bibinfo{year}{2012}.
\newblock \bibinfo{title}{{INTEGRAL observations of gamma-ray polarized
  sources: from the Crab pulsar to Cygnus X-1}}, in:
  \bibinfo{booktitle}{Proceedings of "An INTEGRAL view of the high-energy sky
  (the first 10 years)" - 9th INTEGRAL Workshop and celebration of the 10th
  anniversary of the launch (INTEGRAL 2012). 15-19 October 2012.},
  p.~\bibinfo{pages}{5}.
\bibitem[{Lazzati(2006)}]{2006NJPh....8..131L}
\bibinfo{author}{Lazzati, D.}, \bibinfo{year}{2006}.
\newblock \bibinfo{title}{{Polarization in the prompt emission of gamma-ray
  bursts and their afterglows}}.
\newblock \bibinfo{journal}{New Journal of Physics} \bibinfo{volume}{8},
  \bibinfo{pages}{131}.
\bibitem[{Lazzati et~al.(2004)Lazzati, Rossi, Ghisellini and
  Rees}]{2004MNRAS.347L...1L}
\bibinfo{author}{Lazzati, D.}, \bibinfo{author}{Rossi, E.},
  \bibinfo{author}{Ghisellini, G.}, \bibinfo{author}{Rees, M.J.},
  \bibinfo{year}{2004}.
\newblock \bibinfo{title}{{Compton drag as a mechanism for very high linear
  polarization in gamma-ray bursts}}.
\newblock \bibinfo{journal}{Monthly Notices of the Royal Astronomical Society}
  \bibinfo{volume}{347}, \bibinfo{pages}{L1--L5}.
\bibitem[{Lebrun et~al.(2003)Lebrun, Leray, Lavocat, Cr{\'e}tolle, Arqu{\`e}s,
  Blondel, Bonnin, Bou{\`e}re, Cara, Chaleil, Daly, Desages, Dzitko, Horeau,
  Laurent, Limousin, Mathy, Mauguen, Meignier, Molini{\'e}, Poindron, Rouger,
  Sauvageon and Tourrette}]{2003A&A...411L.141L}
\bibinfo{author}{Lebrun, F.}, \bibinfo{author}{Leray, J.P.},
  \bibinfo{author}{Lavocat, P.}, \bibinfo{author}{Cr{\'e}tolle, J.},
  \bibinfo{author}{Arqu{\`e}s, M.}, \bibinfo{author}{Blondel, C.},
  \bibinfo{author}{Bonnin, C.}, \bibinfo{author}{Bou{\`e}re, A.},
  \bibinfo{author}{Cara, C.}, \bibinfo{author}{Chaleil, T.},
  \bibinfo{author}{Daly, F.}, \bibinfo{author}{Desages, F.},
  \bibinfo{author}{Dzitko, H.}, \bibinfo{author}{Horeau, B.},
  \bibinfo{author}{Laurent, P.}, \bibinfo{author}{Limousin, O.},
  \bibinfo{author}{Mathy, F.}, \bibinfo{author}{Mauguen, V.},
  \bibinfo{author}{Meignier, F.}, \bibinfo{author}{Molini{\'e}, F.},
  \bibinfo{author}{Poindron, E.}, \bibinfo{author}{Rouger, M.},
  \bibinfo{author}{Sauvageon, A.}, \bibinfo{author}{Tourrette, T.},
  \bibinfo{year}{2003}.
\newblock \bibinfo{title}{{ISGRI: The INTEGRAL Soft Gamma-Ray Imager}}.
\newblock \bibinfo{journal}{Astronomy and Astrophysics} \bibinfo{volume}{411},
  \bibinfo{pages}{L141--L148}.
\bibitem[{Lei et~al.(1997)Lei, Dean and Hills}]{1997SSRv...82..309L}
\bibinfo{author}{Lei, F.}, \bibinfo{author}{Dean, A.J.J.},
  \bibinfo{author}{Hills, G.L.}, \bibinfo{year}{1997}.
\newblock \bibinfo{title}{{Compton Polarimetry in Gamma-Ray Astronomy}}.
\newblock \bibinfo{journal}{Space Science Reviews} \bibinfo{volume}{82},
  \bibinfo{pages}{309--388}.
\bibitem[{Lei et~al.(1996)Lei, Hills, Dean and Swinyard}]{1996A&AS..120C.695L}
\bibinfo{author}{Lei, F.}, \bibinfo{author}{Hills, G.L.},
  \bibinfo{author}{Dean, A.J.J.}, \bibinfo{author}{Swinyard, B.M.},
  \bibinfo{year}{1996}.
\newblock \bibinfo{title}{{Characteristics of COMPTEL as a polarimeter and its
  data analysis.}}
\newblock \bibinfo{journal}{Astronomy and Astrophysics Supplement Series}
  \bibinfo{volume}{120}, \bibinfo{pages}{695}.
\bibitem[{Lin et~al.(2002)Lin, Dennis, Hurford, Smith, Zehnder, Harvey, Curtis,
  Pankow, Turin, Bester, Csillaghy, Lewis, Madden, van Beek, Appleby, Raudorf,
  McTiernan, Ramaty, Schmahl, Schwartz, Krucker, Abiad, Quinn, Berg, Hashii,
  Sterling, Jackson, Pratt, Campbell, Malone, Landis, Barrington-Leigh,
  Slassi-Sennou, Cork, Clark, Amato, Orwig, Boyle, Banks, Shirey, Tolbert,
  Zarro, Snow, Thomsen, Henneck, McHedlishvili, Ming, Fivian, Jordan, Wanner,
  Crubb, Preble, Matranga, Benz, Hudson, Canfield, Holman, Crannell, Kosugi,
  Emslie, Vilmer, Brown, Johns-Krull, Aschwanden, Metcalf and
  Conway}]{2002SoPh..210....3L}
\bibinfo{author}{Lin, R.P.}, \bibinfo{author}{Dennis, B.R.},
  \bibinfo{author}{Hurford, G.J.}, \bibinfo{author}{Smith, D.M.},
  \bibinfo{author}{Zehnder, A.}, \bibinfo{author}{Harvey, P.R.},
  \bibinfo{author}{Curtis, D.W.}, \bibinfo{author}{Pankow, D.},
  \bibinfo{author}{Turin, P.}, \bibinfo{author}{Bester, M.},
  \bibinfo{author}{Csillaghy, A.}, \bibinfo{author}{Lewis, M.},
  \bibinfo{author}{Madden, N.}, \bibinfo{author}{van Beek, H.F.},
  \bibinfo{author}{Appleby, M.}, \bibinfo{author}{Raudorf, T.},
  \bibinfo{author}{McTiernan, J.M.}, \bibinfo{author}{Ramaty, R.},
  \bibinfo{author}{Schmahl, E.}, \bibinfo{author}{Schwartz, R.},
  \bibinfo{author}{Krucker, S.}, \bibinfo{author}{Abiad, R.},
  \bibinfo{author}{Quinn, T.}, \bibinfo{author}{Berg, P.},
  \bibinfo{author}{Hashii, M.}, \bibinfo{author}{Sterling, R.},
  \bibinfo{author}{Jackson, R.}, \bibinfo{author}{Pratt, R.},
  \bibinfo{author}{Campbell, R.D.}, \bibinfo{author}{Malone, D.},
  \bibinfo{author}{Landis, D.}, \bibinfo{author}{Barrington-Leigh, C.P.},
  \bibinfo{author}{Slassi-Sennou, S.}, \bibinfo{author}{Cork, C.},
  \bibinfo{author}{Clark, D.}, \bibinfo{author}{Amato, D.},
  \bibinfo{author}{Orwig, L.}, \bibinfo{author}{Boyle, R.},
  \bibinfo{author}{Banks, I.S.}, \bibinfo{author}{Shirey, K.},
  \bibinfo{author}{Tolbert, A.K.}, \bibinfo{author}{Zarro, D.},
  \bibinfo{author}{Snow, F.}, \bibinfo{author}{Thomsen, K.},
  \bibinfo{author}{Henneck, R.}, \bibinfo{author}{McHedlishvili, A.},
  \bibinfo{author}{Ming, P.}, \bibinfo{author}{Fivian, M.D.},
  \bibinfo{author}{Jordan, J.}, \bibinfo{author}{Wanner, R.},
  \bibinfo{author}{Crubb, J.}, \bibinfo{author}{Preble, J.},
  \bibinfo{author}{Matranga, M.}, \bibinfo{author}{Benz, A.},
  \bibinfo{author}{Hudson, H.}, \bibinfo{author}{Canfield, R.C.},
  \bibinfo{author}{Holman, G.D.}, \bibinfo{author}{Crannell, C.J.},
  \bibinfo{author}{Kosugi, T.}, \bibinfo{author}{Emslie, A.G.},
  \bibinfo{author}{Vilmer, N.}, \bibinfo{author}{Brown, J.C.},
  \bibinfo{author}{Johns-Krull, C.}, \bibinfo{author}{Aschwanden, M.J.},
  \bibinfo{author}{Metcalf, T.R.}, \bibinfo{author}{Conway, A.},
  \bibinfo{year}{2002}.
\newblock \bibinfo{title}{{The Reuven Ramaty High-Energy Solar Spectroscopic
  Imager (RHESSI)}}.
\newblock \bibinfo{journal}{Solar Physics} \bibinfo{volume}{210},
  \bibinfo{pages}{3--32}.
\bibitem[{Lundman et~al.(2014)Lundman, Peer and Ryde}]{2014MNRAS.440.3292L}
\bibinfo{author}{Lundman, C.}, \bibinfo{author}{Peer, A.},
  \bibinfo{author}{Ryde, F.}, \bibinfo{year}{2014}.
\newblock \bibinfo{title}{{Polarization properties of photospheric emission
  from relativistic, collimated outflows}}.
\newblock \bibinfo{journal}{Monthly Notices of the Royal Astronomical Society}
  \bibinfo{volume}{440}, \bibinfo{pages}{3292--3308}.
\bibitem[{Lyutikov et~al.(2003)Lyutikov, Pariev and
  Blandford}]{2003ApJ...597..998L}
\bibinfo{author}{Lyutikov, M.}, \bibinfo{author}{Pariev, V.I.},
  \bibinfo{author}{Blandford, R.D.}, \bibinfo{year}{2003}.
\newblock \bibinfo{title}{{Polarization of Prompt Gamma-Ray Burst Emission:
  Evidence for Electromagnetically Dominated Outflow}}.
\newblock \bibinfo{journal}{Astrophysical Journal} \bibinfo{volume}{597},
  \bibinfo{pages}{998--1009}.
\bibitem[{McBreen et~al.(2006)McBreen, Hanlon, McGlynn, McBreen, Foley, Preece,
  von Kienlin and Williams}]{2006A&A...455..433M}
\bibinfo{author}{McBreen, S.}, \bibinfo{author}{Hanlon, L.O.},
  \bibinfo{author}{McGlynn, S.}, \bibinfo{author}{McBreen, B.},
  \bibinfo{author}{Foley, S.}, \bibinfo{author}{Preece, R.D.},
  \bibinfo{author}{von Kienlin, A.}, \bibinfo{author}{Williams, O.R.},
  \bibinfo{year}{2006}.
\newblock \bibinfo{title}{{Observations of the intense and ultra-long burst GRB
  041219a with the Germanium spectrometer on INTEGRAL}}.
\newblock \bibinfo{journal}{Astronomy and Astrophysics} \bibinfo{volume}{455},
  \bibinfo{pages}{433--440}.
\bibitem[{McConnell et~al.(2009)McConnell, Angelini, Baring, Barthelmy, Black,
  Bloser, Dennis, Emslie, Greiner, Hajdas, Harding, Hartmann, Hill, Ioka,
  Kaaret, Kanbach, Kniffen, Legere, Macri, Morris, Nakamura, Produit, Ryan,
  Sakamoto, Toma, Wu, Yamazaki and Zhang}]{2009AIPC.1133...64M}
\bibinfo{author}{McConnell, M.L.}, \bibinfo{author}{Angelini, L.},
  \bibinfo{author}{Baring, M.G.}, \bibinfo{author}{Barthelmy, S.D.},
  \bibinfo{author}{Black, J.K.}, \bibinfo{author}{Bloser, P.F.},
  \bibinfo{author}{Dennis, B.R.}, \bibinfo{author}{Emslie, A.G.},
  \bibinfo{author}{Greiner, J.}, \bibinfo{author}{Hajdas, W.},
  \bibinfo{author}{Harding, A.K.}, \bibinfo{author}{Hartmann, D.H.},
  \bibinfo{author}{Hill, J.E.}, \bibinfo{author}{Ioka, K.},
  \bibinfo{author}{Kaaret, P.E.}, \bibinfo{author}{Kanbach, G.},
  \bibinfo{author}{Kniffen, D.}, \bibinfo{author}{Legere, J.S.},
  \bibinfo{author}{Macri, J.R.}, \bibinfo{author}{Morris, R.},
  \bibinfo{author}{Nakamura, T.}, \bibinfo{author}{Produit, N.},
  \bibinfo{author}{Ryan, J.M.}, \bibinfo{author}{Sakamoto, T.},
  \bibinfo{author}{Toma, K.}, \bibinfo{author}{Wu, X.},
  \bibinfo{author}{Yamazaki, R.}, \bibinfo{author}{Zhang, B.},
  \bibinfo{year}{2009}.
\newblock \bibinfo{title}{{GRB Polarimetry with POET}}.
\newblock \bibinfo{journal}{AIP Conf. Proc.} \bibinfo{volume}{1133},
  \bibinfo{pages}{64--66}.
\bibitem[{McConnell et~al.(2014a)McConnell, Baring, Bloser, Dwyer, Emslie,
  Ertley, Greiner, Harding, Hartmann, Hill, Kaaret, Kippen, Mattingly, McBreen,
  Pearce, Produit, Ryan, Ryde, Sakamoto, Toma, Vestrand and
  Zhang}]{2014SPIE.9144E..0OM}
\bibinfo{author}{McConnell, M.L.}, \bibinfo{author}{Baring, M.G.},
  \bibinfo{author}{Bloser, P.F.}, \bibinfo{author}{Dwyer, J.F.},
  \bibinfo{author}{Emslie, A.G.}, \bibinfo{author}{Ertley, C.D.},
  \bibinfo{author}{Greiner, J.}, \bibinfo{author}{Harding, A.K.},
  \bibinfo{author}{Hartmann, D.H.}, \bibinfo{author}{Hill, J.E.},
  \bibinfo{author}{Kaaret, P.E.}, \bibinfo{author}{Kippen, R.M.},
  \bibinfo{author}{Mattingly, D.}, \bibinfo{author}{McBreen, S.},
  \bibinfo{author}{Pearce, M.}, \bibinfo{author}{Produit, N.},
  \bibinfo{author}{Ryan, J.M.}, \bibinfo{author}{Ryde, F.},
  \bibinfo{author}{Sakamoto, T.}, \bibinfo{author}{Toma, K.},
  \bibinfo{author}{Vestrand, W.T.}, \bibinfo{author}{Zhang, B.},
  \bibinfo{year}{2014}a.
\newblock \bibinfo{title}{{POET: a SMEX mission for gamma ray burst
  polarimetry}}.
\newblock \bibinfo{journal}{Proceedings of SPIE} \bibinfo{volume}{9144},
  \bibinfo{pages}{91440O--91440O--8}.
\bibitem[{McConnell et~al.(2014b)McConnell, Bloser, Ertley, Legere, Ryan and
  Wasti}]{2014SPIE.9144E..3PM}
\bibinfo{author}{McConnell, M.L.}, \bibinfo{author}{Bloser, P.F.},
  \bibinfo{author}{Ertley, C.}, \bibinfo{author}{Legere, J.S.},
  \bibinfo{author}{Ryan, J.M.}, \bibinfo{author}{Wasti, S.K.},
  \bibinfo{year}{2014}b.
\newblock \bibinfo{title}{{Current status of the GRAPE balloon program}}.
\newblock \bibinfo{journal}{Proceedings of SPIE} \bibinfo{volume}{9144},
  \bibinfo{pages}{91443P--91443P--10}.
\bibitem[{McConnell et~al.(2016)McConnell, Bloser, Legere and
  Ryan}]{McConnell:2016gd}
\bibinfo{author}{McConnell, M.L.}, \bibinfo{author}{Bloser, P.F.},
  \bibinfo{author}{Legere, J.S.}, \bibinfo{author}{Ryan, J.M.},
  \bibinfo{year}{2016}.
\newblock \bibinfo{title}{{ The development of a low energy Compton imager for
  GRB polarization studies}}.
\newblock \bibinfo{journal}{Proceedings of SPIE} \bibinfo{volume}{9905},
  \bibinfo{pages}{99052O--99052O--9}.
\bibitem[{McConnell and Collmar(2016)}]{2016HEAD...1511204M}
\bibinfo{author}{McConnell, M.L.}, \bibinfo{author}{Collmar, W.},
  \bibinfo{year}{2016}.
\newblock \bibinfo{title}{{GRB Polarization Measurements with CGRO/COMPTEL}}.
\newblock \bibinfo{journal}{Bulletin of the American Astronomical Society}
  \bibinfo{volume}{15}, \bibinfo{pages}{112.04}.
\bibitem[{McConnell et~al.(1996)McConnell, Forrest, Vestrand and
  Finger}]{1996AIPC..384..851M}
\bibinfo{author}{McConnell, M.L.}, \bibinfo{author}{Forrest, D.J.},
  \bibinfo{author}{Vestrand, W.T.}, \bibinfo{author}{Finger, M.H.},
  \bibinfo{year}{1996}.
\newblock \bibinfo{title}{{Using BATSE to measure gamma-ray burst
  polarization}}.
\newblock \bibinfo{journal}{AIP Conference Proceedings} \bibinfo{volume}{384},
  \bibinfo{pages}{851--855}.
\bibitem[{McConnell and Kippen(2004)}]{2004HEAD....8.4101M}
\bibinfo{author}{McConnell, M.L.}, \bibinfo{author}{Kippen, R.M.},
  \bibinfo{year}{2004}.
\newblock \bibinfo{title}{{The GLEPS Package for Simulating Polarized Gamma
  Rays with GEANT3}}.
\newblock \bibinfo{journal}{Bulletin of the American Astronomical Society}
  \bibinfo{volume}{36}, \bibinfo{pages}{1206}.
\bibitem[{McConnell et~al.(2002)McConnell, Ryan, Smith, Lin and
  Emslie}]{2002SoPh..210..125M}
\bibinfo{author}{McConnell, M.L.}, \bibinfo{author}{Ryan, J.M.},
  \bibinfo{author}{Smith, D.M.}, \bibinfo{author}{Lin, R.P.},
  \bibinfo{author}{Emslie, A.G.}, \bibinfo{year}{2002}.
\newblock \bibinfo{title}{{RHESSI as a Hard X-Ray Polarimeter}}.
\newblock \bibinfo{journal}{Solar Physics} \bibinfo{volume}{210},
  \bibinfo{pages}{125--142}.
\bibitem[{McGlynn et~al.(2007)McGlynn, Clark, Dean, Hanlon, McBreen, Willis,
  McBreen, Bird and Foley}]{2007A&A...466..895M}
\bibinfo{author}{McGlynn, S.}, \bibinfo{author}{Clark, D.J.},
  \bibinfo{author}{Dean, A.J.J.}, \bibinfo{author}{Hanlon, L.O.},
  \bibinfo{author}{McBreen, S.}, \bibinfo{author}{Willis, D.R.},
  \bibinfo{author}{McBreen, B.}, \bibinfo{author}{Bird, A.J.},
  \bibinfo{author}{Foley, S.}, \bibinfo{year}{2007}.
\newblock \bibinfo{title}{{Polarisation studies of the prompt gamma-ray
  emission from GRB 041219a using the spectrometer aboard INTEGRAL}}.
\newblock \bibinfo{journal}{Astronomy and Astrophysics} \bibinfo{volume}{466},
  \bibinfo{pages}{895--904}.
\bibitem[{McGlynn et~al.(2009)McGlynn, Foley, McBreen, Hanlon, McBreen, Clark,
  Dean, Martin-Carrillo and O'Connor}]{2009A&A...499..465M}
\bibinfo{author}{McGlynn, S.}, \bibinfo{author}{Foley, S.},
  \bibinfo{author}{McBreen, B.}, \bibinfo{author}{Hanlon, L.O.},
  \bibinfo{author}{McBreen, S.}, \bibinfo{author}{Clark, D.J.},
  \bibinfo{author}{Dean, A.J.J.}, \bibinfo{author}{Martin-Carrillo, A.},
  \bibinfo{author}{O'Connor, R.}, \bibinfo{year}{2009}.
\newblock \bibinfo{title}{{High energy emission and polarisation limits for the
  INTEGRAL burst GRB 061122}}.
\newblock \bibinfo{journal}{Astronomy and Astrophysics} \bibinfo{volume}{499},
  \bibinfo{pages}{465--472}.
\bibitem[{M{\'e}sz{\'a}ros(2006)}]{2006RPPh...69.2259M}
\bibinfo{author}{M{\'e}sz{\'a}ros, P.}, \bibinfo{year}{2006}.
\newblock \bibinfo{title}{{Gamma-ray bursts}}.
\newblock \bibinfo{journal}{Reports on Progress in Physics}
  \bibinfo{volume}{69}, \bibinfo{pages}{2259--2321}.
\bibitem[{Mizuno et~al.(2005)Mizuno, Kamae, Ng, Tajima, Mitchell, Streitmatter,
  Fernholz, Groth and Fukazawa}]{2005NIMPA.540..158M}
\bibinfo{author}{Mizuno, T.}, \bibinfo{author}{Kamae, T.}, \bibinfo{author}{Ng,
  J.S.T.}, \bibinfo{author}{Tajima, H.}, \bibinfo{author}{Mitchell, J.W.},
  \bibinfo{author}{Streitmatter, R.}, \bibinfo{author}{Fernholz, R.C.},
  \bibinfo{author}{Groth, E.}, \bibinfo{author}{Fukazawa, Y.},
  \bibinfo{year}{2005}.
\newblock \bibinfo{title}{{Beam test of a prototype detector array for the PoGO
  astronomical hard X-ray/soft gamma-ray polarimeter}}.
\newblock \bibinfo{journal}{Nuclear Instruments and Methods in Physics Research
  Section A} \bibinfo{volume}{540}, \bibinfo{pages}{158--168}.
\bibitem[{Murakami et~al.(2010)Murakami, Yonetoku, Sakashita, Morihara,
  Kikuchi, Gunji, Tokairin and Mihara}]{2010AIPC.1279..227M}
\bibinfo{author}{Murakami, T.}, \bibinfo{author}{Yonetoku, D.},
  \bibinfo{author}{Sakashita, T.}, \bibinfo{author}{Morihara, Y.},
  \bibinfo{author}{Kikuchi, Y.}, \bibinfo{author}{Gunji, S.},
  \bibinfo{author}{Tokairin, N.}, \bibinfo{author}{Mihara, T.},
  \bibinfo{year}{2010}.
\newblock \bibinfo{title}{{Gamma-Ray Burst Polarimeter aboard IKAROS}}.
\newblock \bibinfo{journal}{AIP Conf. Proc.} \bibinfo{volume}{1279},
  \bibinfo{pages}{227--230}.
\bibitem[{Novick(1975)}]{1975SSRv...18..389N}
\bibinfo{author}{Novick, R.}, \bibinfo{year}{1975}.
\newblock \bibinfo{title}{{Stellar and Solar X-Ray Polarimetry}}.
\newblock \bibinfo{journal}{Space Science Reviews} \bibinfo{volume}{18},
  \bibinfo{pages}{389--408}.
\bibitem[{Orsi et~al.(2011)Orsi, Haas, Hajdas, Honkim{\"a}ki, Lamanna,
  Lechanoine-Leluc, Marcinkowski, Pohl, Produit, Rapin, Suarez-Garcia, Rybka
  and Vialle}]{2011NIMPA.648..139O}
\bibinfo{author}{Orsi, S.}, \bibinfo{author}{Haas, D.},
  \bibinfo{author}{Hajdas, W.}, \bibinfo{author}{Honkim{\"a}ki, V.},
  \bibinfo{author}{Lamanna, G.}, \bibinfo{author}{Lechanoine-Leluc, C.},
  \bibinfo{author}{Marcinkowski, R.}, \bibinfo{author}{Pohl, M.},
  \bibinfo{author}{Produit, N.}, \bibinfo{author}{Rapin, D.},
  \bibinfo{author}{Suarez-Garcia, E.}, \bibinfo{author}{Rybka, D.},
  \bibinfo{author}{Vialle, J.P.}, \bibinfo{year}{2011}.
\newblock \bibinfo{title}{{Response of the Compton polarimeter POLAR to
  polarized hard X-rays}}.
\newblock \bibinfo{journal}{Nuclear Instruments and Methods in Physics Research
  Section A} \bibinfo{volume}{648}, \bibinfo{pages}{139--154}.
\bibitem[{Paciesas et~al.(2012)Paciesas, Meegan, von Kienlin, Bhat, Bissaldi,
  Briggs, Burgess, Chaplin, Connaughton, Diehl, Fishman, Fitzpatrick, Foley,
  Gibby, Giles, Goldstein, Greiner, Gruber, Guiriec, van~der Horst, Kippen,
  Kouveliotou, Lichti, Lin, McBreen, Preece, Rau, Tierney and
  Wilson-Hodge}]{2012ApJS..199...18P}
\bibinfo{author}{Paciesas, W.S.}, \bibinfo{author}{Meegan, C.A.},
  \bibinfo{author}{von Kienlin, A.}, \bibinfo{author}{Bhat, P.N.},
  \bibinfo{author}{Bissaldi, E.}, \bibinfo{author}{Briggs, M.S.},
  \bibinfo{author}{Burgess, J.M.}, \bibinfo{author}{Chaplin, V.},
  \bibinfo{author}{Connaughton, V.}, \bibinfo{author}{Diehl, R.},
  \bibinfo{author}{Fishman, G.J.}, \bibinfo{author}{Fitzpatrick, G.},
  \bibinfo{author}{Foley, S.}, \bibinfo{author}{Gibby, M.},
  \bibinfo{author}{Giles, M.}, \bibinfo{author}{Goldstein, A.},
  \bibinfo{author}{Greiner, J.}, \bibinfo{author}{Gruber, D.},
  \bibinfo{author}{Guiriec, S.}, \bibinfo{author}{van~der Horst, A.J.},
  \bibinfo{author}{Kippen, R.M.}, \bibinfo{author}{Kouveliotou, C.},
  \bibinfo{author}{Lichti, G.G.}, \bibinfo{author}{Lin, L.},
  \bibinfo{author}{McBreen, S.}, \bibinfo{author}{Preece, R.D.},
  \bibinfo{author}{Rau, A.}, \bibinfo{author}{Tierney, D.},
  \bibinfo{author}{Wilson-Hodge, C.}, \bibinfo{year}{2012}.
\newblock \bibinfo{title}{{The Fermi GBM Gamma-Ray Burst Catalog: The First Two
  Years}}.
\newblock \bibinfo{journal}{Astrophysical Journal Supplement Series}
  \bibinfo{volume}{199}, \bibinfo{pages}{18}.
\bibitem[{Paciesas et~al.(1999)Paciesas, Meegan, Pendleton, Briggs,
  Kouveliotou, Koshut, Lestrade, McCollough, Brainerd, Hakkila, Henze, Preece,
  Connaughton, Kippen, Mallozzi, Fishman, Richardson and
  Sahi}]{1999ApJS..122..465P}
\bibinfo{author}{Paciesas, W.S.}, \bibinfo{author}{Meegan, C.A.},
  \bibinfo{author}{Pendleton, G.N.}, \bibinfo{author}{Briggs, M.S.},
  \bibinfo{author}{Kouveliotou, C.}, \bibinfo{author}{Koshut, T.M.},
  \bibinfo{author}{Lestrade, J.P.}, \bibinfo{author}{McCollough, M.L.},
  \bibinfo{author}{Brainerd, J.J.}, \bibinfo{author}{Hakkila, J.},
  \bibinfo{author}{Henze, W.}, \bibinfo{author}{Preece, R.D.},
  \bibinfo{author}{Connaughton, V.}, \bibinfo{author}{Kippen, R.M.},
  \bibinfo{author}{Mallozzi, R.S.}, \bibinfo{author}{Fishman, G.J.},
  \bibinfo{author}{Richardson, G.A.}, \bibinfo{author}{Sahi, M.},
  \bibinfo{year}{1999}.
\newblock \bibinfo{title}{{The Fourth BATSE Gamma-Ray Burst Catalog
  (Revised)}}.
\newblock \bibinfo{journal}{Astrophysical Journal Supplement Series}
  \bibinfo{volume}{122}, \bibinfo{pages}{465--495}.
\bibitem[{Paciesas et~al.(1989)Paciesas, Pendleton, Lestrade, Fishman, Meegan,
  Wilson, Parnell, Austin, Berry and Horack}]{1989SPIE.1159..156P}
\bibinfo{author}{Paciesas, W.S.}, \bibinfo{author}{Pendleton, G.N.},
  \bibinfo{author}{Lestrade, J.P.}, \bibinfo{author}{Fishman, G.J.},
  \bibinfo{author}{Meegan, C.A.}, \bibinfo{author}{Wilson, R.B.},
  \bibinfo{author}{Parnell, T.A.}, \bibinfo{author}{Austin, R.W.},
  \bibinfo{author}{Berry, F.A.J.}, \bibinfo{author}{Horack, J.M.},
  \bibinfo{year}{1989}.
\newblock \bibinfo{title}{{Performance of the large-area detectors for the
  Burst and Transient Source Experiment (BATSE) on the Gamma Ray Observatory}}.
\newblock \bibinfo{journal}{Proceedings of SPIE} \bibinfo{volume}{1159},
  \bibinfo{pages}{156--164}.
\bibitem[{Piran(2005)}]{Piran:2005cs}
\bibinfo{author}{Piran, T.}, \bibinfo{year}{2005}.
\newblock \bibinfo{title}{{The physics of gamma-ray bursts}}.
\newblock \bibinfo{journal}{Reviews of Modern Physics} \bibinfo{volume}{76},
  \bibinfo{pages}{1143}.
\bibitem[{Produit et~al.(2005)Produit, Barao, Deluit, Hajdas, Leluc, Pohl,
  Rapin, Vialle, Walter and Wigger}]{2005NIMPA.550..616P}
\bibinfo{author}{Produit, N.}, \bibinfo{author}{Barao, F.},
  \bibinfo{author}{Deluit, S.}, \bibinfo{author}{Hajdas, W.},
  \bibinfo{author}{Leluc, C.}, \bibinfo{author}{Pohl, M.},
  \bibinfo{author}{Rapin, D.}, \bibinfo{author}{Vialle, J.P.},
  \bibinfo{author}{Walter, R.}, \bibinfo{author}{Wigger, C.},
  \bibinfo{year}{2005}.
\newblock \bibinfo{title}{{POLAR, a compact detector for gamma-ray bursts
  photon polarization measurements}}.
\newblock \bibinfo{journal}{Nuclear Instruments and Methods in Physics Research
  Section A} \bibinfo{volume}{550}, \bibinfo{pages}{616--625}.
\bibitem[{Rao et~al.(2016)Rao, Chand, Hingar, Iyyani, Khanna, Kutty, Malkar,
  Paul, Bhalerao, Bhattacharya, Dewangan, Pawar, Vibhute, Chattopadhyay,
  Mithun, Vadawale, Vagshette, Basak, Pradeep, Samuel, Sreekumar, Vinod,
  Navalgund, Pandiyan, Sarma, Seetha and Subbarao}]{Rao:2016tt}
\bibinfo{author}{Rao, A.R.}, \bibinfo{author}{Chand, V.},
  \bibinfo{author}{Hingar, M.K.}, \bibinfo{author}{Iyyani, S.},
  \bibinfo{author}{Khanna, R.}, \bibinfo{author}{Kutty, A.P.K.},
  \bibinfo{author}{Malkar, J.P.}, \bibinfo{author}{Paul, D.},
  \bibinfo{author}{Bhalerao, V.B.}, \bibinfo{author}{Bhattacharya, D.},
  \bibinfo{author}{Dewangan, G.C.}, \bibinfo{author}{Pawar, P.},
  \bibinfo{author}{Vibhute, A.M.}, \bibinfo{author}{Chattopadhyay, T.},
  \bibinfo{author}{Mithun, N.P.S.}, \bibinfo{author}{Vadawale, S.V.},
  \bibinfo{author}{Vagshette, N.}, \bibinfo{author}{Basak, R.},
  \bibinfo{author}{Pradeep, P.}, \bibinfo{author}{Samuel, E.},
  \bibinfo{author}{Sreekumar, S.}, \bibinfo{author}{Vinod, P.},
  \bibinfo{author}{Navalgund, K.H.}, \bibinfo{author}{Pandiyan, R.},
  \bibinfo{author}{Sarma, K.S.}, \bibinfo{author}{Seetha, S.},
  \bibinfo{author}{Subbarao, K.}, \bibinfo{year}{2016}.
\newblock \bibinfo{title}{{AstroSat CZT Imager observations of GRB 151006A:
  timing, spectroscopy, and polarisation study}}
  \href{http://arxiv.org/abs/1608.07388}{\tt arXiv:1608.07388}.
\bibitem[{Rees and M{\'e}sz{\'a}ros(1994)}]{1994ApJ...430L..93R}
\bibinfo{author}{Rees, M.J.}, \bibinfo{author}{M{\'e}sz{\'a}ros, P.},
  \bibinfo{year}{1994}.
\newblock \bibinfo{title}{{Unsteady outflow models for cosmological gamma-ray
  bursts}}.
\newblock \bibinfo{journal}{Astrophysical Journal} \bibinfo{volume}{430},
  \bibinfo{pages}{L93--L96}.
\bibitem[{Rutledge and Fox(2004)}]{2004MNRAS.350.1288R}
\bibinfo{author}{Rutledge, R.E.}, \bibinfo{author}{Fox, D.B.},
  \bibinfo{year}{2004}.
\newblock \bibinfo{title}{{Re-analysis of polarization in the $\gamma$-ray flux
  of GRB 021206}}.
\newblock \bibinfo{journal}{Monthly Notices of the Royal Astronomical Society}
  \bibinfo{volume}{350}, \bibinfo{pages}{1288--1300}.
\bibitem[{Ryan et~al.(1994)Ryan, Bennett, Collmar, Connors, Fishman, Greiner,
  Hanlon, Hermsen, Kippen, Kouveliotou, Kuiper, Lichti, Macri, Mattox,
  McConnell, McNamara, Meegan, Sch{\"o}nfelder, van Dijk, Varendorff, Webber
  and Winkler}]{1994ApJ...422L..67R}
\bibinfo{author}{Ryan, J.M.}, \bibinfo{author}{Bennett, K.},
  \bibinfo{author}{Collmar, W.}, \bibinfo{author}{Connors, A.},
  \bibinfo{author}{Fishman, G.J.}, \bibinfo{author}{Greiner, J.},
  \bibinfo{author}{Hanlon, L.O.}, \bibinfo{author}{Hermsen, W.},
  \bibinfo{author}{Kippen, R.M.}, \bibinfo{author}{Kouveliotou, C.},
  \bibinfo{author}{Kuiper, L.M.}, \bibinfo{author}{Lichti, G.G.},
  \bibinfo{author}{Macri, J.R.}, \bibinfo{author}{Mattox, J.},
  \bibinfo{author}{McConnell, M.L.}, \bibinfo{author}{McNamara, B.J.},
  \bibinfo{author}{Meegan, C.A.}, \bibinfo{author}{Sch{\"o}nfelder, V.},
  \bibinfo{author}{van Dijk, R.}, \bibinfo{author}{Varendorff, M.},
  \bibinfo{author}{Webber, W.R.}, \bibinfo{author}{Winkler, C.},
  \bibinfo{year}{1994}.
\newblock \bibinfo{title}{{COMPTEL measurements of the gamma-ray burst GRB
  930131}}.
\newblock \bibinfo{journal}{Astrophysical Journal} \bibinfo{volume}{422},
  \bibinfo{pages}{L67--L70}.
\bibitem[{Ryde(2005)}]{2005ApJ...625L..95R}
\bibinfo{author}{Ryde, F.}, \bibinfo{year}{2005}.
\newblock \bibinfo{title}{{Is Thermal Emission in Gamma-Ray Bursts
  Ubiquitous?}}
\newblock \bibinfo{journal}{Astrophysical Journal} \bibinfo{volume}{625},
  \bibinfo{pages}{L95--L98}.
\bibitem[{Sch{\"o}nfelder et~al.(1993)Sch{\"o}nfelder, Aarts, Bennett, de~Boer,
  Clear, Collmar, Connors, Deerenberg, Diehl, von Dordrecht, den Herder,
  Hermsen, Kippen, Kuiper, Lichti, Lockwood, Macri, McConnell, Morris, Much,
  Ryan, Simpson, Snelling, Stacy, Steinle, Strong, Swanenburg, Taylor, de~Vries
  and Winkler}]{1993ApJS...86..657S}
\bibinfo{author}{Sch{\"o}nfelder, V.}, \bibinfo{author}{Aarts, H.J.M.},
  \bibinfo{author}{Bennett, K.}, \bibinfo{author}{de~Boer, H.},
  \bibinfo{author}{Clear, J.}, \bibinfo{author}{Collmar, W.},
  \bibinfo{author}{Connors, A.}, \bibinfo{author}{Deerenberg, A.J.M.},
  \bibinfo{author}{Diehl, R.}, \bibinfo{author}{von Dordrecht, A.},
  \bibinfo{author}{den Herder, J.W.}, \bibinfo{author}{Hermsen, W.},
  \bibinfo{author}{Kippen, R.M.}, \bibinfo{author}{Kuiper, L.M.},
  \bibinfo{author}{Lichti, G.G.}, \bibinfo{author}{Lockwood, J.A.},
  \bibinfo{author}{Macri, J.R.}, \bibinfo{author}{McConnell, M.L.},
  \bibinfo{author}{Morris, D.}, \bibinfo{author}{Much, R.},
  \bibinfo{author}{Ryan, J.M.}, \bibinfo{author}{Simpson, G.},
  \bibinfo{author}{Snelling, M.}, \bibinfo{author}{Stacy, G.},
  \bibinfo{author}{Steinle, H.}, \bibinfo{author}{Strong, A.W.},
  \bibinfo{author}{Swanenburg, B.N.}, \bibinfo{author}{Taylor, B.},
  \bibinfo{author}{de~Vries, C.}, \bibinfo{author}{Winkler, C.},
  \bibinfo{year}{1993}.
\newblock \bibinfo{title}{{Instrument description and performance of the
  Imaging Gamma-Ray Telescope COMPTEL aboard the Compton Gamma-Ray
  Observatory}}.
\newblock \bibinfo{journal}{Astrophysical Journal Supplement Series}
  \bibinfo{volume}{86}, \bibinfo{pages}{657--692}.
\bibitem[{Shaviv and Dar(1995)}]{1995ApJ...447..863S}
\bibinfo{author}{Shaviv, N.J.}, \bibinfo{author}{Dar, A.},
  \bibinfo{year}{1995}.
\newblock \bibinfo{title}{{Gamma-Ray Bursts from Minijets}}.
\newblock \bibinfo{journal}{Astrophysical Journal} \bibinfo{volume}{447},
  \bibinfo{pages}{863}.
\bibitem[{Singh et~al.(2014)Singh, Tandon, {Agrawal, P. C.}, Antia, Manchanda,
  Yadav, Seetha, Ramadevi, Rao, Bhattacharya, Paul, Sreekumar, Bhattacharyya,
  Stewart, Hutchings, Annapurni, Ghosh, Murthy, Pati, Rao, Stalin, Girish,
  Sankarasubramanian, Vadawale, Bhalerao, Dewangan, Dedhia, Hingar, Katoch,
  Kothare, Mirza, Mukerjee, Shah, Shah, Mohan, Sangal, Nagabhusana, Sriram,
  Malkar, Sreekumar, Abbey, Hansford, Beardmore, Sharma, Murthy, Kulkarni,
  Meena, Babu and Postma}]{2014SPIE.9144E..1SS}
\bibinfo{author}{Singh, K.P.}, \bibinfo{author}{Tandon, S.N.},
  \bibinfo{author}{{Agrawal, P. C.}}, \bibinfo{author}{Antia, H.M.},
  \bibinfo{author}{Manchanda, R.K.}, \bibinfo{author}{Yadav, J.S.},
  \bibinfo{author}{Seetha, S.}, \bibinfo{author}{Ramadevi, M.C.},
  \bibinfo{author}{Rao, A.R.}, \bibinfo{author}{Bhattacharya, D.},
  \bibinfo{author}{Paul, B.}, \bibinfo{author}{Sreekumar, P.},
  \bibinfo{author}{Bhattacharyya, S.}, \bibinfo{author}{Stewart, G.C.},
  \bibinfo{author}{Hutchings, J.}, \bibinfo{author}{Annapurni, S.A.},
  \bibinfo{author}{Ghosh, S.K.}, \bibinfo{author}{Murthy, J.},
  \bibinfo{author}{Pati, A.}, \bibinfo{author}{Rao, N.K.},
  \bibinfo{author}{Stalin, C.S.}, \bibinfo{author}{Girish, V.},
  \bibinfo{author}{Sankarasubramanian, K.}, \bibinfo{author}{Vadawale, S.},
  \bibinfo{author}{Bhalerao, V.B.}, \bibinfo{author}{Dewangan, G.C.},
  \bibinfo{author}{Dedhia, D.K.}, \bibinfo{author}{Hingar, M.K.},
  \bibinfo{author}{Katoch, T.B.}, \bibinfo{author}{Kothare, A.T.},
  \bibinfo{author}{Mirza, I.}, \bibinfo{author}{Mukerjee, K.},
  \bibinfo{author}{Shah, H.}, \bibinfo{author}{Shah, P.},
  \bibinfo{author}{Mohan, R.}, \bibinfo{author}{Sangal, A.K.},
  \bibinfo{author}{Nagabhusana, S.}, \bibinfo{author}{Sriram, S.},
  \bibinfo{author}{Malkar, J.P.}, \bibinfo{author}{Sreekumar, S.},
  \bibinfo{author}{Abbey, A.F.}, \bibinfo{author}{Hansford, G.M.},
  \bibinfo{author}{Beardmore, A.P.}, \bibinfo{author}{Sharma, M.R.},
  \bibinfo{author}{Murthy, S.}, \bibinfo{author}{Kulkarni, R.},
  \bibinfo{author}{Meena, G.}, \bibinfo{author}{Babu, V.C.},
  \bibinfo{author}{Postma, J.}, \bibinfo{year}{2014}.
\newblock \bibinfo{title}{{ASTROSAT mission}}.
\newblock \bibinfo{journal}{Proceedings of SPIE} \bibinfo{volume}{9144},
  \bibinfo{pages}{91441S--91441S--15}.
\bibitem[{Smith et~al.(2002)Smith, Lin, Turin, Curtis, Primbsch, Campbell,
  Abiad, Schroeder, Cork, Hull, Landis, Madden, Malone, Pehl, Raudorf,
  Sangsingkeow, Boyle, Banks, Shirey and Schwartz}]{2002SoPh..210...33S}
\bibinfo{author}{Smith, D.M.}, \bibinfo{author}{Lin, R.P.},
  \bibinfo{author}{Turin, P.}, \bibinfo{author}{Curtis, D.W.},
  \bibinfo{author}{Primbsch, J.H.}, \bibinfo{author}{Campbell, R.D.},
  \bibinfo{author}{Abiad, R.}, \bibinfo{author}{Schroeder, P.},
  \bibinfo{author}{Cork, C.P.}, \bibinfo{author}{Hull, E.L.},
  \bibinfo{author}{Landis, D.A.}, \bibinfo{author}{Madden, N.W.},
  \bibinfo{author}{Malone, D.}, \bibinfo{author}{Pehl, R.H.},
  \bibinfo{author}{Raudorf, T.}, \bibinfo{author}{Sangsingkeow, P.},
  \bibinfo{author}{Boyle, R.}, \bibinfo{author}{Banks, I.S.},
  \bibinfo{author}{Shirey, K.}, \bibinfo{author}{Schwartz, R.},
  \bibinfo{year}{2002}.
\newblock \bibinfo{title}{{The RHESSI Spectrometer}}.
\newblock \bibinfo{journal}{Solar Physics} \bibinfo{volume}{210},
  \bibinfo{pages}{33--60}.
\bibitem[{Sommer et~al.(1994)Sommer, Bertsch, Dingus, Fichtel, Fishman,
  Harding, Hartman, Hunter, Hurley, Kanbach, Kniffen, Kouveliotou, Lin, Mattox,
  Mayer-Hasselwander, Michelson, von Montigny, Nolan, Schneid, Sreekumar and
  Thompson}]{1994ApJ...422L..63S}
\bibinfo{author}{Sommer, M.}, \bibinfo{author}{Bertsch, D.L.},
  \bibinfo{author}{Dingus, B.L.}, \bibinfo{author}{Fichtel, C.E.},
  \bibinfo{author}{Fishman, G.J.}, \bibinfo{author}{Harding, A.K.},
  \bibinfo{author}{Hartman, R.C.}, \bibinfo{author}{Hunter, S.D.},
  \bibinfo{author}{Hurley, K.C.}, \bibinfo{author}{Kanbach, G.},
  \bibinfo{author}{Kniffen, D.A.}, \bibinfo{author}{Kouveliotou, C.},
  \bibinfo{author}{Lin, Y.C.}, \bibinfo{author}{Mattox, J.R.},
  \bibinfo{author}{Mayer-Hasselwander, H.A.}, \bibinfo{author}{Michelson,
  P.F.}, \bibinfo{author}{von Montigny, C.}, \bibinfo{author}{Nolan, P.L.},
  \bibinfo{author}{Schneid, E.}, \bibinfo{author}{Sreekumar, P.},
  \bibinfo{author}{Thompson, D.J.}, \bibinfo{year}{1994}.
\newblock \bibinfo{title}{{High-energy gamma rays from the intense 1993 January
  31 gamma-ray burst}}.
\newblock \bibinfo{journal}{Astrophysical Journal} \bibinfo{volume}{422},
  \bibinfo{pages}{L63--L66}.
\bibitem[{Sturner et~al.(2003)Sturner, Shrader, Weidenspointner, Teegarden,
  Atti{\'e}, Cordier, Diehl, Ferguson, Jean, von Kienlin, Paul, S{\'a}nchez,
  Schanne, Sizun, Skinner and Wunderer}]{2003A&A...411L..81S}
\bibinfo{author}{Sturner, S.J.}, \bibinfo{author}{Shrader, C.R.},
  \bibinfo{author}{Weidenspointner, G.}, \bibinfo{author}{Teegarden, B.J.},
  \bibinfo{author}{Atti{\'e}, D.}, \bibinfo{author}{Cordier, B.},
  \bibinfo{author}{Diehl, R.}, \bibinfo{author}{Ferguson, C.},
  \bibinfo{author}{Jean, P.}, \bibinfo{author}{von Kienlin, A.},
  \bibinfo{author}{Paul, P.}, \bibinfo{author}{S{\'a}nchez, F.},
  \bibinfo{author}{Schanne, S.}, \bibinfo{author}{Sizun, P.},
  \bibinfo{author}{Skinner, G.}, \bibinfo{author}{Wunderer, C.B.},
  \bibinfo{year}{2003}.
\newblock \bibinfo{title}{{Monte Carlo simulations and generation of the SPI
  response}}.
\newblock \bibinfo{journal}{Astronomy and Astrophysics} \bibinfo{volume}{411},
  \bibinfo{pages}{L81--L84}.
\bibitem[{Suarez-Garcia et~al.(2010)Suarez-Garcia, Haas, Hajdas, Lamanna,
  Lechanoine-Leluc, Marcinkowski, Orsi, Pohl, Produit, Rapin and
  Vialle}]{2010NIMPA.624..624S}
\bibinfo{author}{Suarez-Garcia, E.}, \bibinfo{author}{Haas, D.},
  \bibinfo{author}{Hajdas, W.}, \bibinfo{author}{Lamanna, G.},
  \bibinfo{author}{Lechanoine-Leluc, C.}, \bibinfo{author}{Marcinkowski, R.},
  \bibinfo{author}{Orsi, S.}, \bibinfo{author}{Pohl, M.},
  \bibinfo{author}{Produit, N.}, \bibinfo{author}{Rapin, D.},
  \bibinfo{author}{Vialle, J.P.}, \bibinfo{year}{2010}.
\newblock \bibinfo{title}{{A method to localize gamma-ray bursts using POLAR}}.
\newblock \bibinfo{journal}{Nuclear Instruments and Methods in Physics Research
  Section A} \bibinfo{volume}{624}, \bibinfo{pages}{624--634}.
\bibitem[{Toma et~al.(2009)Toma, Sakamoto, Zhang, Hill, McConnell, Bloser,
  Yamazaki, Ioka and Nakamura}]{Toma:2009fv}
\bibinfo{author}{Toma, K.}, \bibinfo{author}{Sakamoto, T.},
  \bibinfo{author}{Zhang, B.}, \bibinfo{author}{Hill, J.E.},
  \bibinfo{author}{McConnell, M.L.}, \bibinfo{author}{Bloser, P.F.},
  \bibinfo{author}{Yamazaki, R.}, \bibinfo{author}{Ioka, K.},
  \bibinfo{author}{Nakamura, T.}, \bibinfo{year}{2009}.
\newblock \bibinfo{title}{{Statistical Properties of Gamma-Ray Burst
  Polarization}}.
\newblock \bibinfo{journal}{Astrophysical Journal} \bibinfo{volume}{698},
  \bibinfo{pages}{1042--1053}.
\bibitem[{Tsuda et~al.(2013)Tsuda, Mori, Funase, Sawada, Yamamoto, Saiki, Endo,
  Yonekura, Hoshino and Kawaguchi}]{2013AcAau..82..183T}
\bibinfo{author}{Tsuda, Y.}, \bibinfo{author}{Mori, O.},
  \bibinfo{author}{Funase, R.}, \bibinfo{author}{Sawada, H.},
  \bibinfo{author}{Yamamoto, T.}, \bibinfo{author}{Saiki, T.},
  \bibinfo{author}{Endo, T.}, \bibinfo{author}{Yonekura, K.},
  \bibinfo{author}{Hoshino, H.}, \bibinfo{author}{Kawaguchi, J.},
  \bibinfo{year}{2013}.
\newblock \bibinfo{title}{{Achievement of IKAROS {\textemdash} Japanese deep
  space solar sail demonstration mission}}.
\newblock \bibinfo{journal}{Acta Astronautica} \bibinfo{volume}{82},
  \bibinfo{pages}{183--188}.
\bibitem[{Ubertini et~al.(2003)Ubertini, Lebrun, Di~Cocco, Bazzano, Bird,
  Broenstad, Goldwurm, La~Rosa, Labanti, Laurent, Mirabel, Quadrini, Ramsey,
  Reglero, Sabau, Sacco, Staubert, Vigroux, Weisskopf and
  Zdziarski}]{2003A&A...411L.131U}
\bibinfo{author}{Ubertini, P.}, \bibinfo{author}{Lebrun, F.},
  \bibinfo{author}{Di~Cocco, G.}, \bibinfo{author}{Bazzano, A.},
  \bibinfo{author}{Bird, A.J.}, \bibinfo{author}{Broenstad, K.},
  \bibinfo{author}{Goldwurm, A.}, \bibinfo{author}{La~Rosa, G.},
  \bibinfo{author}{Labanti, C.}, \bibinfo{author}{Laurent, P.},
  \bibinfo{author}{Mirabel, I.F.}, \bibinfo{author}{Quadrini, E.M.},
  \bibinfo{author}{Ramsey, B.}, \bibinfo{author}{Reglero, V.},
  \bibinfo{author}{Sabau, L.}, \bibinfo{author}{Sacco, B.},
  \bibinfo{author}{Staubert, R.}, \bibinfo{author}{Vigroux, L.},
  \bibinfo{author}{Weisskopf, M.C.}, \bibinfo{author}{Zdziarski, A.A.},
  \bibinfo{year}{2003}.
\newblock \bibinfo{title}{{IBIS: The Imager on-board INTEGRAL}}.
\newblock \bibinfo{journal}{Astronomy and Astrophysics} \bibinfo{volume}{411},
  \bibinfo{pages}{L131--L139}.
\bibitem[{Vadawale et~al.(2016)Vadawale, Chattopadhyay, Mithun, Rao,
  Bhattacharya and Bhalerao}]{2016GCN..19011...1V}
\bibinfo{author}{Vadawale, S.V.}, \bibinfo{author}{Chattopadhyay, T.},
  \bibinfo{author}{Mithun, N.P.S.}, \bibinfo{author}{Rao, A.R.},
  \bibinfo{author}{Bhattacharya, D.}, \bibinfo{author}{Bhalerao, V.},
  \bibinfo{year}{2016}.
\newblock \bibinfo{title}{{GRB160131A: detection of polarisation by Astrosat
  CZTI.}}
\newblock \bibinfo{journal}{GCN Circular} \bibinfo{volume}{19011}.
\bibitem[{Vadawale et~al.(2015)Vadawale, Chattopadhyay, Rao, Bhattacharya,
  Bhalerao, Vagshette, Pawar and Sreekumar}]{Vadawale:2015bk}
\bibinfo{author}{Vadawale, S.V.}, \bibinfo{author}{Chattopadhyay, T.},
  \bibinfo{author}{Rao, A.R.}, \bibinfo{author}{Bhattacharya, D.},
  \bibinfo{author}{Bhalerao, V.B.}, \bibinfo{author}{Vagshette, N.},
  \bibinfo{author}{Pawar, P.}, \bibinfo{author}{Sreekumar, S.},
  \bibinfo{year}{2015}.
\newblock \bibinfo{title}{{Hard X-ray polarimetry with Astrosat-CZTI}}.
\newblock \bibinfo{journal}{Astronomy and Astrophysics} \bibinfo{volume}{578},
  \bibinfo{pages}{A73}.
\bibitem[{Vedrenne et~al.(2003)Vedrenne, Roques, Sch{\"o}nfelder, Mandrou,
  Lichti, von Kienlin, Cordier, Schanne, Kn{\"o}dlseder, Skinner, Jean,
  S{\'a}nchez, Caraveo, Teegarden, von Ballmoos, Bouchet, Paul, Matteson,
  Boggs, Wunderer, Leleux, Weidenspointner, Durouchoux, Diehl, Strong,
  Cass{\'e}, Clair and Andr{\'e}}]{2003A&A...411L..63V}
\bibinfo{author}{Vedrenne, G.}, \bibinfo{author}{Roques, J.P.},
  \bibinfo{author}{Sch{\"o}nfelder, V.}, \bibinfo{author}{Mandrou, P.},
  \bibinfo{author}{Lichti, G.G.}, \bibinfo{author}{von Kienlin, A.},
  \bibinfo{author}{Cordier, B.}, \bibinfo{author}{Schanne, S.},
  \bibinfo{author}{Kn{\"o}dlseder, J.}, \bibinfo{author}{Skinner, G.},
  \bibinfo{author}{Jean, P.}, \bibinfo{author}{S{\'a}nchez, F.},
  \bibinfo{author}{Caraveo, P.A.}, \bibinfo{author}{Teegarden, B.},
  \bibinfo{author}{von Ballmoos, P.}, \bibinfo{author}{Bouchet, L.},
  \bibinfo{author}{Paul, P.}, \bibinfo{author}{Matteson, J.L.},
  \bibinfo{author}{Boggs, S.E.}, \bibinfo{author}{Wunderer, C.},
  \bibinfo{author}{Leleux, P.}, \bibinfo{author}{Weidenspointner, G.},
  \bibinfo{author}{Durouchoux, P.}, \bibinfo{author}{Diehl, R.},
  \bibinfo{author}{Strong, A.W.}, \bibinfo{author}{Cass{\'e}, M.},
  \bibinfo{author}{Clair, M.A.}, \bibinfo{author}{Andr{\'e}, Y.},
  \bibinfo{year}{2003}.
\newblock \bibinfo{title}{{SPI: The spectrometer aboard INTEGRAL}}.
\newblock \bibinfo{journal}{Astronomy and Astrophysics} \bibinfo{volume}{411},
  \bibinfo{pages}{L63--L70}.
\bibitem[{Waxman(2003)}]{2003Natur.423..388W}
\bibinfo{author}{Waxman, E.}, \bibinfo{year}{2003}.
\newblock \bibinfo{title}{{Astronomy: New direction for $\gamma$-rays}}.
\newblock \bibinfo{journal}{Nature} \bibinfo{volume}{423},
  \bibinfo{pages}{388--389}.
\bibitem[{Weisskopf et~al.(2010)Weisskopf, Elsner and
  O'dell}]{2010SPIE.7732E..11W}
\bibinfo{author}{Weisskopf, M.C.}, \bibinfo{author}{Elsner, R.F.},
  \bibinfo{author}{O'dell, S.L.}, \bibinfo{year}{2010}.
\newblock \bibinfo{title}{{On understanding the figures of merit for detection
  and measurement of x-ray polarization}}.
\newblock \bibinfo{journal}{Proceedings of SPIE} \bibinfo{volume}{7732},
  \bibinfo{pages}{11}.
\bibitem[{Wigger et~al.(2004)Wigger, Hajdas, Arzner, G{\"u}del and
  Zehnder}]{2004ApJ...613.1088W}
\bibinfo{author}{Wigger, C.}, \bibinfo{author}{Hajdas, W.},
  \bibinfo{author}{Arzner, K.}, \bibinfo{author}{G{\"u}del, M.},
  \bibinfo{author}{Zehnder, A.}, \bibinfo{year}{2004}.
\newblock \bibinfo{title}{{Gamma-Ray Burst Polarization: Limits from RHESSI
  Measurements}}.
\newblock \bibinfo{journal}{Astrophysical Journal} \bibinfo{volume}{613},
  \bibinfo{pages}{1088--1100}.
\bibitem[{Willis et~al.(2005)Willis, Barlow, Bird, Clark, Dean, McConnell,
  Moran, Shaw and Sguera}]{2005A&A...439..245W}
\bibinfo{author}{Willis, D.R.}, \bibinfo{author}{Barlow, E.J.},
  \bibinfo{author}{Bird, A.J.}, \bibinfo{author}{Clark, D.J.},
  \bibinfo{author}{Dean, A.J.J.}, \bibinfo{author}{McConnell, M.L.},
  \bibinfo{author}{Moran, L.}, \bibinfo{author}{Shaw, S.E.},
  \bibinfo{author}{Sguera, V.}, \bibinfo{year}{2005}.
\newblock \bibinfo{title}{{Evidence of polarisation in the prompt gamma-ray
  emission from GRB 930131 and GRB 960924}}.
\newblock \bibinfo{journal}{Astronomy and Astrophysics} \bibinfo{volume}{439},
  \bibinfo{pages}{245--253}.
\bibitem[{Xiao et~al.(2016)Xiao, Hajdas, Wu, Produit, Bao, Batsch, Cadoux,
  Chai, Dong, Kong, Kong, Rybka, Leluc, Li, Liu, Liu, Marcinkowski, Paniccia,
  Pohl, Rapin, Shi, Song, Sun, Szabelski, Wang, Wen, Xu, Zhang, Zhang, Zhang,
  Zhang, Zhang and Zwolinska}]{2016APh....83....6X}
\bibinfo{author}{Xiao, H.}, \bibinfo{author}{Hajdas, W.}, \bibinfo{author}{Wu,
  B.}, \bibinfo{author}{Produit, N.}, \bibinfo{author}{Bao, T.},
  \bibinfo{author}{Batsch, T.}, \bibinfo{author}{Cadoux, F.},
  \bibinfo{author}{Chai, J.}, \bibinfo{author}{Dong, Y.},
  \bibinfo{author}{Kong, M.}, \bibinfo{author}{Kong, S.},
  \bibinfo{author}{Rybka, D.K.}, \bibinfo{author}{Leluc, C.},
  \bibinfo{author}{Li, L.}, \bibinfo{author}{Liu, J.}, \bibinfo{author}{Liu,
  X.}, \bibinfo{author}{Marcinkowski, R.}, \bibinfo{author}{Paniccia, M.},
  \bibinfo{author}{Pohl, M.}, \bibinfo{author}{Rapin, D.},
  \bibinfo{author}{Shi, H.}, \bibinfo{author}{Song, L.}, \bibinfo{author}{Sun,
  J.}, \bibinfo{author}{Szabelski, J.}, \bibinfo{author}{Wang, R.},
  \bibinfo{author}{Wen, X.}, \bibinfo{author}{Xu, H.}, \bibinfo{author}{Zhang,
  L.}, \bibinfo{author}{Zhang, L.}, \bibinfo{author}{Zhang, S.},
  \bibinfo{author}{Zhang, X.}, \bibinfo{author}{Zhang, Y.},
  \bibinfo{author}{Zwolinska, A.}, \bibinfo{year}{2016}.
\newblock \bibinfo{title}{{A crosstalk and non-uniformity correction method for
  the space-borne Compton polarimeter POLAR}}.
\newblock \bibinfo{journal}{Astroparticle Physics} \bibinfo{volume}{83},
  \bibinfo{pages}{6--12}.
\bibitem[{Yatsu et~al.(2011)Yatsu, Enomoto, Kawakami, Tokoyoda, Toizumi, Kawai,
  Ishizaka, Matsunaga, Nakamori, Kataoka and Kubo}]{2011SPIE.8145E..08Y}
\bibinfo{author}{Yatsu, Y.}, \bibinfo{author}{Enomoto, T.},
  \bibinfo{author}{Kawakami, K.}, \bibinfo{author}{Tokoyoda, K.},
  \bibinfo{author}{Toizumi, T.}, \bibinfo{author}{Kawai, N.},
  \bibinfo{author}{Ishizaka, K.}, \bibinfo{author}{Matsunaga, S.},
  \bibinfo{author}{Nakamori, T.}, \bibinfo{author}{Kataoka, J.},
  \bibinfo{author}{Kubo, S.}, \bibinfo{year}{2011}.
\newblock \bibinfo{title}{{Development micro-satellite TSUBAME for polarimetry
  of gamma-ray bursts}}, in: \bibinfo{editor}{Siegmund, O.H.} (Ed.),
  \bibinfo{booktitle}{Proceedings of the SPIE}, \bibinfo{publisher}{SPIE}. pp.
  \bibinfo{pages}{814508--814508--11}.
\bibitem[{Yatsu et~al.(2014)Yatsu, Ito, Kurita, Arimoto, Kawai, Matsushita,
  Kawajiri, Kitamura, Matunaga, Kimura, Kataoka, Nakamori and
  Kubo}]{2014SPIE.9144E..0LY}
\bibinfo{author}{Yatsu, Y.}, \bibinfo{author}{Ito, K.},
  \bibinfo{author}{Kurita, S.}, \bibinfo{author}{Arimoto, M.},
  \bibinfo{author}{Kawai, N.}, \bibinfo{author}{Matsushita, M.},
  \bibinfo{author}{Kawajiri, S.}, \bibinfo{author}{Kitamura, S.},
  \bibinfo{author}{Matunaga, S.}, \bibinfo{author}{Kimura, S.},
  \bibinfo{author}{Kataoka, J.}, \bibinfo{author}{Nakamori, T.},
  \bibinfo{author}{Kubo, S.}, \bibinfo{year}{2014}.
\newblock \bibinfo{title}{{Pre-flight performance of a micro-satellite TSUBAME
  for X-ray polarimetry of gamma-ray bursts}}.
\newblock \bibinfo{journal}{Proceedings of SPIE} \bibinfo{volume}{9144},
  \bibinfo{pages}{91440L}.
\bibitem[{Yonetoku et~al.(2011a)Yonetoku, Murakami, Gunji, Mihara, Sakashita,
  Morihara, Kikuchi, Takahashi, Fujimoto, Toukairin, Kodama, Kubo and
  Team}]{2011PASJ...63..625Y}
\bibinfo{author}{Yonetoku, D.}, \bibinfo{author}{Murakami, T.},
  \bibinfo{author}{Gunji, S.}, \bibinfo{author}{Mihara, T.},
  \bibinfo{author}{Sakashita, T.}, \bibinfo{author}{Morihara, Y.},
  \bibinfo{author}{Kikuchi, Y.}, \bibinfo{author}{Takahashi, T.},
  \bibinfo{author}{Fujimoto, H.}, \bibinfo{author}{Toukairin, N.},
  \bibinfo{author}{Kodama, Y.}, \bibinfo{author}{Kubo, S.},
  \bibinfo{author}{Team, I.D.}, \bibinfo{year}{2011}a.
\newblock \bibinfo{title}{{Gamma-Ray Burst Polarimeter (GAP) aboard the Small
  Solar Power Sail Demonstrator IKAROS}}.
\newblock \bibinfo{journal}{Publications of the Astronomical Society of Japan}
  \bibinfo{volume}{63}, \bibinfo{pages}{625--638}.
\bibitem[{Yonetoku et~al.(2012)Yonetoku, Murakami, Gunji, Mihara, Toma,
  Morihara, Takahashi, Wakashima, Yonemochi, Sakashita, Toukairin, Fujimoto and
  Kodama}]{2012ApJ...758L...1Y}
\bibinfo{author}{Yonetoku, D.}, \bibinfo{author}{Murakami, T.},
  \bibinfo{author}{Gunji, S.}, \bibinfo{author}{Mihara, T.},
  \bibinfo{author}{Toma, K.}, \bibinfo{author}{Morihara, Y.},
  \bibinfo{author}{Takahashi, T.}, \bibinfo{author}{Wakashima, Y.},
  \bibinfo{author}{Yonemochi, H.}, \bibinfo{author}{Sakashita, T.},
  \bibinfo{author}{Toukairin, N.}, \bibinfo{author}{Fujimoto, H.},
  \bibinfo{author}{Kodama, Y.}, \bibinfo{year}{2012}.
\newblock \bibinfo{title}{{Magnetic Structures in Gamma-Ray Burst Jets Probed
  by Gamma-Ray Polarization}}.
\newblock \bibinfo{journal}{The Astrophysical Journal Letters}
  \bibinfo{volume}{758}, \bibinfo{pages}{L1}.
\bibitem[{Yonetoku et~al.(2011b)Yonetoku, Murakami, Gunji, Mihara, Toma,
  Sakashita, Morihara, Takahashi, Toukairin, Fujimoto, Kodama, Kubo and
  Team}]{2011ApJ...743L..30Y}
\bibinfo{author}{Yonetoku, D.}, \bibinfo{author}{Murakami, T.},
  \bibinfo{author}{Gunji, S.}, \bibinfo{author}{Mihara, T.},
  \bibinfo{author}{Toma, K.}, \bibinfo{author}{Sakashita, T.},
  \bibinfo{author}{Morihara, Y.}, \bibinfo{author}{Takahashi, T.},
  \bibinfo{author}{Toukairin, N.}, \bibinfo{author}{Fujimoto, H.},
  \bibinfo{author}{Kodama, Y.}, \bibinfo{author}{Kubo, S.},
  \bibinfo{author}{Team, I.D.}, \bibinfo{year}{2011}b.
\newblock \bibinfo{title}{{Detection of Gamma-Ray Polarization in Prompt
  Emission of GRB 100826A}}.
\newblock \bibinfo{journal}{The Astrophysical Journal Letters}
  \bibinfo{volume}{743}, \bibinfo{pages}{L30}.
\bibitem[{Zhang and Yan(2011)}]{2011ApJ...726...90Z}
\bibinfo{author}{Zhang, B.}, \bibinfo{author}{Yan, H.}, \bibinfo{year}{2011}.
\newblock \bibinfo{title}{{The Internal-collision-induced Magnetic Reconnection
  and Turbulence (ICMART) Model of Gamma-ray Bursts}}.
\newblock \bibinfo{journal}{Astrophysical Journal} \bibinfo{volume}{726},
  \bibinfo{pages}{90}.

\end{thebibliography}

\end{document}